\documentclass[10pt]{article}
\pdfoutput=1
\usepackage[left=2.54cm,top=2.54cm,right=2.54cm,bottom=2.54cm]{geometry}
\usepackage{fancyhdr}
\usepackage{afterpage}
\usepackage{setspace}
\usepackage{bibspacing}
\usepackage{amsmath}
\usepackage{float}
\usepackage{graphycx}
\usepackage{caption}
\usepackage{subcaption}
\newfont{\toto}{msbm10 at 12 pt}
\newfont{\ithd}{cmr9}

\usepackage{amsmath,amsthm,amsfonts,amssymb}
\usepackage[pdftex]{graphicx}
\usepackage[T1]{fontenc}

\title{
\bf A-posteriori diffusion analysis of higher-order 
numerical schemes for application to propagating linear waves }
\author{
	S.~M.~Joshi\footnote{email: smjoshi@aero.iitb.ac.in} ~\& A.~Chatterjee\\\\
Department of Aerospace Engineering\\
Indian Institute of Technology, Bombay\\
Mumbai 400076, India}

\date{}

\begin{document}

\maketitle
\afterpage{\fancyhead{}}

\centerline{
\begin{minipage}[t]{150mm}
{\bf Abstract:} 
We propose a new technique for 
a-posteriori diffusion analysis of numerical schemes. 
The scalar linear advection equation
with a broadband signal as initial conditions 
is numerically solved to simulate a traveling linear wave.
A diffusion analysis is performed based on 
modal energy content and time evolution of total energy of the broadband signal. 
Onset of numerical instability can also be detected at an early stage using this technique.
This technique is used for analyzing modern linear and
nonlinear as well as
space-time coupled numerical schemes.
This analysis is particularly useful for analyzing numerical schemes used in
simulation of traveling linear
waves such as in computational electromagnetics (CEM) and computational
aeroacoustics(CAA).
\end{minipage}
}
\vskip0.5cm

\section{Introduction}
Simulation of traveling linear waves such as in computational aeroacoustics (CAA)
and computational electromagnetics (CEM) requires numerical schemes with high fidelity.
Higher-order schemes in space and time are commonly used for such applications. Higher-order schemes 
introduce less dispersion and diffusion errors over wider wavenumber range 
and have reduced
points-per wavelength (PPW) requirement for longer simulation time \cite{Tam2012, Zingg2000}. 
In addition to the formal order of accuracy in space and time, a more precise understanding of
the behavior of dispersion and diffusion errors becomes important 
while choosing numerical schemes for simulating traveling linear waves. 
The
dispersion-relation-preserving (DRP) scheme for example has a lower formal order of accuracy in space
compared to a standard finite difference (FD) scheme using a similar stencil \cite{Tam1993}, 
but results in smaller dispersion error in case of signals containing mainly large wavenumbers
\cite{Zingg2000}.
Nonlinear higher-order scheme such as the 
weighted essentially non oscillatory (WENO) scheme
may show different order of spatial accuracy for varying scales
due to inter-modal energy transfer resulting in reduced order of accuracy
for higher scales \cite{Fauconnier2011}.
Thus, a critical analysis of dispersion and diffusion characteristics in 
wavenumber space is an important aspect in the evolution of modern numerical schemes dealing with 
wave propagation problems.

Linear hyperbolic partial differential equations (PDEs) govern physics of traveling linear waves. 
The Fourier transform of a broadband signal projects the signal on each of the Fourier modes.
The relationship between the wavenumber ($k$) and the angular frequency ($\omega$) of a Fourier mode,
the {\emph{dispersion relation}}, depends on the governing PDE.
In case of linear hyperbolic PDEs, the phase velocity and the group velocity of a 
broadband signal are available from the dispersion relation. 
Numerical approximation of a PDE results in a modified dispersion relation 
with a modified wavenumber ($\tilde k$) 
corresponding to each wavenumber $k$. 
The dispersion and diffusion errors can in turn be found out from the modified dispersion relation.
For a spectrally accurate scheme, the modified wavenumber is identical to the actual wavenumber
of the Fourier mode. 
For simple numerical schemes such as conventional linear FD or finite volume (FV) schemes,
the modified wavenumber can be easily obtained by substituting the expression of a 
Fourier mode in the discrete
approximation of the governing PDE. 
However, such an analysis is not straightforward for nonlinear numerical schemes
or for the space-time coupled numerical schemes such as the Arbitrary DERivatives (ADER) scheme. 
Pirozzoli \cite{Pirozzoli2006} suggested an approximate dispersion relation (ADR)
analysis for shock capturing numerical schemes. In this analysis,
an approximate dispersion relation is obtained from spectral representation of a monochromatic
signal. The discrete Fourier transform (DFT) 
technique is used for obtaining frequency representation of 
the signal from the space domain. 
Suman et.al. \cite{Suman2017} have also used the DFT technique 
to compare different numerical schemes on the basis of the diffusion error.
In the ADR technique, the temporal error is minimized by using a small
time-step size. The ADR analysis is 
effective even in case of nonlinear schemes like the WENO scheme. 
However, the ADR estimates may be erroneous for high wavenumber components (`wake' of the solution
in Fourier domain) \cite{Pirozzoli2006}. 
Moreover, multiple simulations for different wavenumbers are required
in order to perform the ADR analysis and 
the analysis does not take into account the complex interactions of waves with 
different wavenumbers in nonlinear schemes\cite{Jia2015}. 
In nonlinear schemes, different modes interact with each other
and new spurious modes get created even in the cutoff wavenumber range \cite{Fauconnier2011}.
To study inter-modal energy dynamics in nonlinear numerical schemes,
a nonlinear spectral analysis was proposed \cite{Fauconnier2011,Jia2015}.
In this analysis, statistical behavior of the full spectrum of modified wavenumbers was
studied.

In the current work,
we analyze diffusion characteristics of numerical schemes based on energy of the signal
using the DFT technique.
In literature, we find examples of turbulent kinetic energy (TKE) being used as a 
measure to evaluate performance of numerical schemes for application to turbulent flows. 
For example in Ref.\cite{Modesti2017}, time evolution of TKE was used 
for comparing different solvers for application to turbulent flows. 
Similarly, time evolution as well as 
spectra of TKE and internal energy were used for studying effects of higher-order accurate
nonlinear schemes for turbulent flows \cite{Ladeinde2001}.
In this work we show that time evolution of the energy of a broadband signal can be used 
for analyzing diffusion characteristics of a numerical scheme 
for application to propagating linear waves.
The time evolution of energy of the signal in a specific wavenumber range 
depends on dissipation characteristics and stability of the underlying numerical scheme.
Such a study can also indicate onset of a numerical instability at an early stage. 
This work is a generalization of the `a posteriori analysis in wavenumber space' 
proposed earlier in  Ref.\cite{Joshi2017}.
The current analysis can provide information on numerical 
stability and conservation property of the scheme along with the ability to preserve signal energy
over a long simulation time.
Variation in ``maximum well-resolved wavenumber" with time is also studied 
for different numerical schemes.
The current analysis is well suited for linear as well as nonlinear spatial and temporal schemes
along with space-time coupled numerical schemes.

\section{A posteriori diffusion analysis}
\subsection{Fourier spectrum}
Consider the $1$D scalar advection equation,
\begin{equation}\label{adv}
	u_t + a u_x = 0, \hspace{4mm} a>0
\end{equation} 
over domain $x\in[x_l,x_u]$ with periodic boundary conditions and $u(x,0)  = u_0(x) $ as the
initial condition. The analytical solution to equation \ref{adv} is given by $u(x,t) = u_0(x-at)$.
Let the solution $u(x,t)$ be sampled at $N$ discrete equidistant points ({$x_0, x_1, ... , x_{N-1}$}) 
such that at time level $t^n$, the value at $j^{th}$ point is given as $u_j^n$. 
The solution $u(x,t)$ can be represented in the frequency domain through a discrete Fourier transform
(DFT) operation. 
For a function $f(x):{\mathbb R} \rightarrow {\mathbb R}$ sampled at $N$ points, the DFT is given as,
\begin{equation}
	\hat f(k) = \frac{1}{N} \sum_{n=0}^{N-1} f(x_n)e^{- i k n} 
\end{equation}
where $k \in {\mathbb R}$ is a wavenumber $\frac{2\pi}{\lambda}$ of a Fourier mode with wavelength $\lambda$. 
The power spectral density (PSD) of a signal is a plot of energy of a Fourier mode 
$| \hat f(k) |^2 $ for a wavenumber ($k$). The total energy of the signal is the area
under the PSD curve.
The largest wavenumber that can be resolved numerically is given as,
$ k_{max} = \frac{2\pi}{\lambda_{min}}$ 
where $\lambda_{min} = 2\Delta x$. Thus,
$ k_{max} = \frac{2\pi}{2\Delta x}, \text{ and } k_{max} \Delta x = \pi $. 
If the wavenumber range $[0,k_{max}\Delta x]$ is divided into $M$ number of points, then the smallest
non-zero wavenumber (scaled with $\Delta x$) which is well resolved is $k_{min} \Delta x= \pi/M$.
Thus,
\begin{equation}\label{kmin}
	k_{min} = \frac{\pi}{M \Delta x}
\end{equation}
Moreover, from the definition of $k_{min}$, 
$\lambda_{max} = 2\pi/k_{min}$.

\begin{figure}[h!]
	\begin{center}
	\begin{subfigure}[t]{0.45\textwidth}
		\begin{center}{\includegraphics[width=\textwidth]{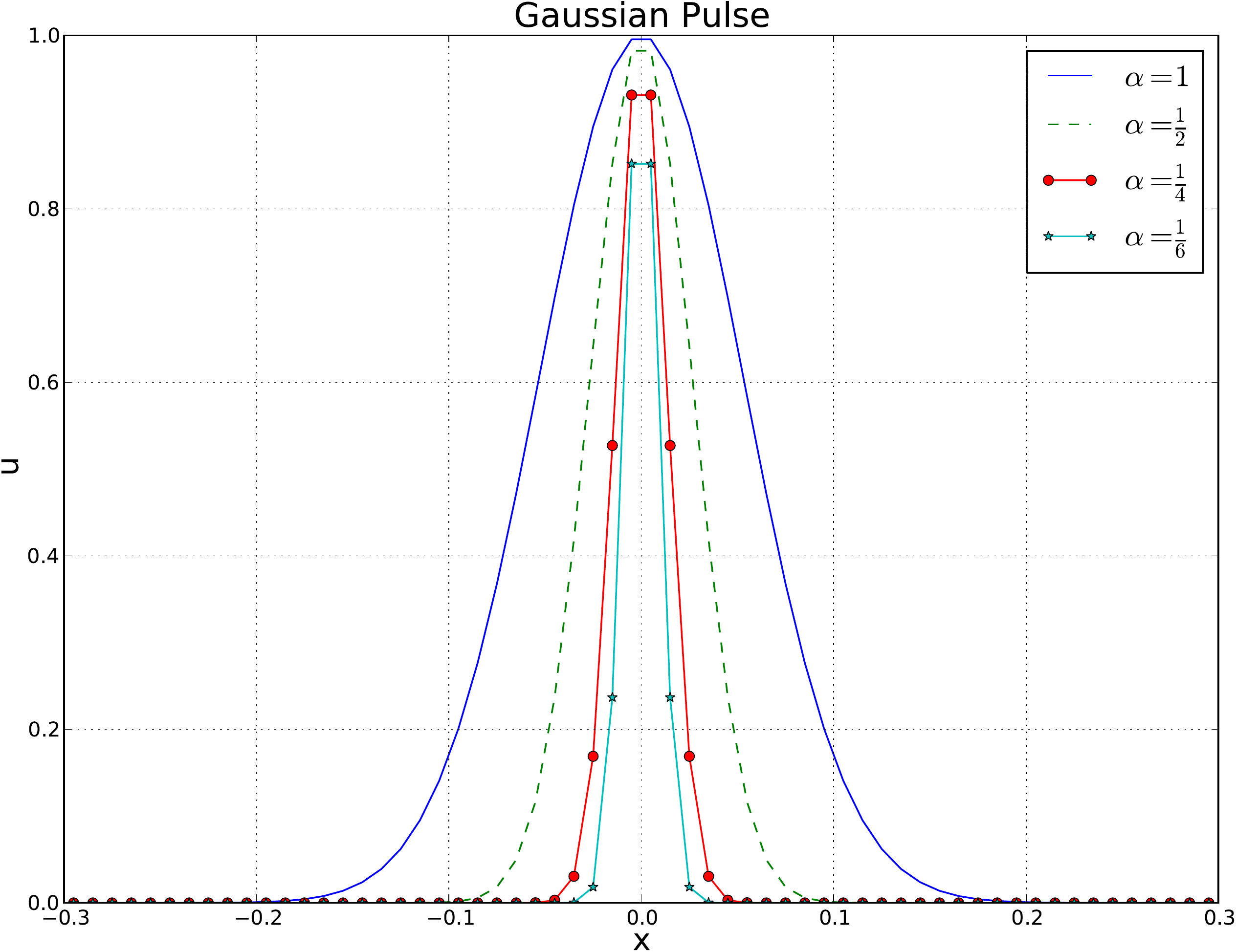}}
		\end{center}
		\caption{\label{fft1} Space domain}
	\end{subfigure}%
	\begin{subfigure}[t]{0.45\textwidth}
		\begin{center}{\includegraphics[width=\textwidth]{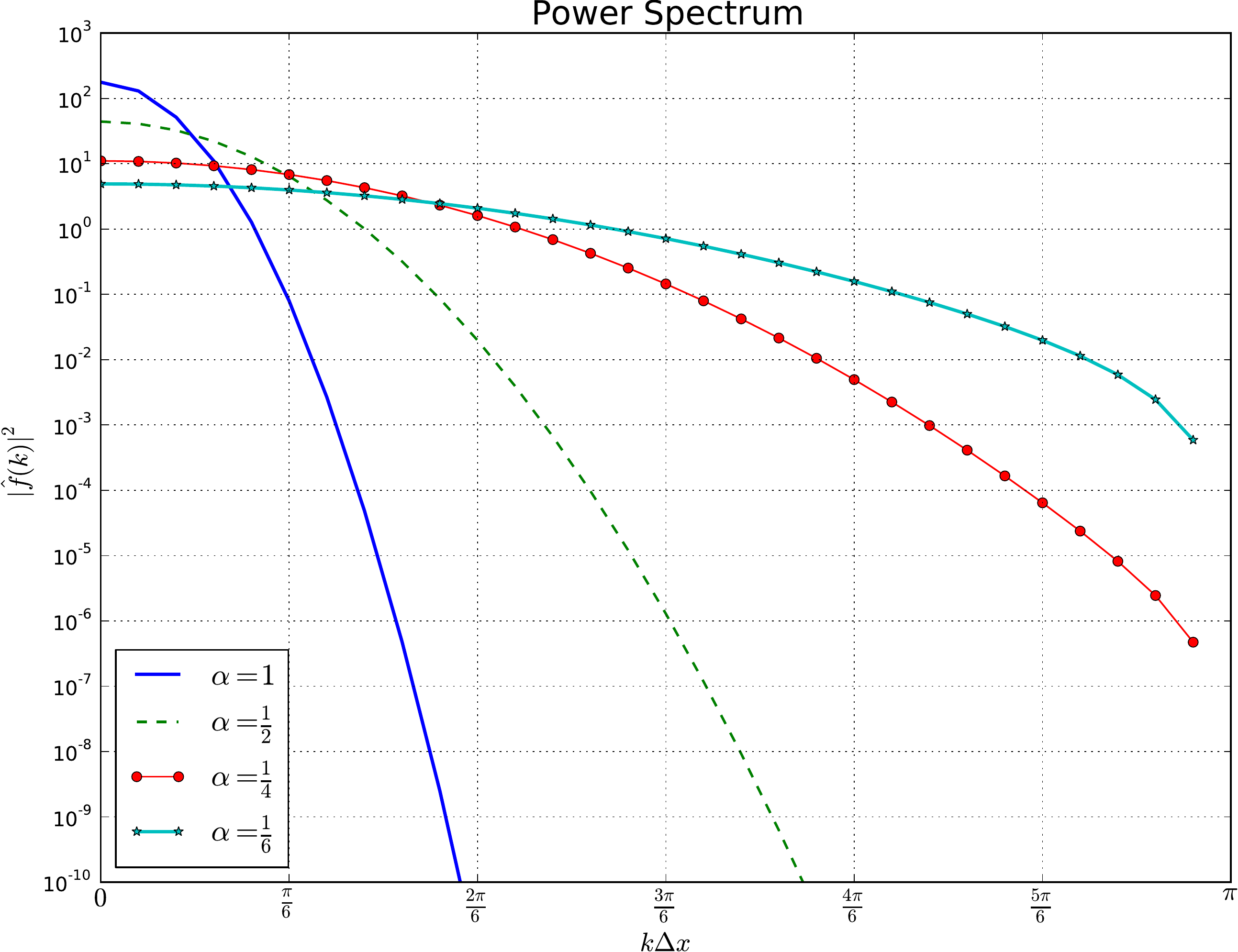}}
		\end{center}
		\caption{\label{fft2} Frequency domain }
	\end{subfigure}
	\newline
	\begin{subfigure}[t]{0.6\textwidth}
		\begin{center}{\includegraphics[width=\textwidth]{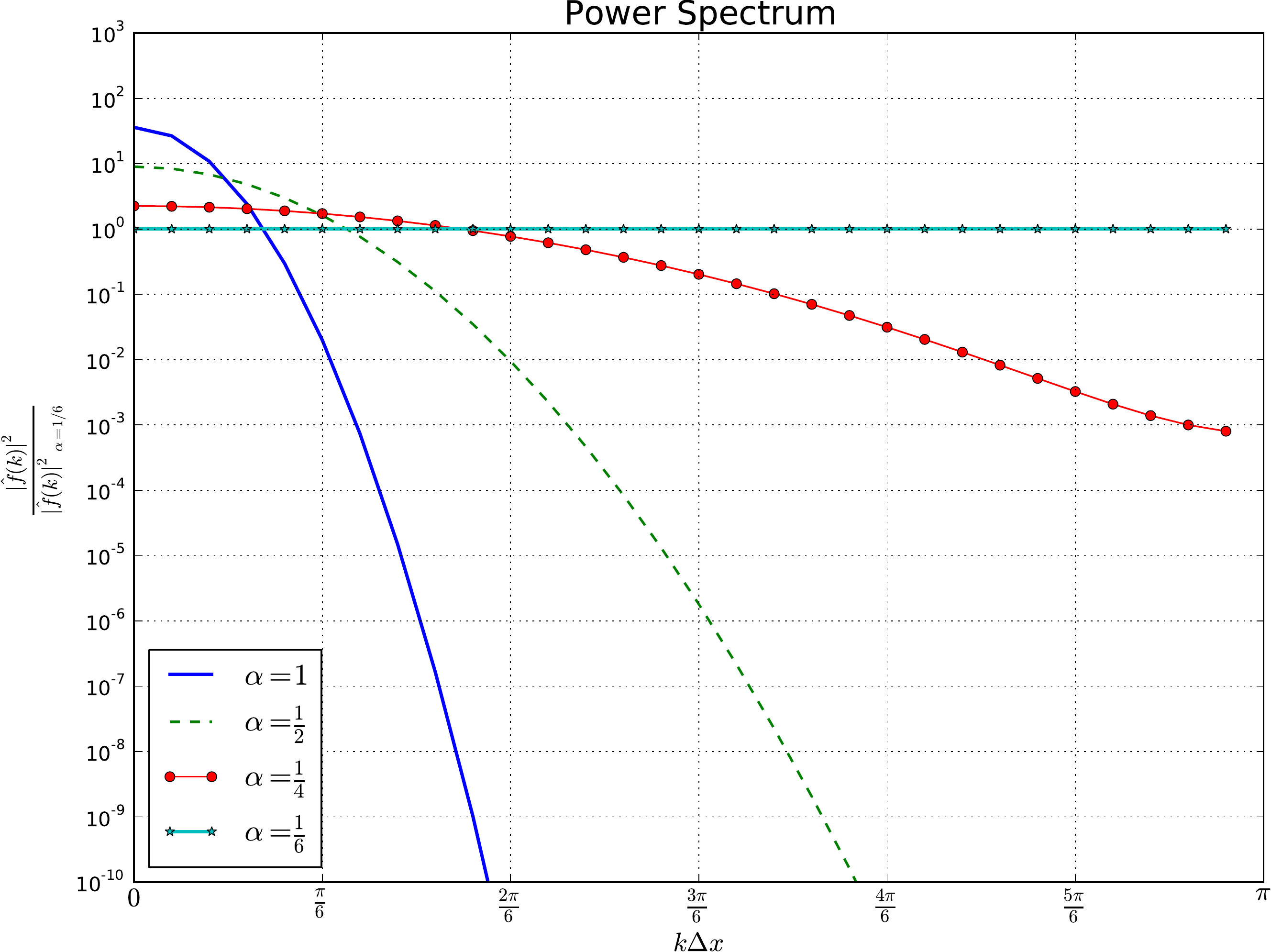}}
		\end{center}
		\caption{\label{fft3} Normalized in frequency domain }
	\end{subfigure}
	\end{center}
	\caption{\label{gauss} Gaussian signal with different values of $\alpha$ in space and
	frequency domains}
\end{figure}

$\lambda_{max}$ is the largest possible wavelength for one complete cycle of a mode  on the domain 
and is equal to the length of the domain. i.e. $\lambda_{max} = L$ where $L = x_u - x_l$.
For convenience, we can take $x_l$ and $x_u$ to be symmetrically located about the origin, 
then the domain required to resolve
$k_{min}$ turns out to be $[-\pi/k_{min} , \pi/k_{min}]$.
Based on this information, the analysis can be done as follows:

\begin{enumerate}
	\item Select the grid/element size in space domain ($\Delta x$) and number of divisions from $0$ to 
		$k_{max} \Delta x$ in frequency domain (i.e. $M : M>1$). 
		These numbers can be arbitrary and
		depend on desired resolution in the space as well as frequency domains.
		For example consider $\Delta x = 0.01$ and $M = 30$.
	\item Find out $k_{min}$ from equation \ref{kmin}. 
		This further yields values of $x_u$ and $x_l$ as discussed above.
		For the current example, $k_{min} = 10.472$ and $x_l = -0.3, \hspace{1mm} x_u = 0.3$.
	\item The total number of points in the space domain is, $N = (x_u-x_l)/\Delta x$ 
		which is the same as $N = 2M$.
		In the current example, $N=60$.
	\item On a $1$D domain $x\in[x_l,x_u]$, define broadband initial conditions given by a Gaussian
		signal 
		\begin{equation}\label{guass}
		u(x) = e^{\left(-\tfrac{8x}{\alpha L }\right)^2}
		\end{equation}
		The parameter $\alpha$ controls the spread of the Gaussian curve thereby controlling 
		energy distribution in the higher Fourier modes  
		as shown in Figures \ref{fft1} and \ref{fft2}. Note that only one part of the Fourier spectrum is shown since it is symmetrical.
	\item Solve equation \ref{adv} numerically for the final time $T$ using a candidate numerical
		scheme.
	\item Take DFT of both the analytical and the numerical results using fast Fourier transform 
		(FFT) to get
		\begin{enumerate}
			\item Power spectrum and normalized power spectrum 
				(normalized with respect to the analytical solution)
			\item Total energy of the signal and its time evolution
			\item Maximum wavenumber corresponding to a well resolved mode and 
				its variation with time
		\end{enumerate}
\end{enumerate}
The zeroth mode of the Fourier spectrum represents the average value of the signal. 
\begin{table}[h]
\begin{center}
\begin{tabular}{ | c | c | c |}
	\hline 
	\multicolumn{3}{|c|}{\bf Average value of the signal} \\ [2 ex]
	\hline 
	\rule{0pt}{4ex}$\alpha$  & Average value $\left(\frac{1}{L}\int_{x_l}^{x_u}q(x)dx\right)$ & $\frac{\hat q(0)}{N}$ \\[2 ex]
	\hline 
	1   & 2.2156 E-01 & 2.2156 E-01  \\
	1/2 & 1.1078 E-01 & 1.1078 E-01  \\
	1/4 & 5.5389 E-02 & 5.5389 E-02  \\
	1/6 & 3.6926 E-02 & 3.6926 E-02  \\
	\hline 
	\end{tabular}
	\caption{\label{avg}Average value of signals computed in space and 
	frequency domain}
\end{center}
\end{table}
Table \ref{avg} shows average value of signals computed in spatial domain and 
frequency domain. In this case, signals have different average value depending on the values of $\alpha$.
If the Zeroth mode of power spectra
of numerically simulated signal is different than the analytical value,
then the numerical scheme is not conservative in nature. 
Addition of dissipation
in the signal due to the numerical scheme 
causes the signal to become `smooth' or more diffused. 
The energy corresponding to the higher modes ($|\hat f(k)|^2$) reduces as a result.
The PSD curve corresponding to the
numerical solution would deviate from the analytical curve in such a case.
However, zeroth Fourier coefficients for numerical and 
analytical spectra will have identical values for a conservative numerical scheme. 
Fig. \ref{fft3} shows all spectra normalized with respect to the spectrum for $\alpha = \frac{1}{6}$.
An important observation from fig.\ref{fft3} is that, smoother signals (larger values of $\alpha$) 
have more energy in the lower wavenumbers whereas signals with a
sharper slope have energy distributed uniformly throughout the wavenumber range.

For an unstable scheme, there should be addition of energy in the 
energy-rich wavenumbers. However, addition of energy in any wavenumber may not 
always indicate onset of an instability. 
Nonlinearity of a numerical scheme may increase slope of the part of
a signal indicating addition of energy in higher wavenumbers. 
Nonlinear schemes such as ENO and WENO schemes are known to show 
spurious modes as well as `numerical turbulence' \cite{Jia2015,Ladeinde2001}.
For this reason, time evolution of `total energy' of the signal is
studied to identify the onset of an instability and to characterize 
numerical diffusion.
Total energy of a discrete signal $q(x): x\in\{x_0,x_1,...,x_{n-1}\}$ is defined as
\begin{equation}
E = \sum_{i=0}^{N-1} (q(x_i))^2 
\end{equation}
According to Parseval's theorem, 
\begin{equation}
\sum_{i=0}^{N-1} (q(x_i))^2 = \frac{1}{N}\sum_{k=0}^{N-1} (\hat q(k))^2
\end{equation}
where $\hat q(k)$ is the $k^{th}$ Fourier coefficient of $q(x)$.
Thus, the area under the PSD curve in the frequency domain gives the total energy of the 
signal. 
Table \ref{enrgy} shows energy of various Gaussian signals computed in spatial domain and
frequency domain. 
\begin{table}[h!]
\begin{center}
\begin{tabular}{ | c | c | c |}
	\hline 
	\multicolumn{3}{|c|}{\bf Energy of the signal} \\ [2 ex]
	\hline 
	\rule{0pt}{4ex}$\alpha$  & Space domain $\sum_{i=0}^{N-1}(q(x_i))^2$ & Frequency domain $\frac{1}{N}\sum_{i=0}^{N-1}(\hat q(k))^2$ \\[2 ex]
	\hline 
	1   & 9.3998 E-00 & 9.3998 E-00  \\
	1/2 & 4.6999 E-00 & 4.6999 E-00  \\
	1/4 & 2.3499 E-00 & 2.3499 E-00  \\
	1/6 & 1.5652 E-00 & 1.5652 E-00  \\
	\hline 
	\end{tabular}
	\caption{\label{enrgy}Energy of signals computed in space and frequency domain}
\end{center}
\end{table}

In this work, diffusion characteristics of several modern higher-order accurate 
spatial and temporal schemes are studied.  
We consider essentially non oscillatory (ENO)\cite{Shu1988}, 
weighted essentially non oscillatory (WENO)\cite{Jiang1995},
discontinuous Galerkin (DG)\cite{Cockburn1998,Cockburn1989a,Cockburn1989}, 
and fixed stencil (FS) finite volume scheme.
For semidiscrete schemes, the Runge Kutta method is used for time integration.
We study effect of time-step size on the solution for all spatial and temporal schemes.
In addition, we compare these schemes with the `arbitrary derivatives (ADER)'
finite volume scheme where space-time coupling takes place as a result of the
Cauchy-Kowalewsky procedure \cite{Titarev2002}. 
The ENO, WENO, FS and ADER schemes are considered in finite volume (FV) form.
Table \ref{schemes} lists candidate numerical schemes in the study, 

\begin{table}[h!]
\begin{center}
\begin{tabular}{ | c | c | c | c | c |}
	\hline 
	\multicolumn{5}{|c|}{\bf Numerical schemes} \\
	\hline 
	Numerical scheme  & Formulation & Linearity & Reconstruction & Type \\
	\hline 
	FS   & FV    & Linear     & Wide stencil  & Semidiscrete scheme          \\
	ENO  & FV & Non linear & Wide stencil  & Semidiscrete scheme           \\
	WENO & FV & Non linear & Wide stencil  & Semidiscrete scheme           \\
	DG   & FE    & Linear     & Subcell stencil& Semidiscrete scheme          \\
	ADER & FV & Linear and Nonlinear      & Wide stencil  & Space-time coupled scheme     \\
	RK   & ODE solver & Linear  & -     & Temporal scheme  \\ 
	\hline 
	\end{tabular}
	\caption{\label{schemes}Difference numerical schemes considered for this study}
\end{center}
\end{table}

In this study even though we focus 
mainly on modern FV and DG schemes,  in principle, this 
analysis can be performed for any numerical scheme used for simulation of traveling linear waves.

\subsection{Effect of spatial order on diffusion}

For this analysis, a Gaussian pulse with $\alpha = 1/6$ is selected as the initial condition
due to a more uniform energy distribution over the wavenumber range. 
The initial conditions are specified for $100$ 
equispaced points over $x\in[-1,1]$. The simulation is run till the Gaussian pulse
travels once round the complete domain and re-centers at the origin. A third order
Runge Kutta scheme is used for integration in time. Courant number $0.1$ is selected.
This requires $1000$ iterations in time for the signal to recenter at the origin.

\subsubsection{Fixed stencil approximation}
\begin{figure}[H]
	\begin{subfigure}[t]{0.5 \textwidth}
		\begin{center}{\includegraphics[width=\textwidth]{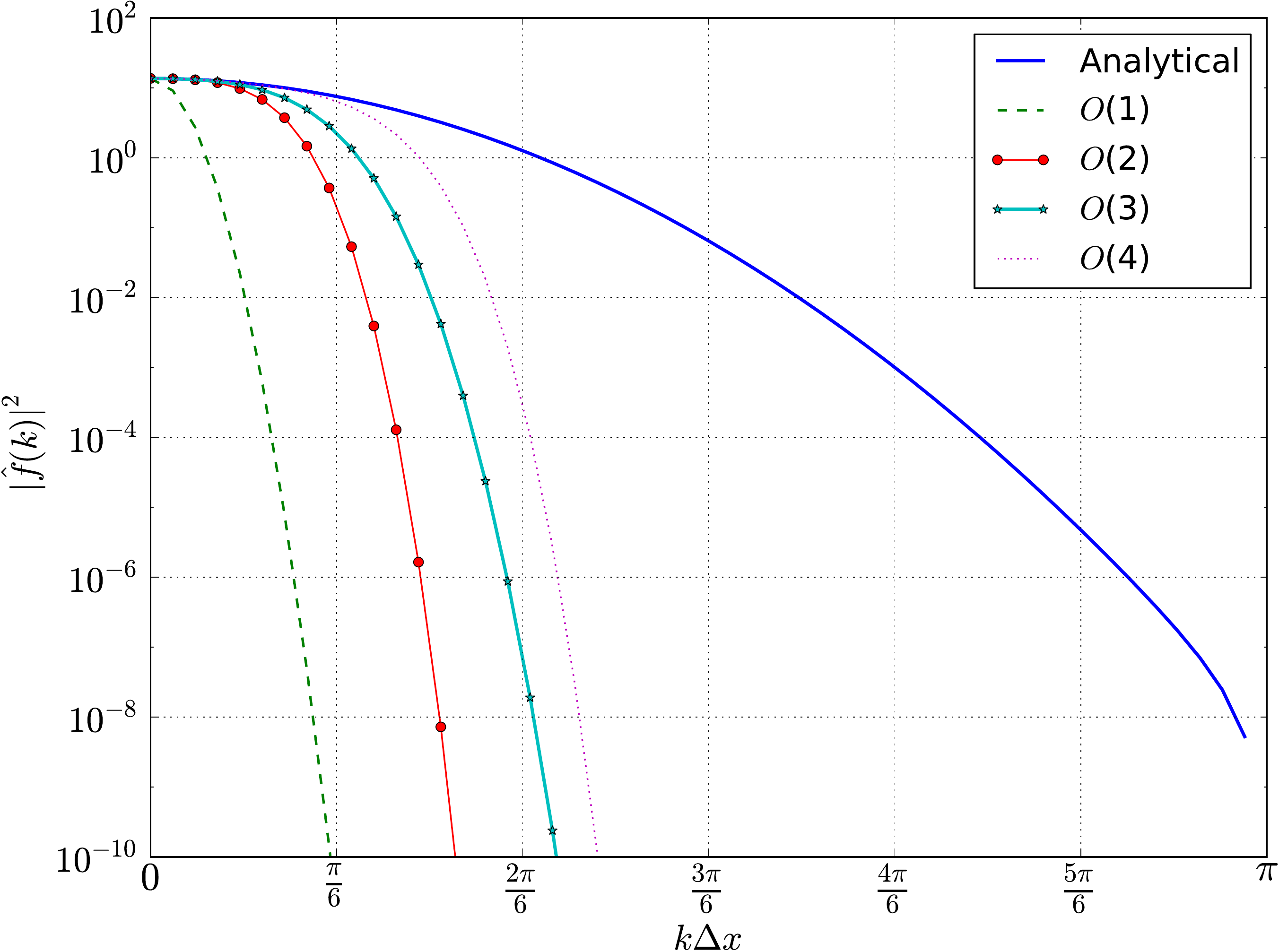}}
		\end{center}
		\caption{}
	\end{subfigure}%
	\begin{subfigure}[t]{0.5 \textwidth}
		\begin{center}{\includegraphics[width=\textwidth]{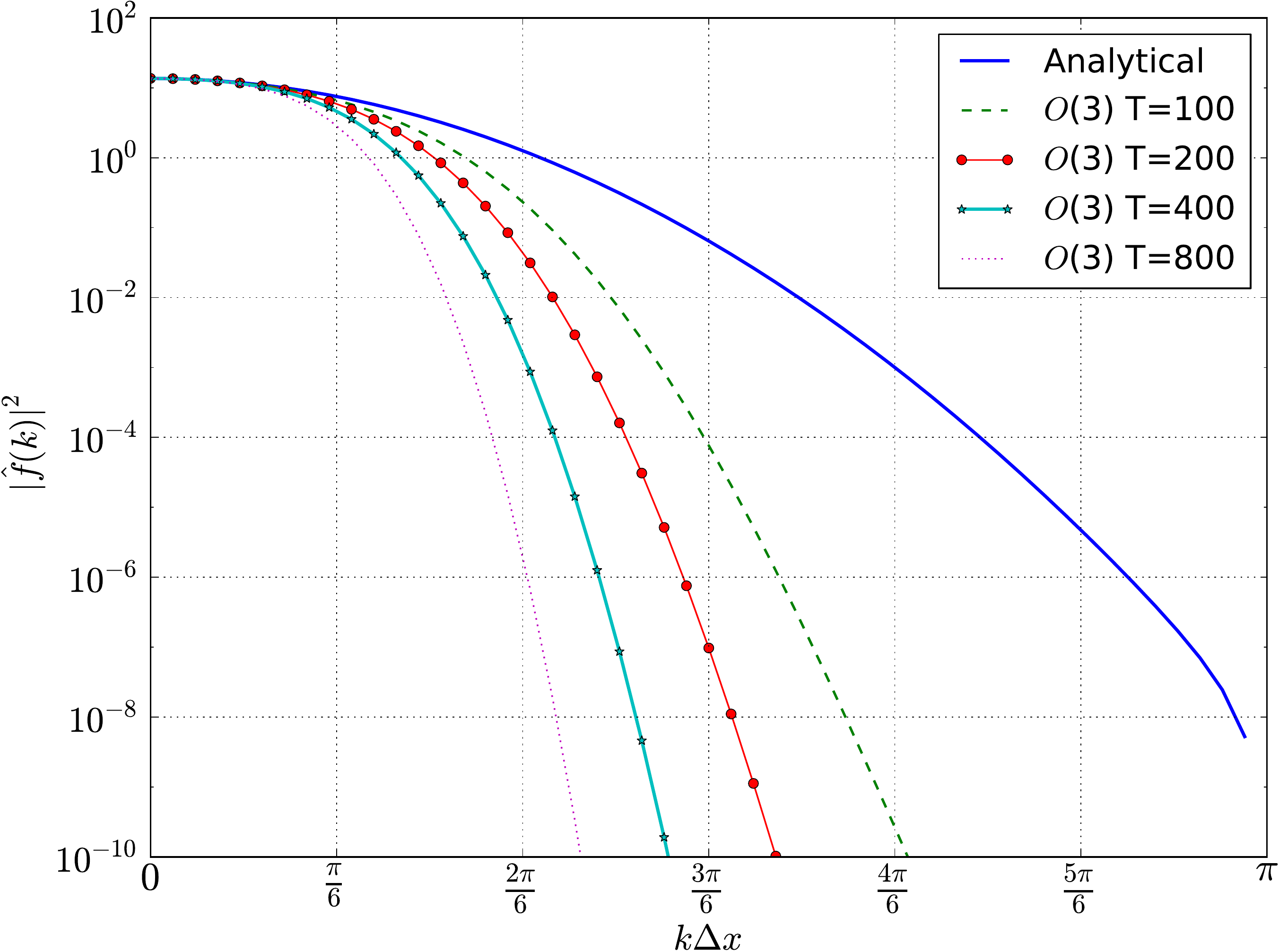}}
		\end{center}
		\caption{}
	\end{subfigure}%
	\caption{\label{freq} Power spectrum for (a) Fixed stencil finite volume schemes (b) Third order scheme over varying simulation time (shown as number of iterations)}
\end{figure}

\begin{figure}[H]
	\begin{subfigure}[t]{0.5 \textwidth}
		\begin{center}{\includegraphics[width=\textwidth]{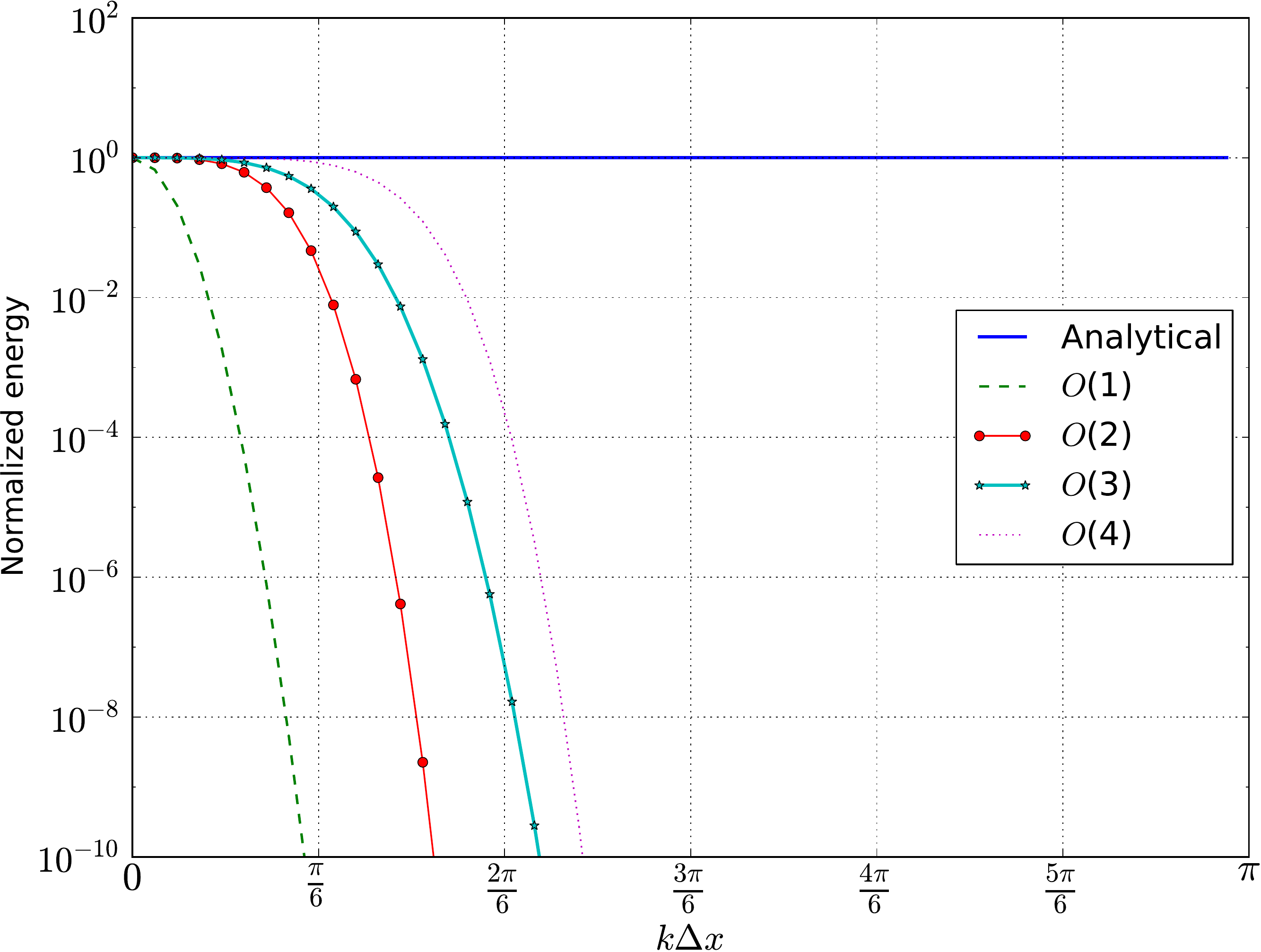}}
		\end{center}
		\caption{}
	\end{subfigure}%
	\begin{subfigure}[t]{0.5 \textwidth}
		\begin{center}{\includegraphics[width=\textwidth]{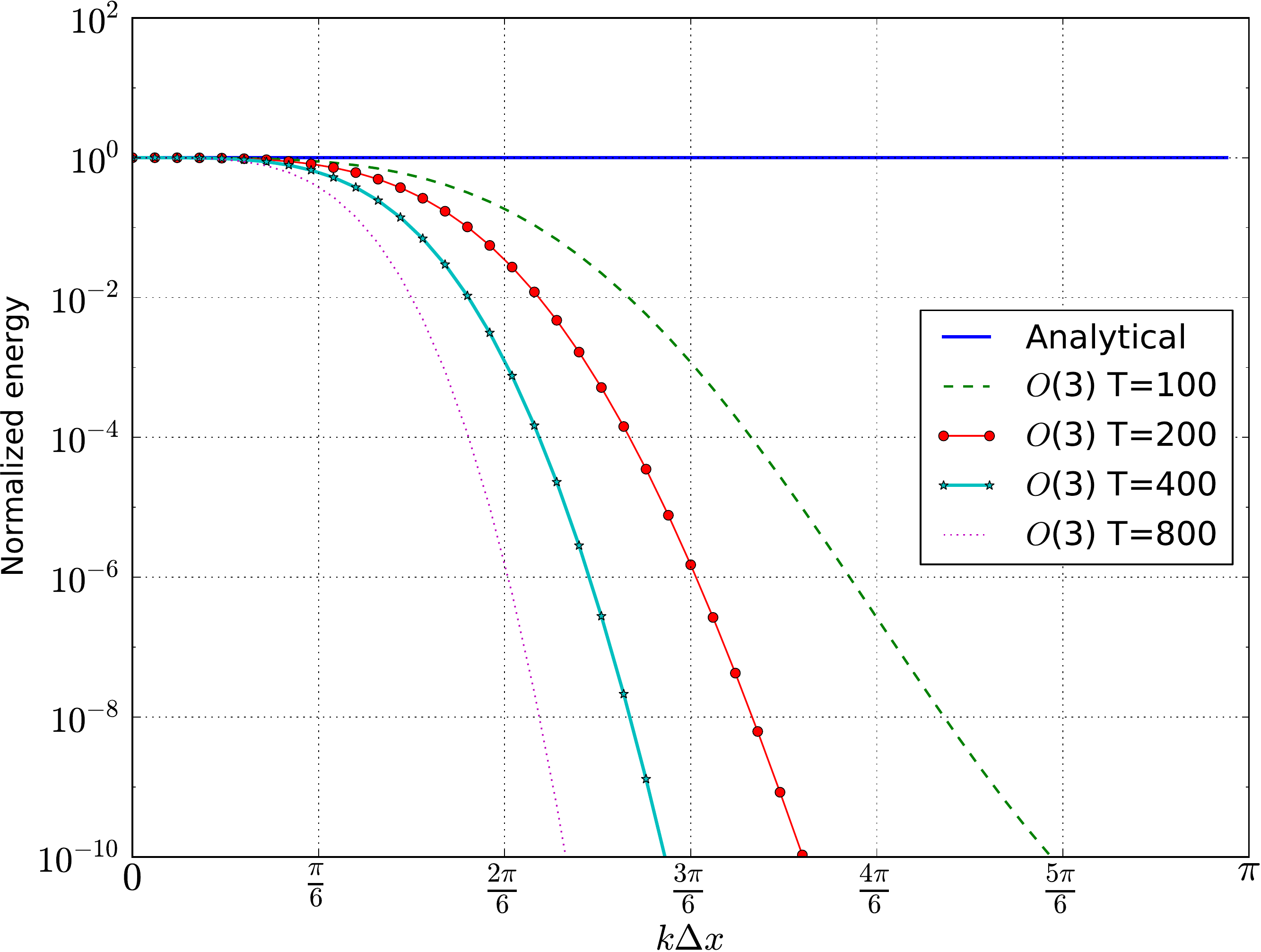}}
		\end{center}
		\caption{}
	\end{subfigure}%
	\caption{\label{freqn} Normalized power spectrum for (a) Fixed stencil finite volume schemes (b) Third order scheme over varying simulation time (shown as number of iterations)}
\end{figure}
Fig. \ref{freq}a. shows comparison of power spectra using fixed-stencil finite volume schemes of 
different spatial orders. Fig. \ref{freqn}a. shows normalized spctra. 
It is seen that, all schemes have same energy corresponding to the zeroth
Fourier mode and further this matches with the analytical value.
This indicates that all of the schemes have same average value of the signal and hence
all schemes are conservative in nature. 
The numerical PSD curve deviates from the analytical curve at a particular wavenumber. 
This wavenumber is higher in higher-order numerical schemes. Energy in lower modes matches 
exactly with analytical curve, whereas diffusion is seen in the higher modes.

Figs. \ref{freq}b. and \ref{freqn}b. show power spectra and the normalized power spectra
respectively corresponding to different simulation times for a third order
scheme. T is number of iterations in time. It is seen that as the time increases,
diffusion is progressively added in the lower wavenumbers. As seen in Figs \ref{freq}b and \ref{freqn}b, 
the energy of the signal 
decreases with time. In addition to this, the threshold wavenumber where the numerical curve
deviates from the analytical curve also shifts towards zero. 
However, the average value of the signal as given by the zeroth Fourier mode
remains fixed, indicating conservative nature over time.

\subsubsection{Discontinuous Galerkin scheme}

\begin{figure}[H]
	\begin{subfigure}[t]{0.5 \textwidth}
		\begin{center}{\includegraphics[width=\textwidth]{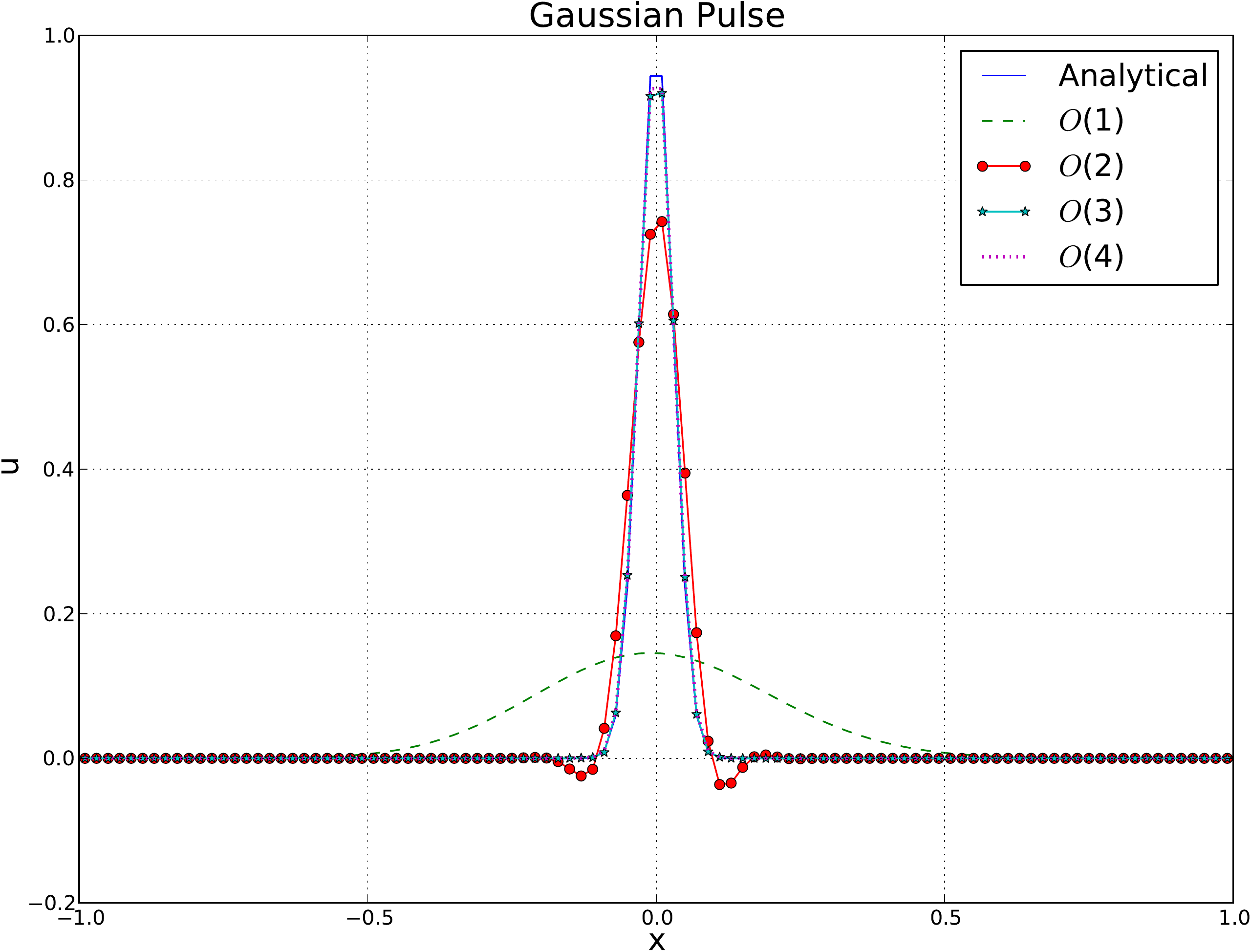}}
		\end{center}
		\caption{Time-domain signal}
	\end{subfigure}%
	\begin{subfigure}[t]{0.5 \textwidth}
		\begin{center}{\includegraphics[width=\textwidth]{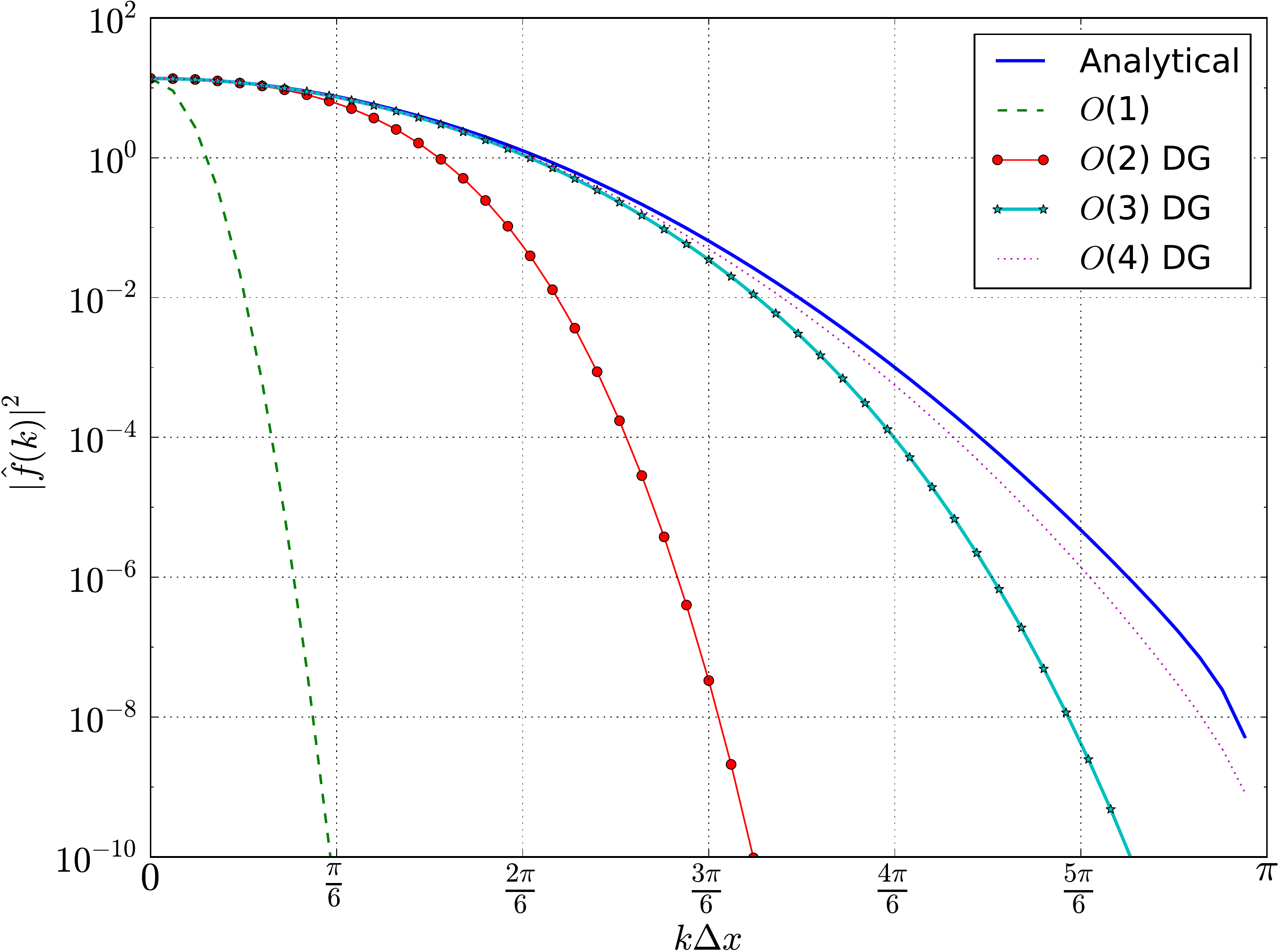}}
		\end{center}
		\caption{PSD curve}
	\end{subfigure}%
	\caption{\label{dgtdfd} Results for discontinuous Galerkin method }
\end{figure}

Fig. \ref{dgtdfd} shows results for $1$D discontinuous Galerkin (DG) method in time domain and 
in frequency domain. 
The initial conditions are implemented by $L_2$ projection of the Gaussian at the location of 
degrees of freedom (DOF). This is carried out by a weak formulation of the interpolation.
Consider the Lagrange interpolation formula for a $n^{th}$ order polynomial approximation,
\begin{equation}
	q^n(x) = \sum_{i=0}^{n-1} q(x_i) \phi_i(x)
\end{equation}
where $\phi_i(x), i=0,1,...$ are Lagrange polynomials.
The interpolation error ($e$) is given as,
\begin{equation}
	e = q(x) - q^n(x) = q(x) - \sum_{i=0}^{n-1} q(x_i) \phi_i(x)
\end{equation}
Forcing the interpolation error to be orthogonal to each one of the basis functions yields
\begin{equation}
	\int_{x_l}^{x_u}\phi_i(x)q(x)dx = \int_{x_l}^{x_u}\phi_i(x)\left(\sum_{j=0}^{n-1} q(x_j) \phi_j(x)\right) dx, \hspace{5mm} i=0,1,...,n-1
\end{equation}
Solving the integration on left hand side using a Gauss quadrature formula,
\begin{equation}
	\sum_{k=0}^{m}\phi_i(x_k)q(x_k)\omega_k = \int_{x_l}^{x_u}\phi_i(x)\left(\sum_{j=0}^{n-1} q(x_j) \phi_j(x)\right) dx, \hspace{5mm} i=0,1,...,n-1
\end{equation}
where $x_k$ are the quadrature points and $\omega_k$ are corresponding weights. 
For exact integration, $m = 2K + 1$ where $K$ is degree of the interpolating polynomial.
As $q(x)$ is known from the initial condition, value of $q(x)$ at the DOF location is obtained as,

\begin{equation}
	\bar{q} = M^{-1} \bar{I} 
\end{equation}
where,
$M_{ij} = \int_{x_l}^{x_u} \phi_i(x) \phi_j(x) dx, \hspace{2mm} \forall i,j$ is the familiar mass matrix,
$I_i = \sum_{k=0}^{m}\phi_i(x_k)q(x_k)\omega_k$ is the $i^{th}$ component of 
$\bar {I}_{n \times 1}$ and
$\bar{q}_{n \times 1}$ contains value of $q(x)$ at $n$ DOF locations.
An open source finite element library DEAL.II (\cite{dealIIa,dealIIb}) was used for DG implementation.

\begin{figure}[H]
	\begin{subfigure}[t]{0.5 \textwidth}
		\begin{center}{\includegraphics[width=\textwidth]{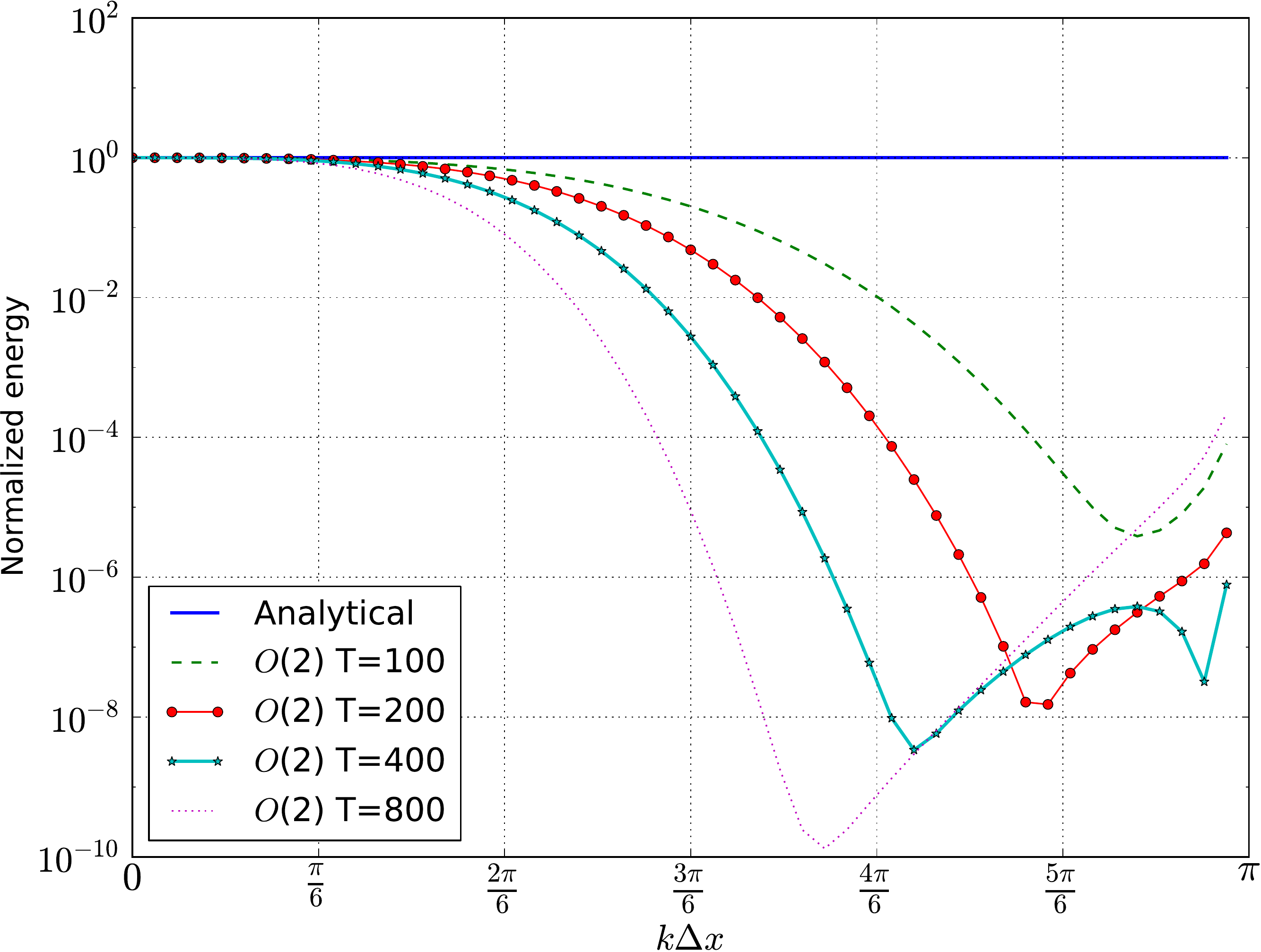}}
		\end{center}
		\caption{}
	\end{subfigure}%
	\begin{subfigure}[t]{0.5 \textwidth}
		\begin{center}{\includegraphics[width=\textwidth]{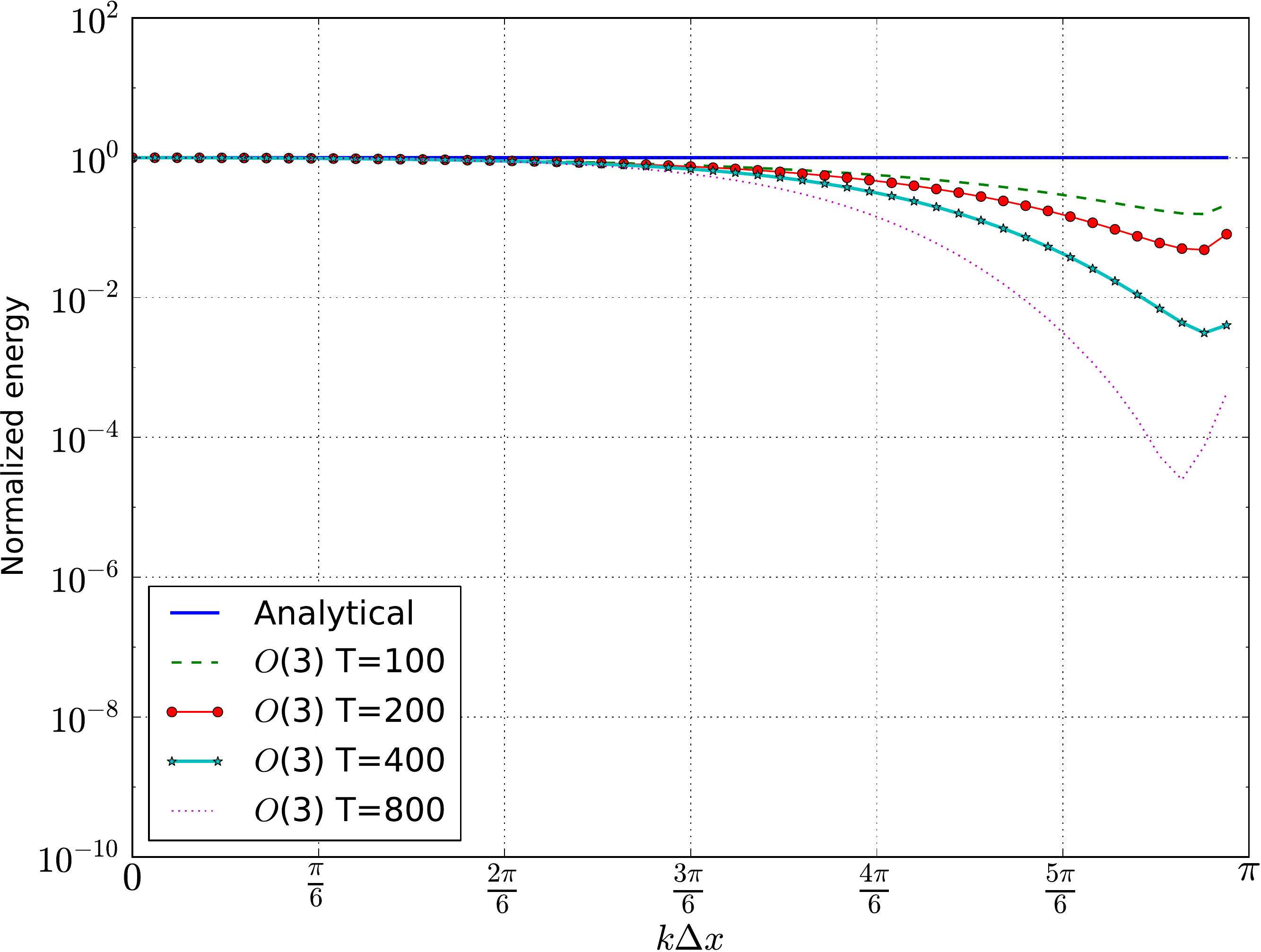}}
		\end{center}
		\caption{}
	\end{subfigure}%
	\caption{\label{dgtvsf} Variation of PSD over simulation time for a (a) second-order (b) third-order DG scheme}
\end{figure}

Figs.\ref{dgtdfd}a. and \ref{dgtdfd}b. show results in space and frequency domain respectively. 
Fig.\ref{dgtvsf} shows energy 
spectra corresponding to different time intervals.
It is observed that the scheme is conservative and higher-order
variants results in reduced diffusion at higher modes. For higher-order schemes, the reduction in 
signal energy is less compared to a lower-order scheme.
It is to be noted that the total DOFs for the scheme in the above results is order $\times$ DOF for
an equivalent higher-order finite volume scheme. This difference is accounted for by considering 
same overall DOF of all schemes for one-to-one comparison in subsequent sections.

\subsubsection{ENO and WENO schemes}

Essentially non oscillatory (ENO) and weighted essentially non oscillatory (WENO) schemes are 
popular high resolution schemes \cite{Shu1988,Jiang1995}. 
These nonlinear schemes are commonly used for simulation of shocks, traveling waves and 
turbulence \cite{Fauconnier2011}. 

Consider the $1$D domain divided in equi-spaced cells each with volume $\Delta x$.
The semidiscrete form of the governing equation at cell $i$ reads as,
\begin{equation}
	\frac{dq_i}{dt} = -\frac{1}{\Delta x}\left( F_{i+1/2}^{(W)ENO} - F_{i-1/2}^{(W)ENO} \right)
\end{equation}
In case of the ENO scheme in finite volume form, 
the numerical flux at a cell interface is obtained by using a stencil with 
smoothest data distribution. 
So, for a $N^{th}$ order ENO scheme,
\begin{equation}
F_{i+1/2}^{ENO} = f(q_{i+1/2,L}^{ENO},q_{i+1/2,R}^{ENO})
\end{equation}
where $(q_{i+1/2,L})$ is the left-interpolated value and $(q_{i+1/2,R})$ is the right-interpolated 
value of the conservative variable $q(x,t)$ at cell interface located at $x_{i+1/2}$.
The interpolation is given as,
\begin{equation}
q_{i+1/2}^{ENO} = \sum_{j=1}^N C_{rj} q_j
\end{equation}
Here $r$ is the left shift used to locate the stencil having smoothest data distribution and $C_{rj}$
is the Lagrange polynomial based on the cell averaged values.
Implementation details of the ENO scheme are widely available in literature including Ref.\cite{Shu1988}. 

\begin{figure}[ht]
	\begin{subfigure}[t]{0.48 \textwidth}
		\begin{center}{\includegraphics[width=\textwidth]{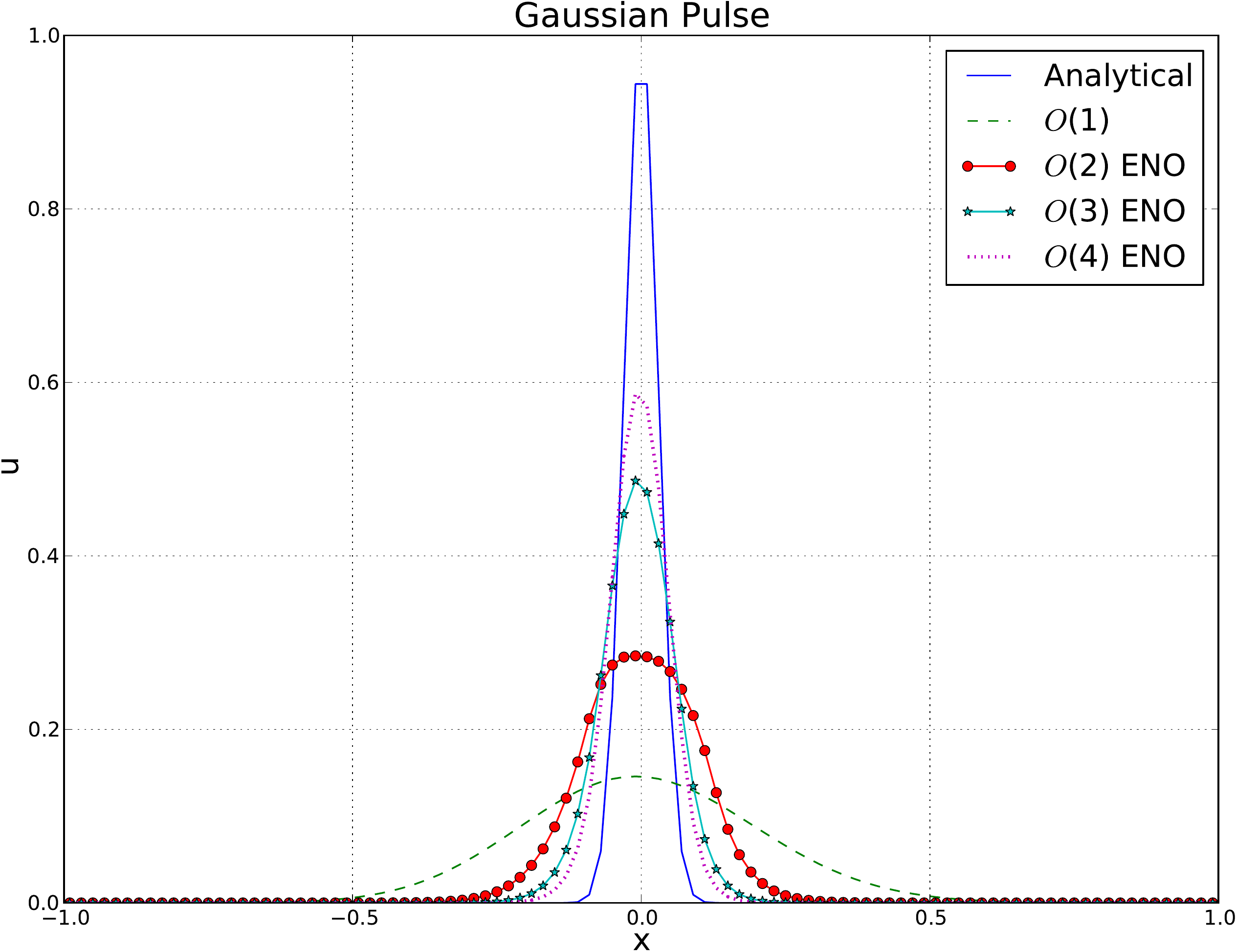}}
		\end{center}
		\caption{ENO: Time domain signal}
	\end{subfigure}%
	\begin{subfigure}[t]{0.48 \textwidth}
		\begin{center}{\includegraphics[width=\textwidth]{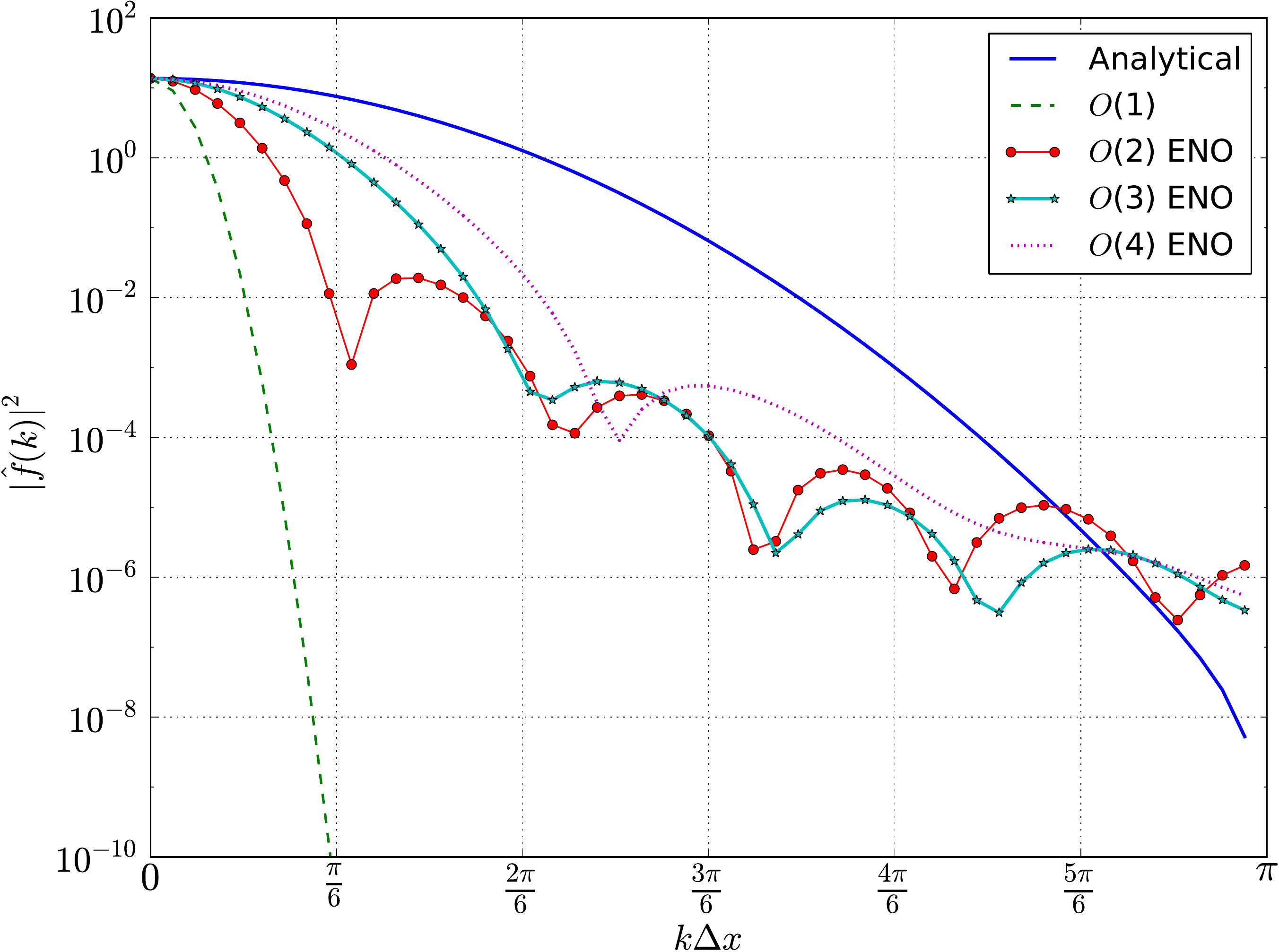}}
		\end{center}
		\caption{\label{enofd}ENO: Power spectrum}
	\end{subfigure}%
	\newline
	\begin{subfigure}[t]{0.48 \textwidth}
		\begin{center}{\includegraphics[width=\textwidth]{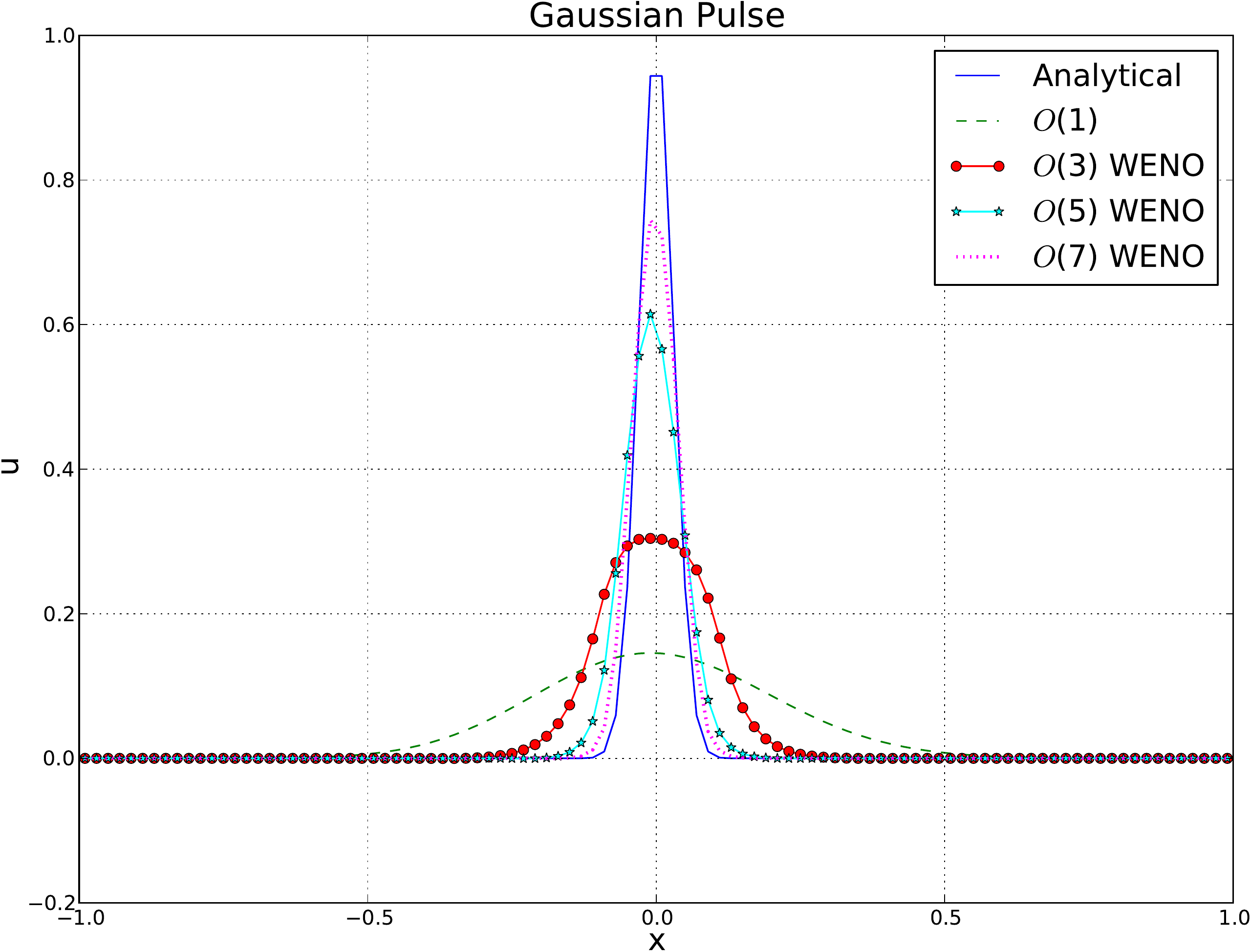}}
		\end{center}
		\caption{WENO: Time-domain signal}
	\end{subfigure}%
	\begin{subfigure}[t]{0.48 \textwidth}
		\begin{center}{\includegraphics[width=\textwidth]{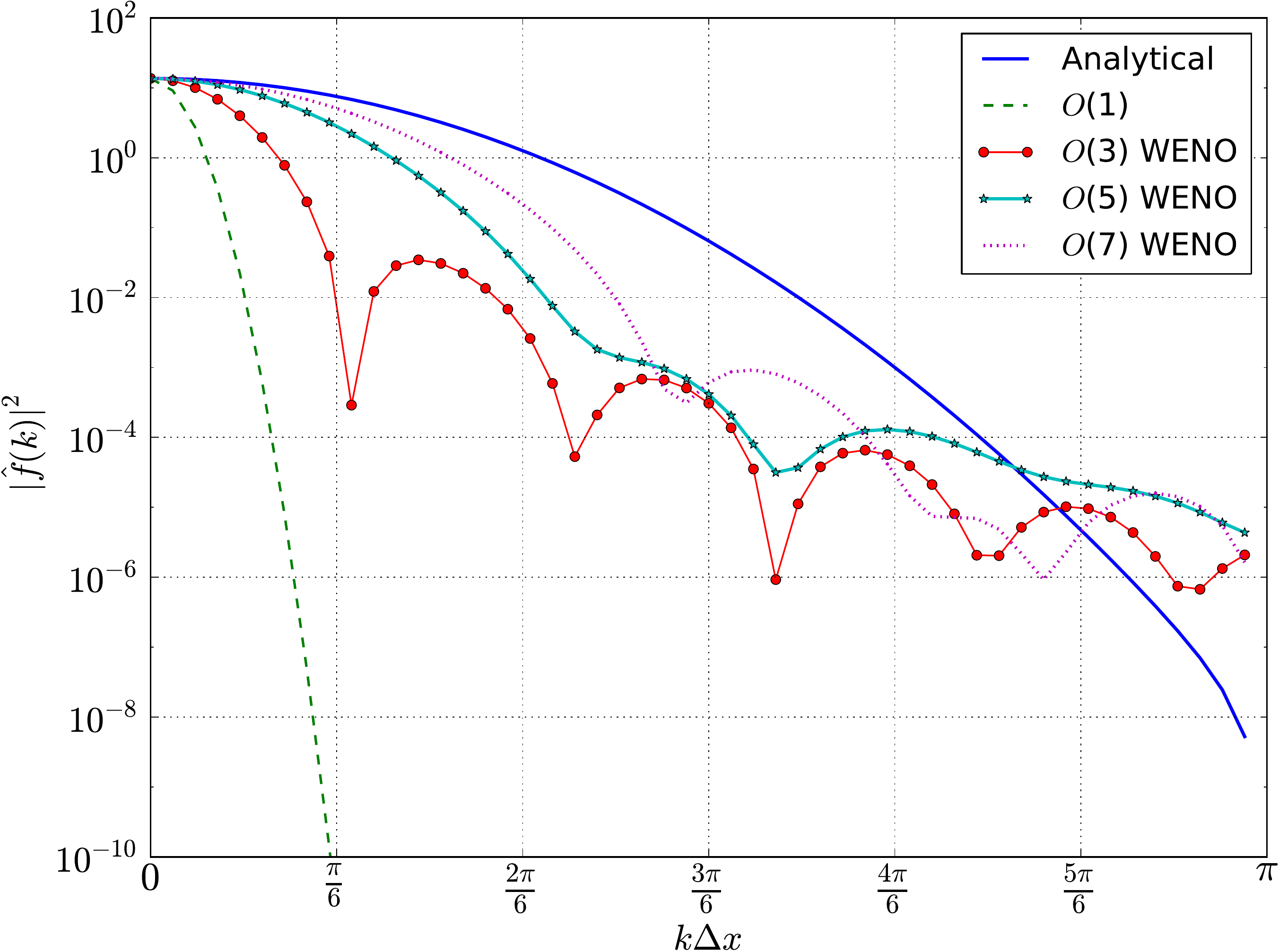}}
		\end{center}
		\caption{\label{wenofd}WENO: Power spectrum}
	\end{subfigure}%
	\caption{\label{enowenotdfd} Results for ENO and WENO scheme with varying spatial order}
\end{figure}

In the WENO scheme, a weighted sum of all possible stencils 
is used for computing fluxes. 
For a $(2N-1)^{th}$ order scheme, $N$ stencils each with  $N$ cells are used for obtaining
$N$ interpolated values at the cell interface,
\begin{equation}
q_{r} = \sum_{j=1}^N C_{rj} q_j, \hspace{4mm} r \in{0,1,...,N-1}
\end{equation}
Each stencil is associated with a nonlinear weight $\omega$ 
based on smoothness of data in that
stencil. 
For smooth data in a stencil, optimum weight is assigned. On the other hand for a stencil containing
a discontinuity, the nonlinear weight reduces to zero.
The interpolated value at the interface is a convex combination of all interpolated values
\cite{Jiang1995}.
\begin{equation}\label{wenoq}
q_{i+1/2}^{WENO} = \sum_{k=1}^N \omega_k q_k 
\end{equation}
Equation \ref{wenoq} gives left-interpolated value at face $i+\frac{1}{2}$.
A similar procedure is used for interpolation of the conserved variable 
from the right hand side of the face.
Solution of the Riemann problem, arising from the left and right interpolated values at the face,
is used for finding the numerical flux at the interface.

\begin{figure}[ht]
	\begin{subfigure}[t]{0.5 \textwidth}
		\begin{center}{\includegraphics[width=\textwidth]{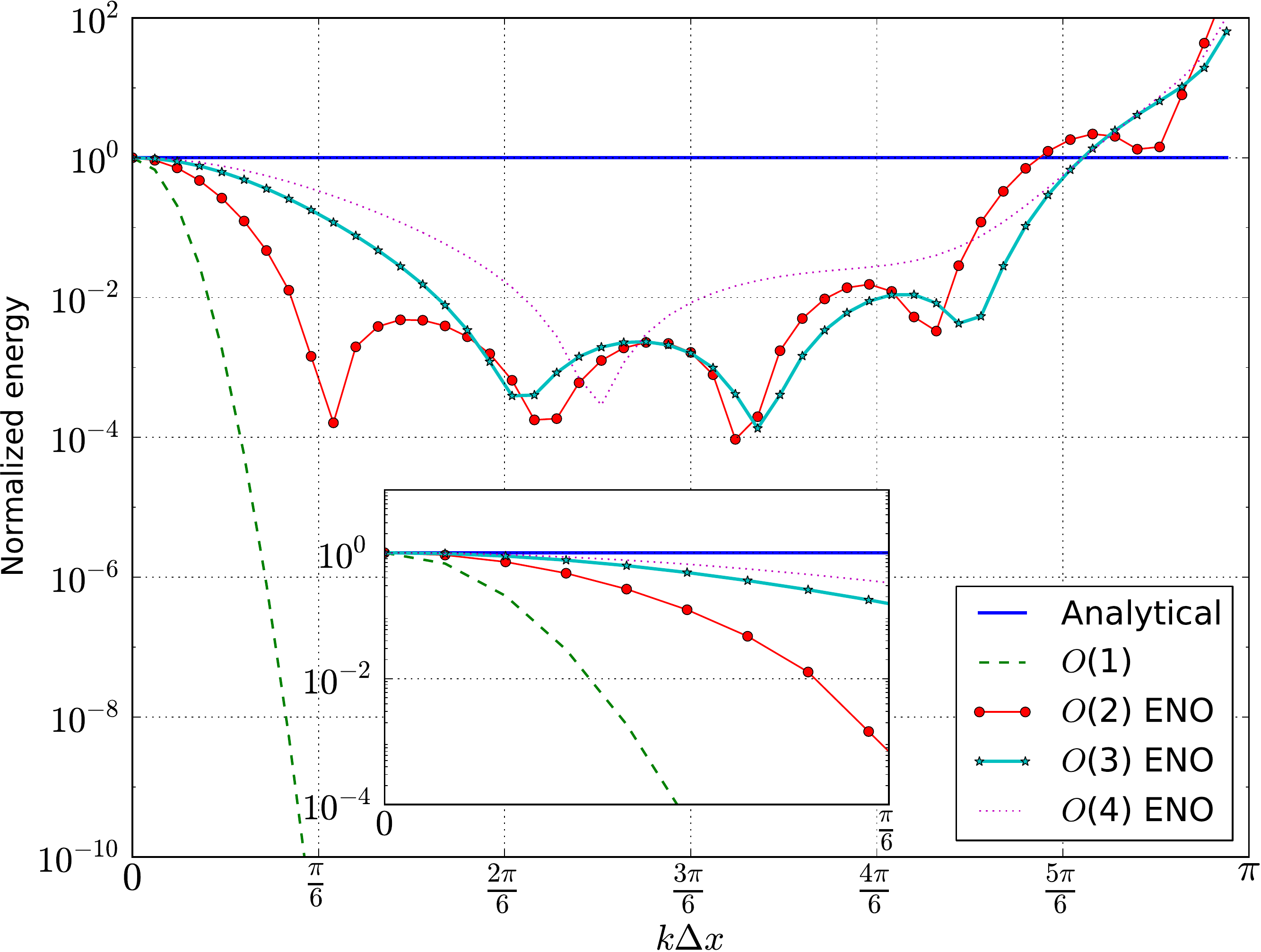}}
		\end{center}
		\caption{ENO}
	\end{subfigure}
	\begin{subfigure}[t]{0.5 \textwidth}
		\begin{center}{\includegraphics[width=\textwidth]{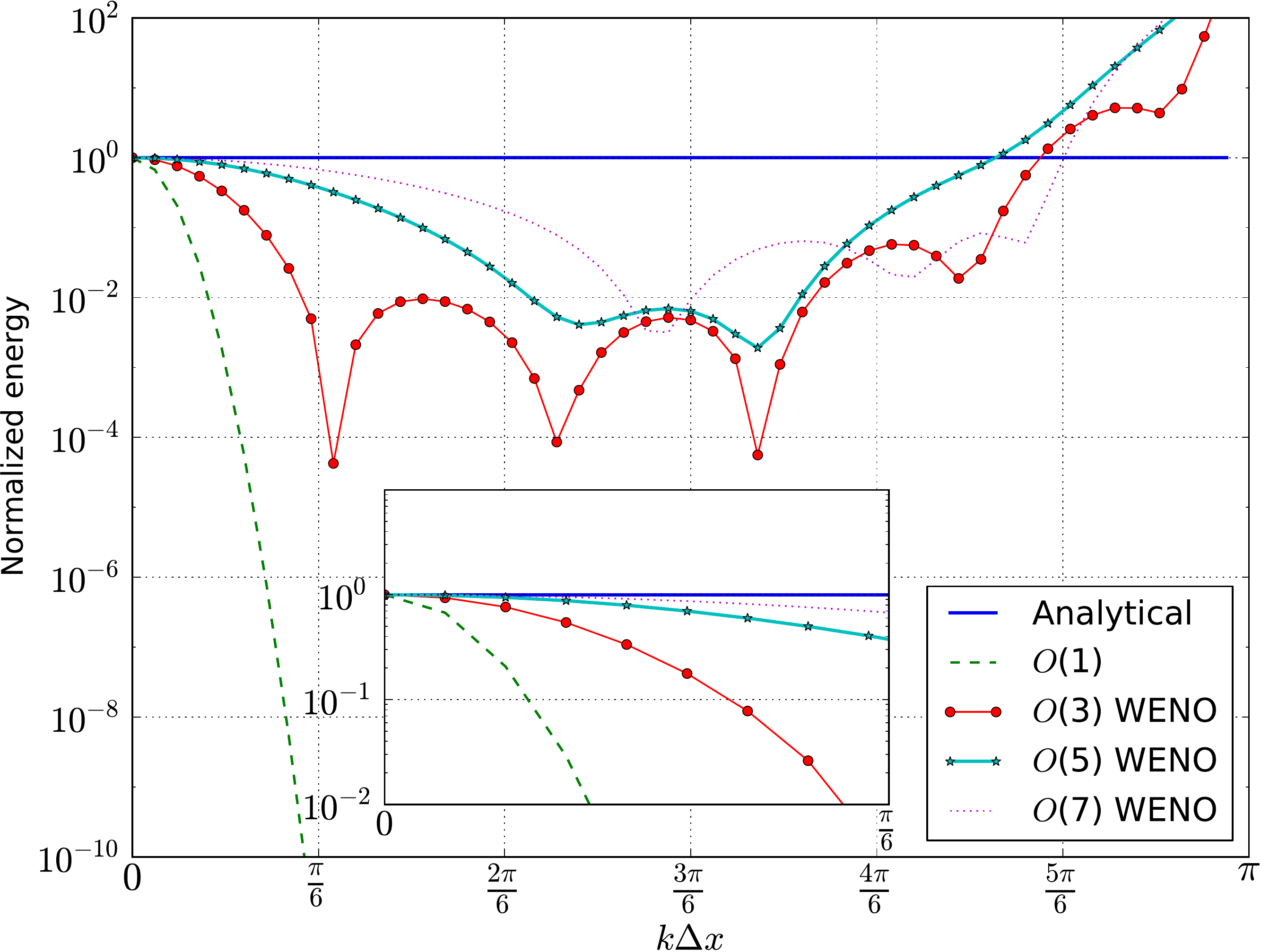}}
		\end{center}
		\caption{WENO}
	\end{subfigure}%
	\caption{\label{enowenons} Normalized PSD curve for ENO and WENO schemes}
\end{figure}

\begin{figure}[ht]
	\begin{center}
	\begin{subfigure}[t]{0.5 \textwidth}
		\begin{center}{\includegraphics[width=\textwidth]{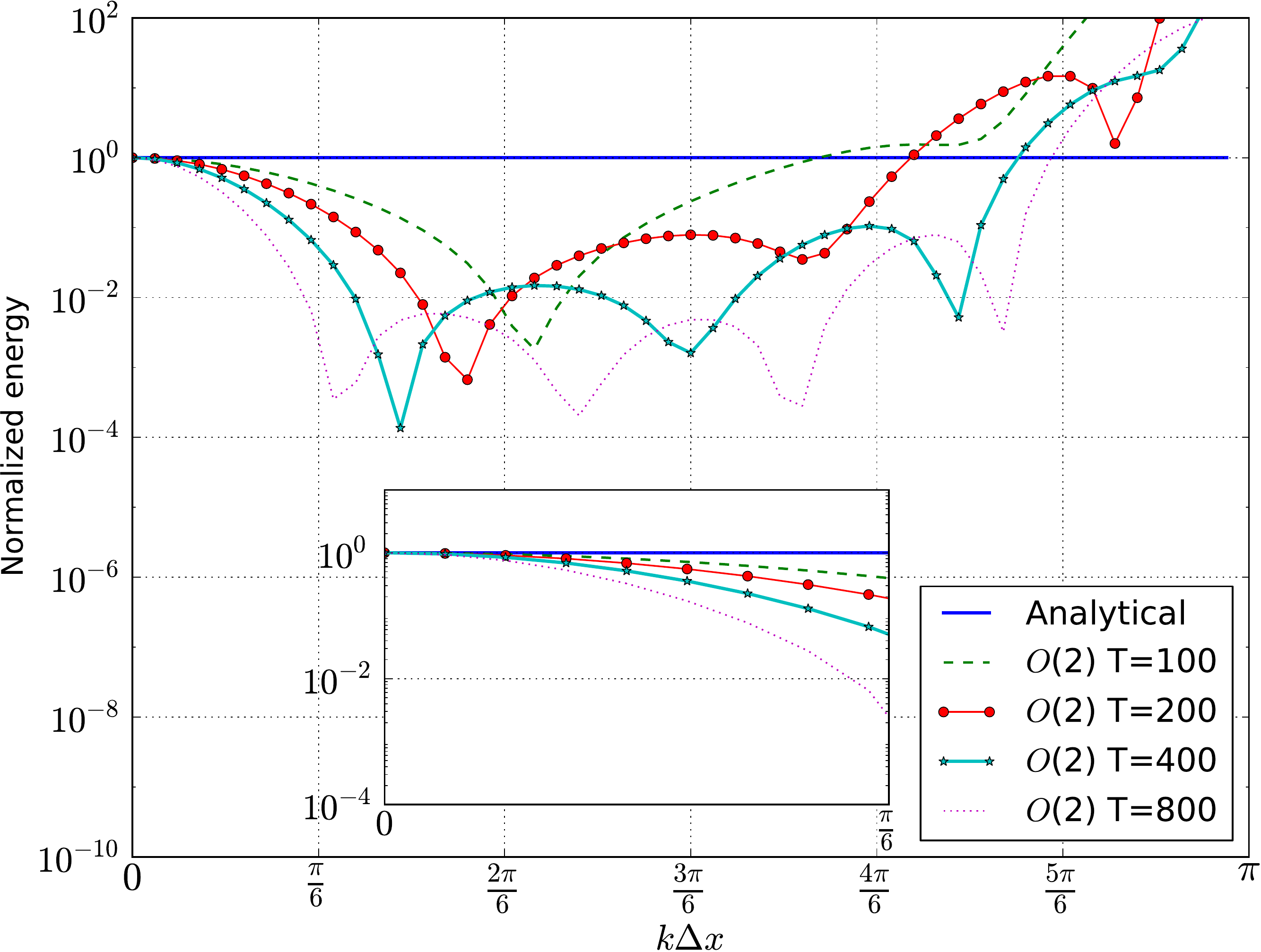}}
		\end{center}
		\caption{Second-order ENO}
	\end{subfigure}%
	\begin{subfigure}[t]{0.5 \textwidth}
		\begin{center}{\includegraphics[width=\textwidth]{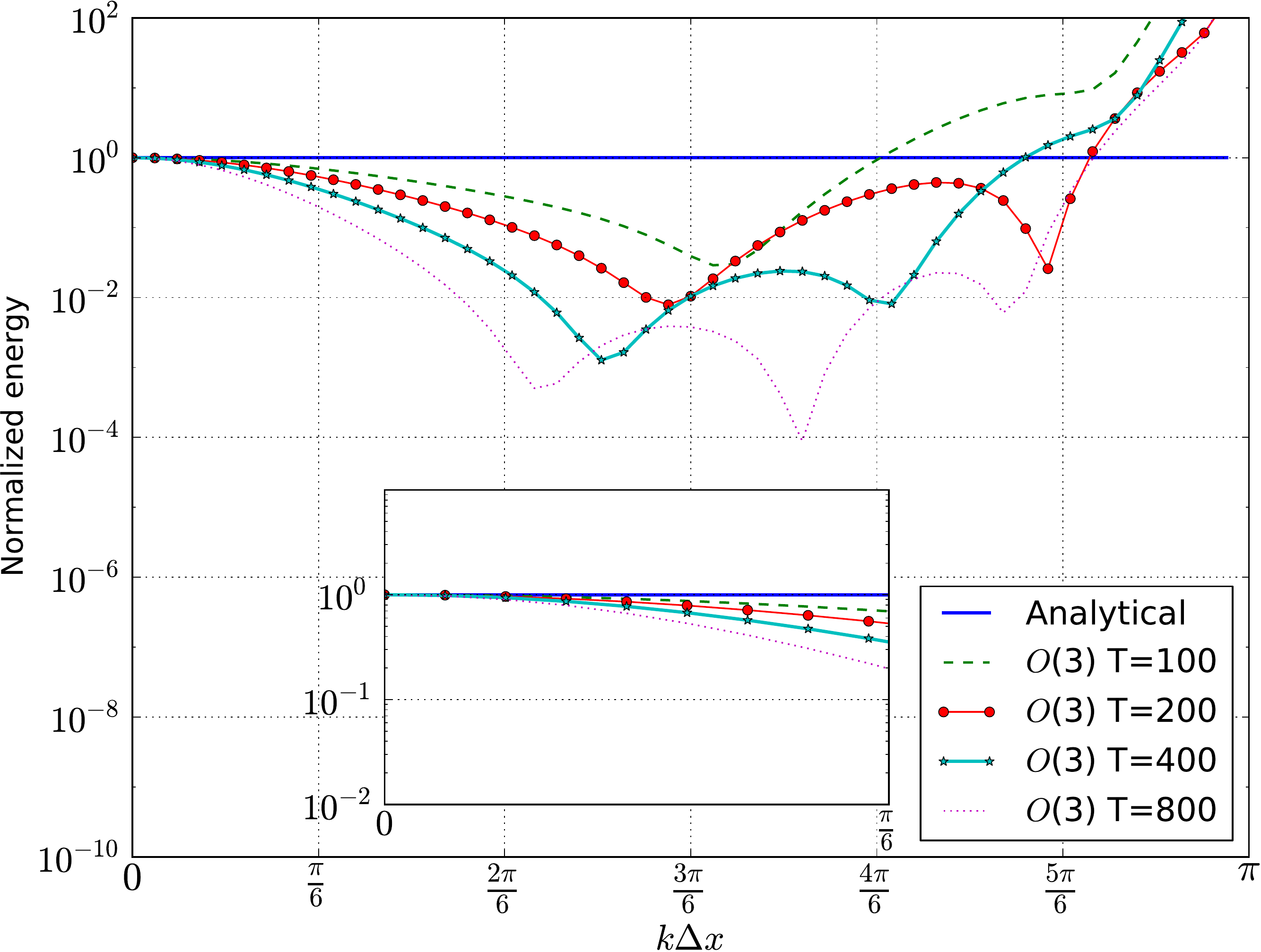}}
		\end{center}
		\caption{Third-order ENO}
	\end{subfigure}%
	\newline
	\begin{subfigure}[t]{0.5 \textwidth}
		\begin{center}{\includegraphics[width=\textwidth]{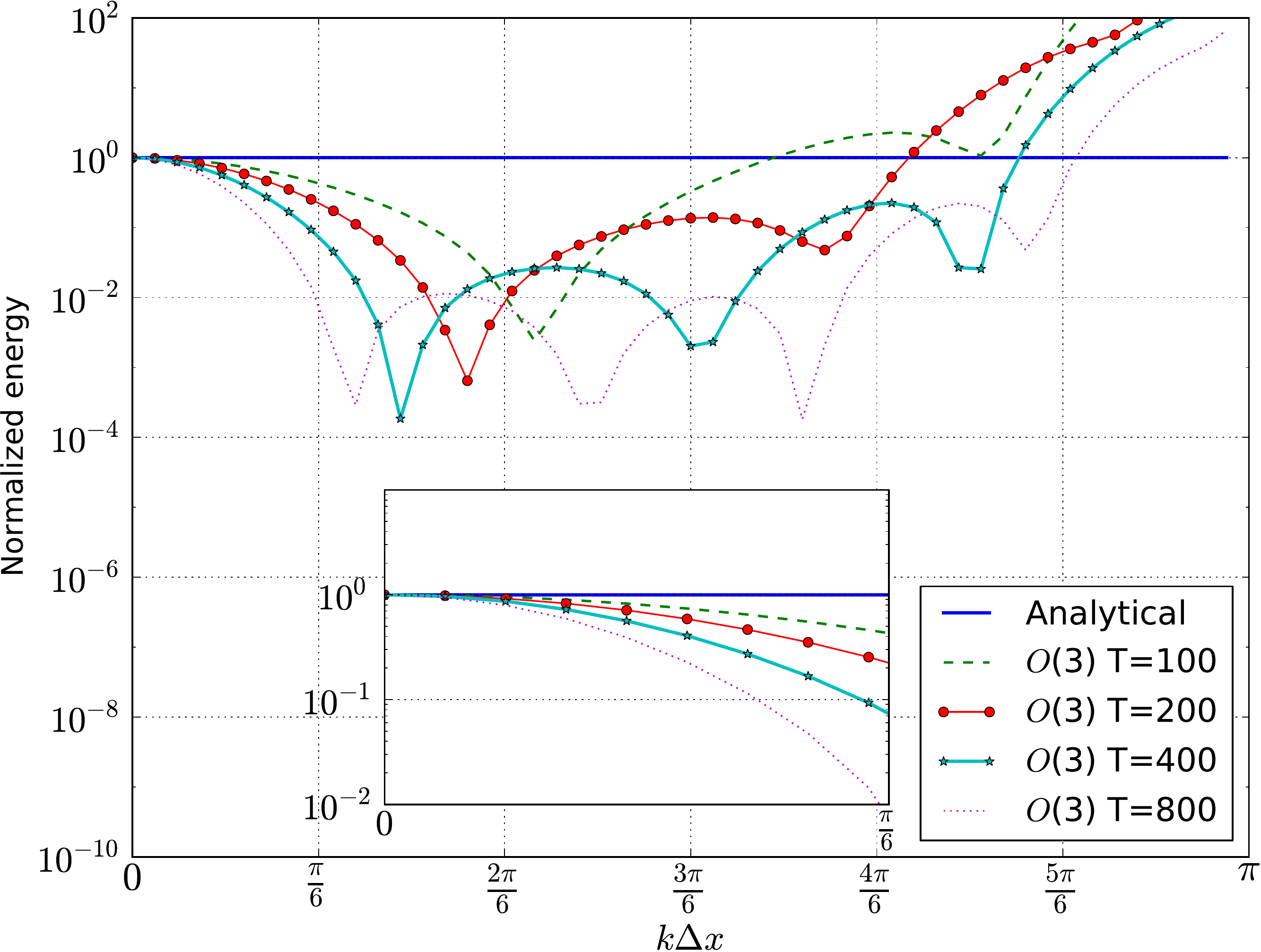}}
		\end{center}
		\caption{Third-order WENO}
	\end{subfigure}%
	\begin{subfigure}[t]{0.5 \textwidth}
		\begin{center}{\includegraphics[width=\textwidth]{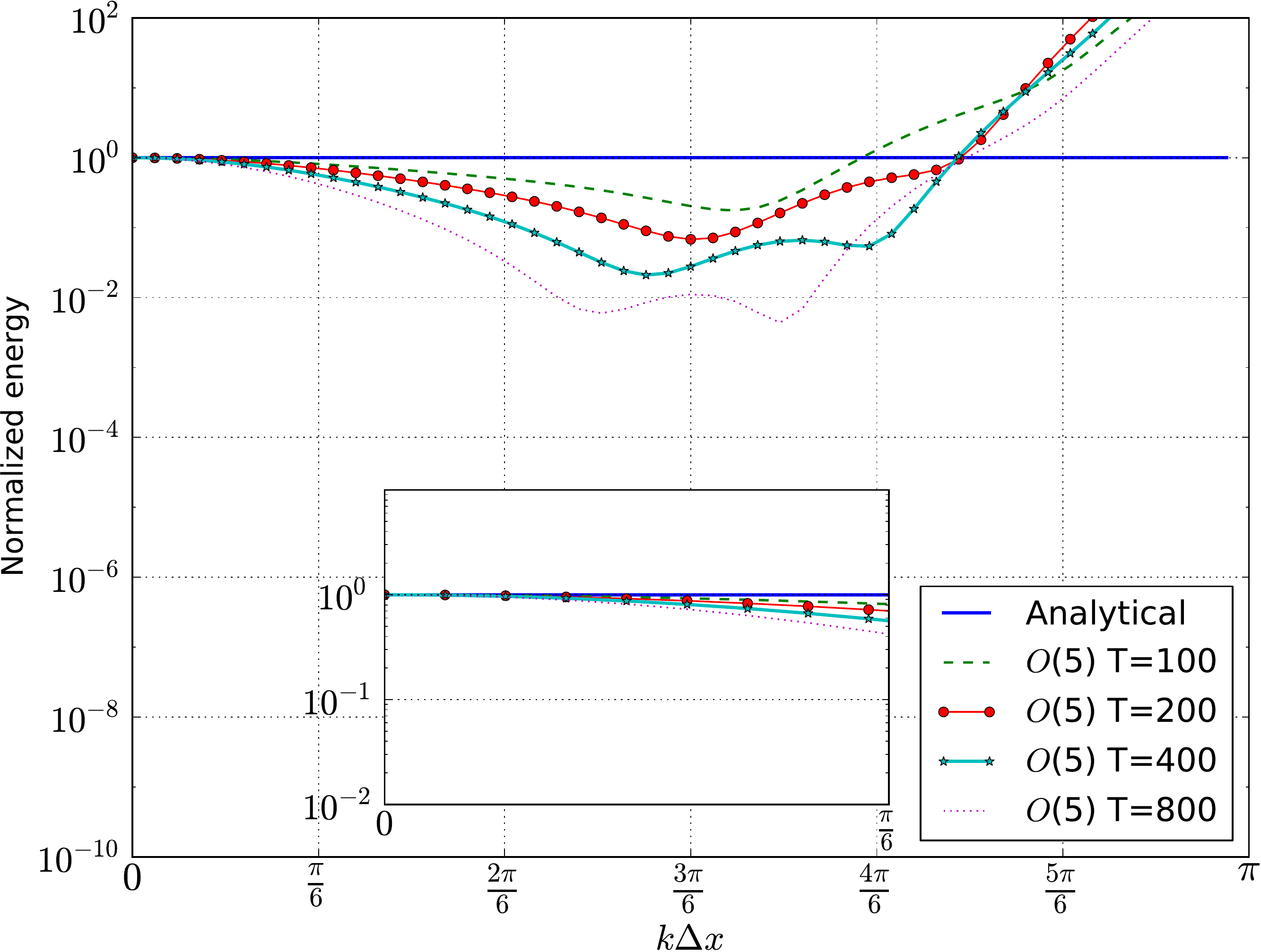}}
		\end{center}
		\caption{Fifth-order WENO}
	\end{subfigure}%
	\end{center}
	\caption{\label{enowenotvsf} Variation of PSD over simulation time }
\end{figure}

Figs. \ref{enowenotdfd} and \ref{enowenons} show results for finite volume ENO and WENO schemes.  
The energy of the zeroth Fourier mode of the numerical solution again matches with that of the
analytical results for both ENO and WENO schemes. This implies both ENO and WENO
schemes to be conservative in nature. Spurious modes are observed in the energy spectra for both 
the schemes. This is consistent with observations reported in the literature \cite{Fauconnier2011}.
Both schemes show anti-dissipation in higher modes. This is indicated by addition of energy
in higher Fourier modes. 
However, this additional energy in the higher modes is several orders of magnitude less
than the reference energy (i.e. zeroth mode energy) as seen in Figs.\ref{enofd} and \ref{wenofd}.
This causes distortion in the original Gaussian signal over time. The distortion is seen
to be less for lower-order ENO and WENO schemes. 
The anti-dissipation in higher modes also gives rise to `numerical turbulence' \cite{Fauconnier2011}.

Fig \ref{enowenotvsf} shows variation of PSD over computing time. Reduction
in energy corresponding to lower wavenumbers is clearly seen. 
This implies that dissipation is progressively added in the larger scales over time \cite{Fauconnier2011}. 
In contrast, the anti-dissipation added in higher wavenumbers does not consistently 
increase or reduce.
Because of nonlinear interactions between modes, it becomes important
to see evolution of the total energy of the signal. 
These results also show that anti-dissipation in higher wavenumber modes does not
necessarily indicate instability in the solution. 
Detection of onset of an instability along with evolution of the
total energy of the signal are studied in subsequent sections.

\subsubsection{ADER scheme}
The space-time coupled Arbitrary DERivatives (ADER) finite volume 
numerical schemes\cite{Titarev2002,Schwartzkopff2002,Schwartzkopff2004,Toro2005, Titarev2005}
are analyzed in this section. 
ADER scheme is derived from the generalized Riemann problem (GRP) \cite{Ben-Artzi1984} and 
modified generalized Riemann problem (MGRP) schemes\cite{Toro1998}. 
In these schemes, spatial and temporal derivatives are coupled through the Cauchy-Kowalewsky 
procedure. The resulting scheme is higher-order accurate in both space and time.
ADER schemes are found to be particularly suitable for wave propagation problems 
\cite{Schwartzkopff2002}.

Apart from the finite volume formulation presented in the above cited papers, a local space-time
DG formulation of the ADER scheme is presented in \cite{Dumbser2008,Dumbser2013,Dumbser2008a}.
This formulation is especially suitable for flows with stiff sources and nonlinear systems
\cite{Dumbser2008,Balsara2009,Balsara2013}.
In this work, we use the finite volume formulation of the ADER scheme.
In this scheme, numerical fluxes at cell interfaces are obtained by solving a derivative Riemann
problem (DRP)\cite{Toro2006}.
A detailed procedure for computing ADER fluxes is found in
\cite{Titarev2002, Schwartzkopff2002,Schwartzkopff2004,Toro2005, Titarev2005, Toro2006}.

\begin{figure}[H]
	\begin{subfigure}[t]{0.5 \textwidth}
		\begin{center}{\includegraphics[width=\textwidth]{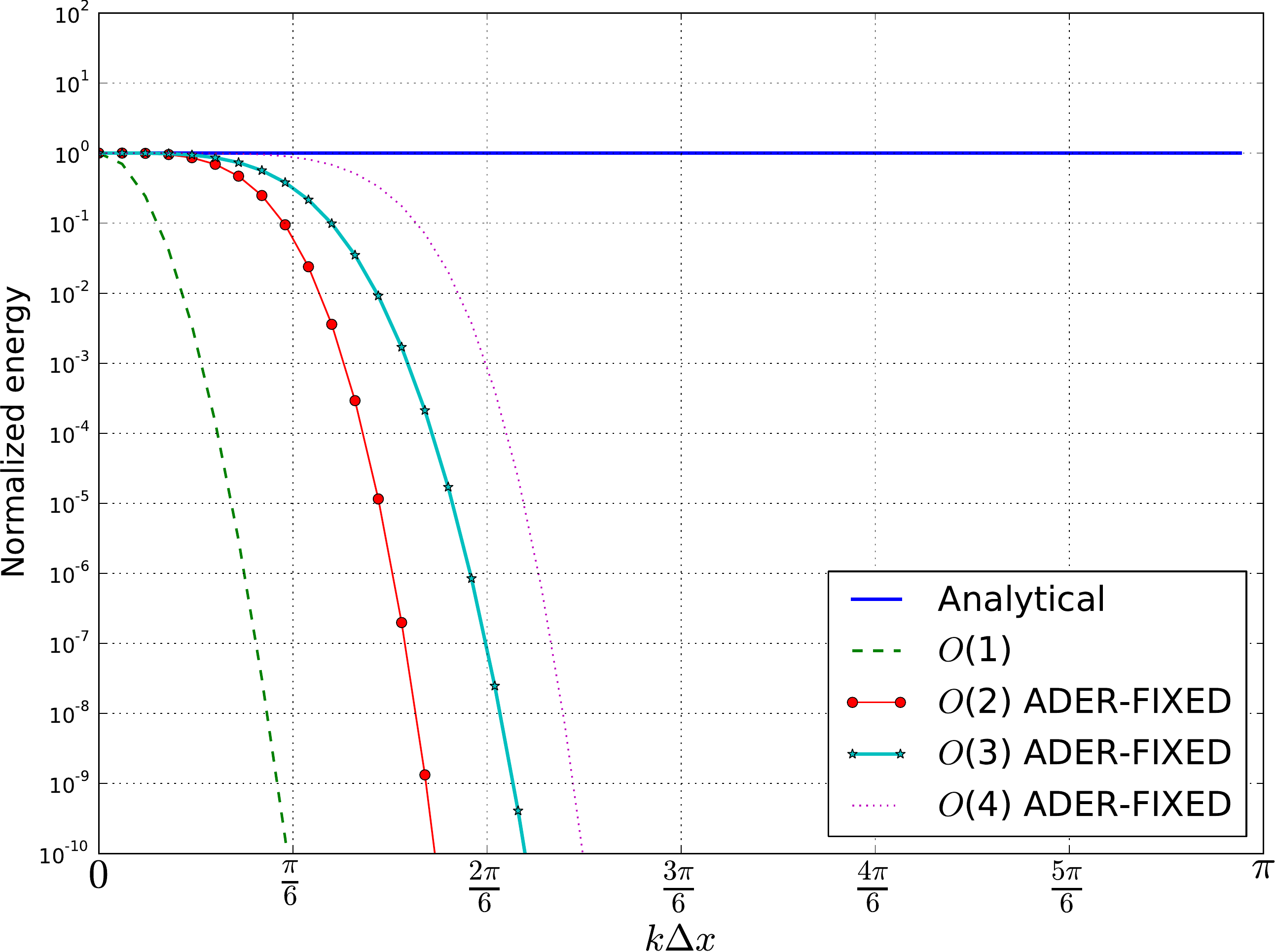}}
		\end{center}
		\caption{}
	\end{subfigure}%
	\begin{subfigure}[t]{0.5 \textwidth}
		\begin{center}{\includegraphics[width=\textwidth]{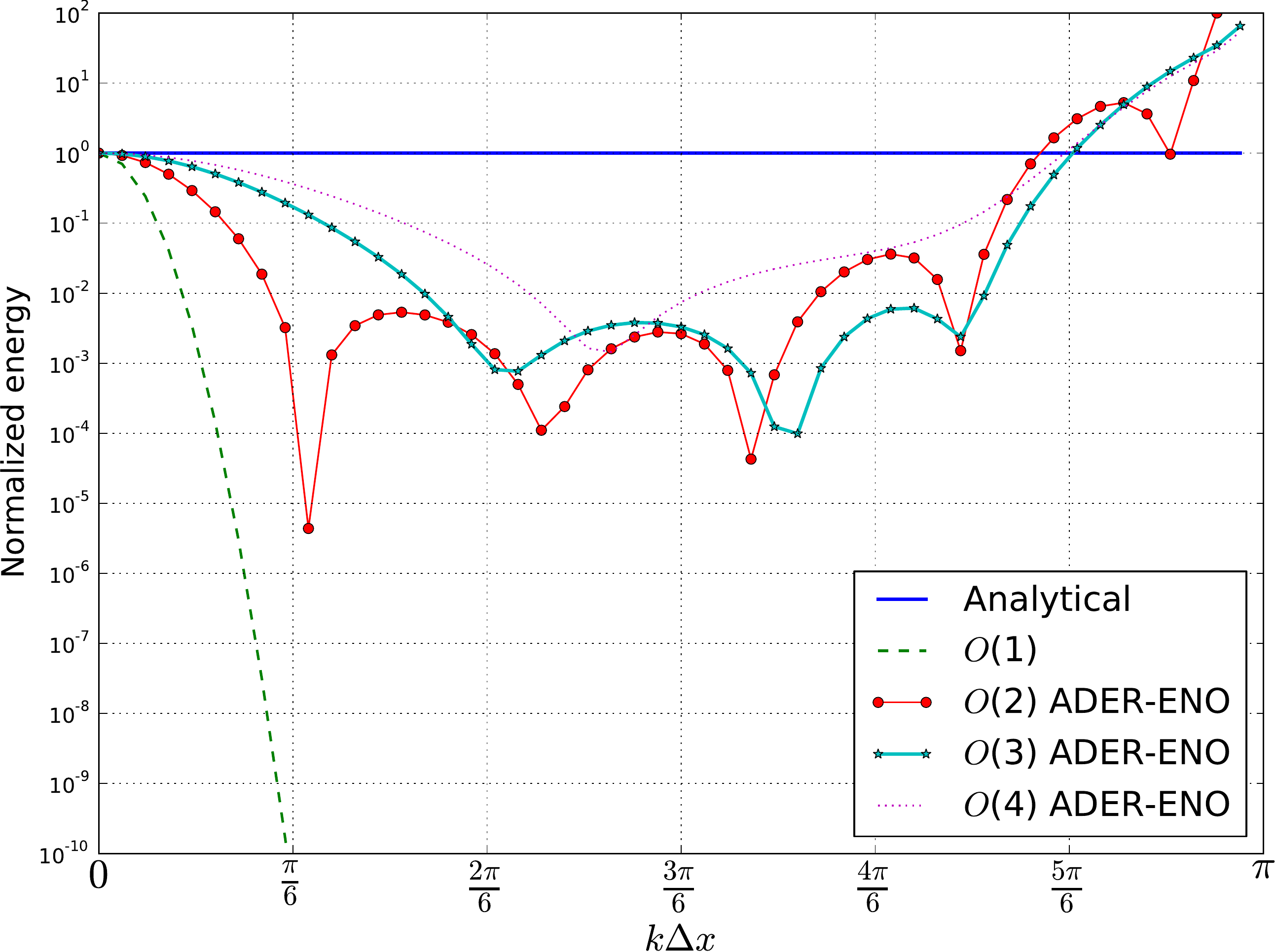}}
		\end{center}
		\caption{}
	\end{subfigure}%
	\caption{\label{ader1} Spectrum for (a) fixed-stencil ADER scheme (b) ADER-ENO scheme}
\end{figure}

\begin{figure}[H]
	\begin{subfigure}[t]{0.5 \textwidth}
		\begin{center}{\includegraphics[width=\textwidth]{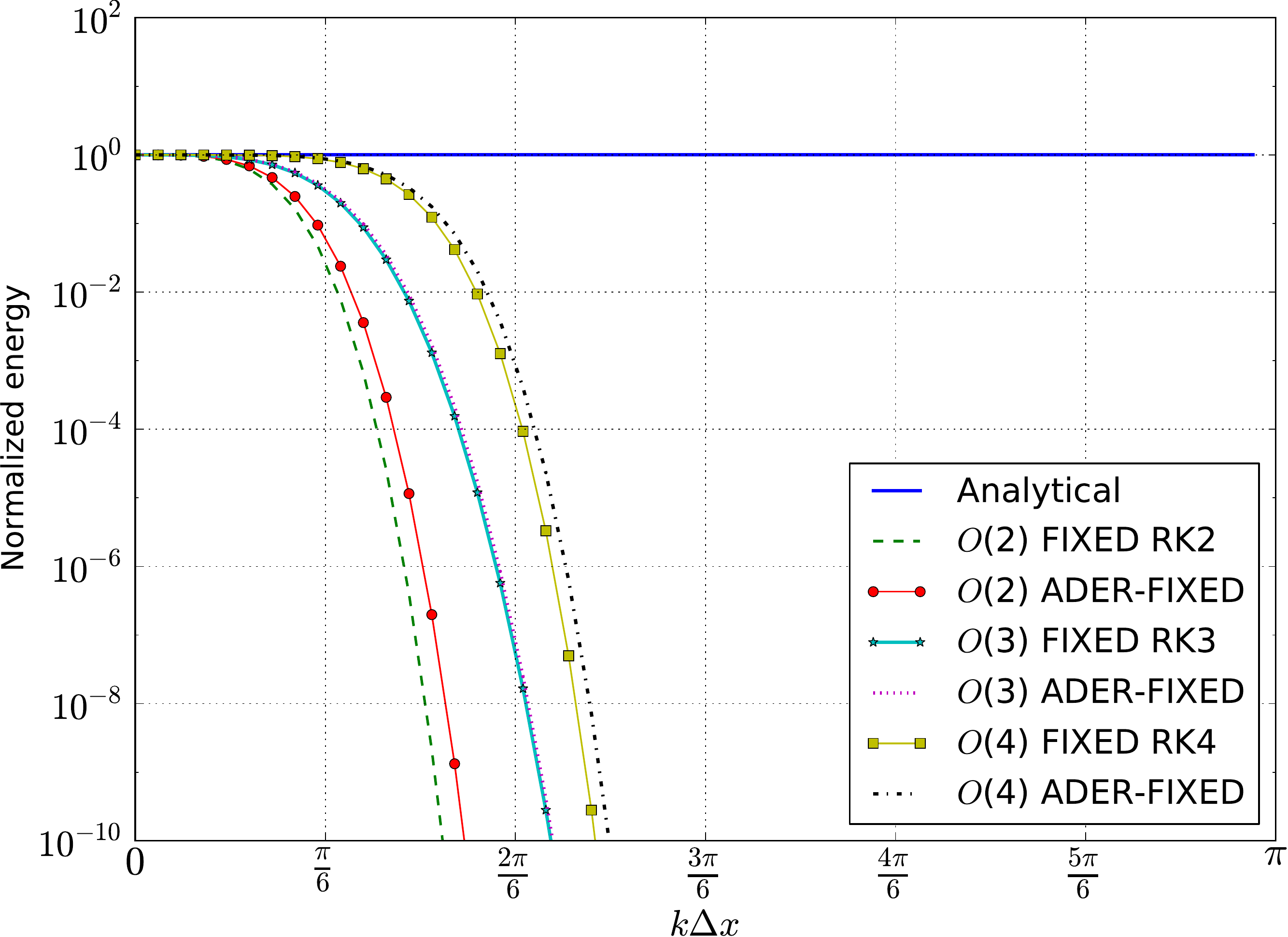}}
		\end{center}
		\caption{Fixed-stencil reconstruction}
	\end{subfigure}%
	\begin{subfigure}[t]{0.5 \textwidth}
		\begin{center}{\includegraphics[width=\textwidth]{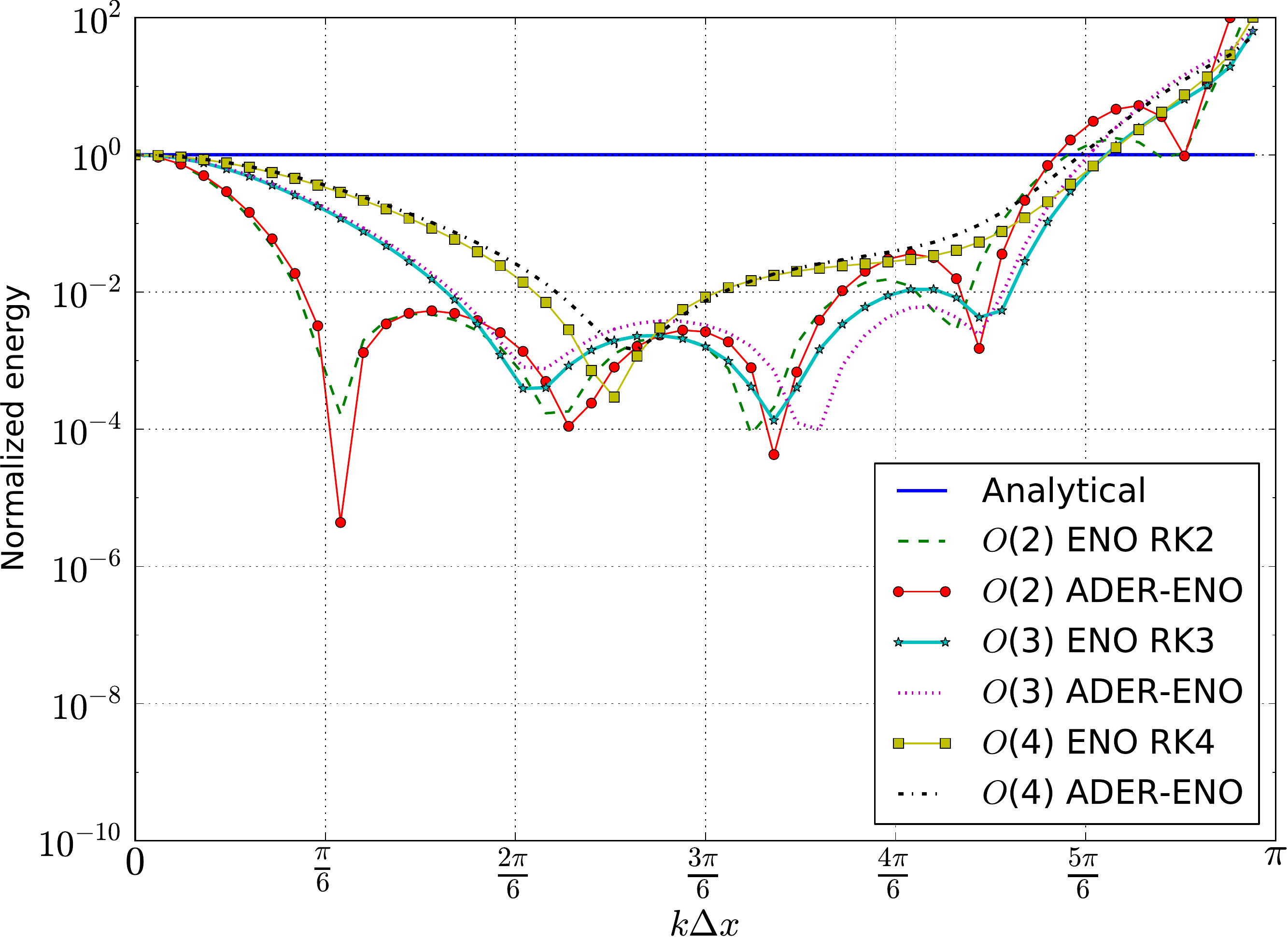}}
		\end{center}
		\caption{ENO reconstruction}
	\end{subfigure}%
	\caption{\label{ader2} Comparison of ADER schemes with semidiscrete schemes }
\end{figure}

Fig.\ref{ader1}a. shows power spectrum of ADER schemes with varying order of accuracy.
A fixed stencil finite volume reconstruction is used for spatial derivatives.
Fig.\ref{ader1}b. shows power spectrum for ADER schemes with ENO reconstruction. 
ADER schemes show similar results as semidiscrete schemes using the same method of 
data reconstruction (ENO and FS).
To compare space-time coupled numerical schemes with the semidiscrete counterparts,
the normalized PSD graph of ADER solution and that of an equivalent semidiscrete scheme
are compared in Fig.\ref{ader2}a and \ref{ader2}b. 
It is observed that, the ADER scheme adds slightly less diffusion in the numerical solution
as compared to a semidiscrete scheme of equal spatial and temporal order.
ADER schemes are known to have superior dispersion characteristics compared to
semidiscrete schemes of equal order of accuracy \cite{Schwartzkopff2002}.
The current analysis indicates that ADER schemes add less numerical dissipation to 
the solution as well.
Thus, ADER schemes may be better suited
for simulating wave propagation over a long computational domain than conventional semidiscrete schemes.

\subsection{Effect of time-step size on diffusion}\label{dtsize}

\begin{figure}[H]
	\begin{subfigure}[t]{0.5 \textwidth}
		\begin{center}{\includegraphics[width=\textwidth]{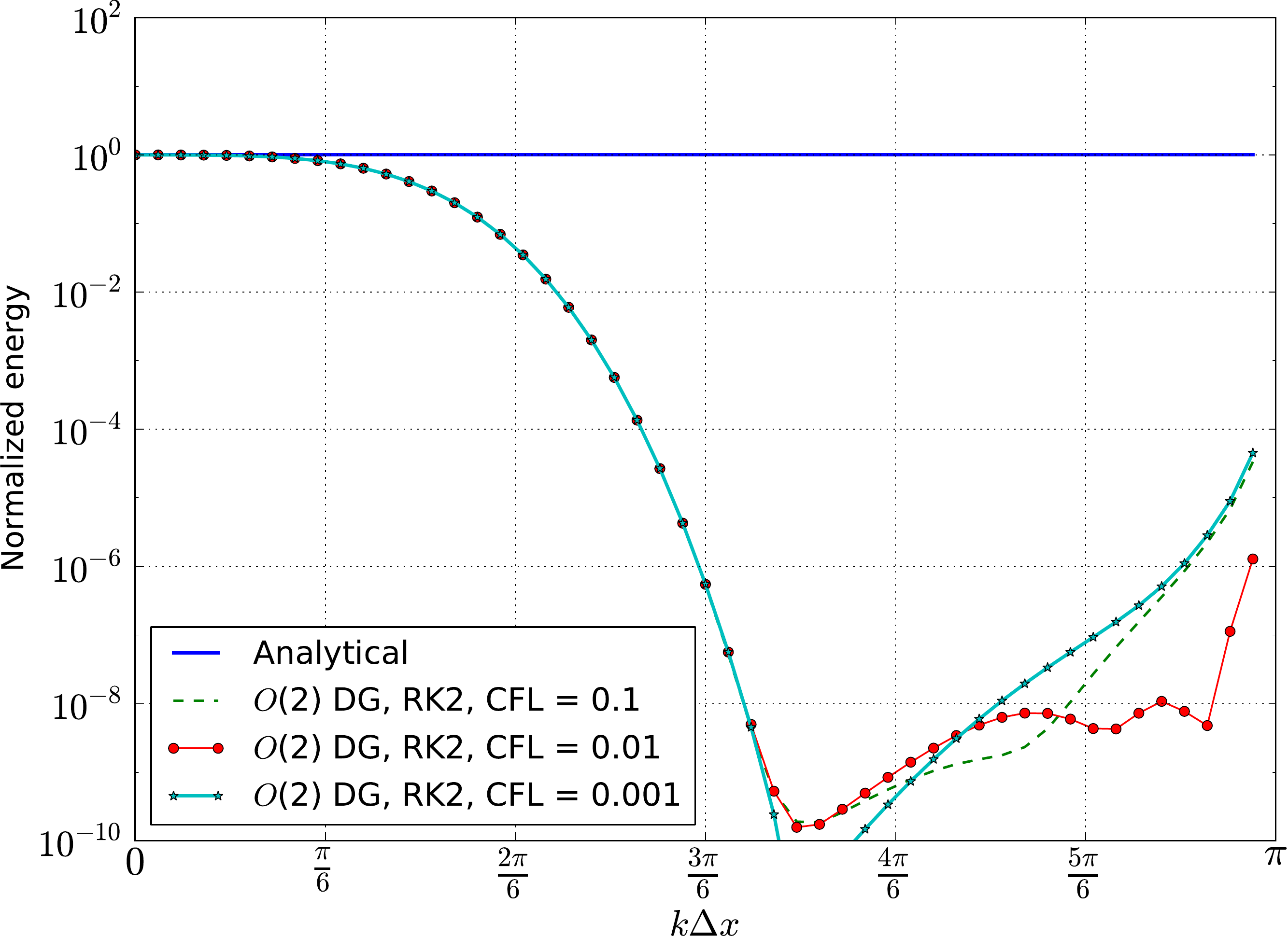}}
		\end{center}
		\caption{\label{dtdg2rk2}$O(2)DG, RK2$}
	\end{subfigure}%
	\begin{subfigure}[t]{0.5 \textwidth}
		\begin{center}{\includegraphics[width=\textwidth]{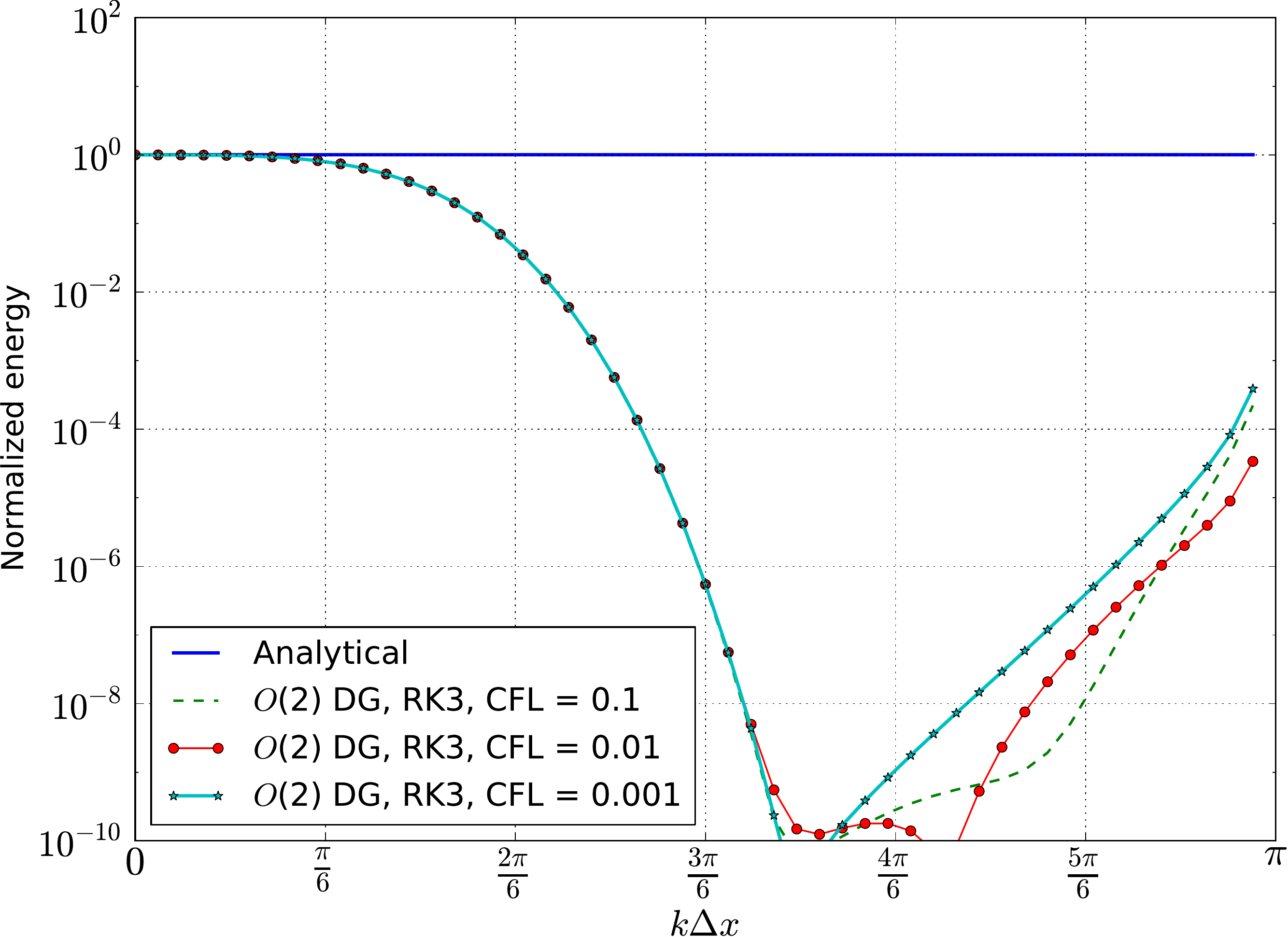}}
		\end{center}
		\caption{\label{dtdg2rk3}$O(2)DG, RK3$}
	\end{subfigure}%
	\newline
	\begin{subfigure}[t]{0.5 \textwidth}
		\begin{center}{\includegraphics[width=\textwidth]{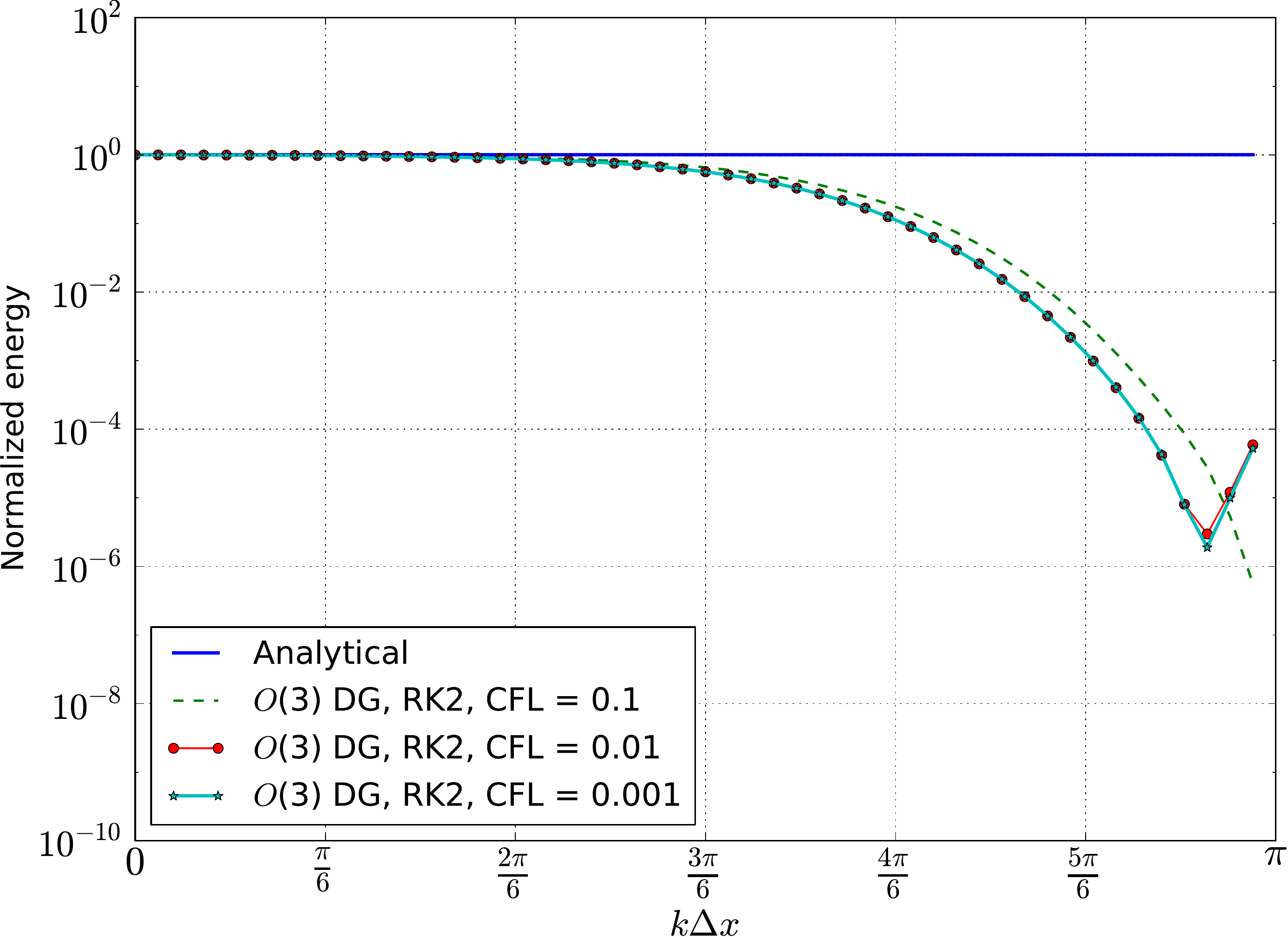}}
		\end{center}
		\caption{\label{dtdg3rk2}$O(3)DG, RK2$}
	\end{subfigure}%
	\begin{subfigure}[t]{0.5 \textwidth}
		\begin{center}{\includegraphics[width=\textwidth]{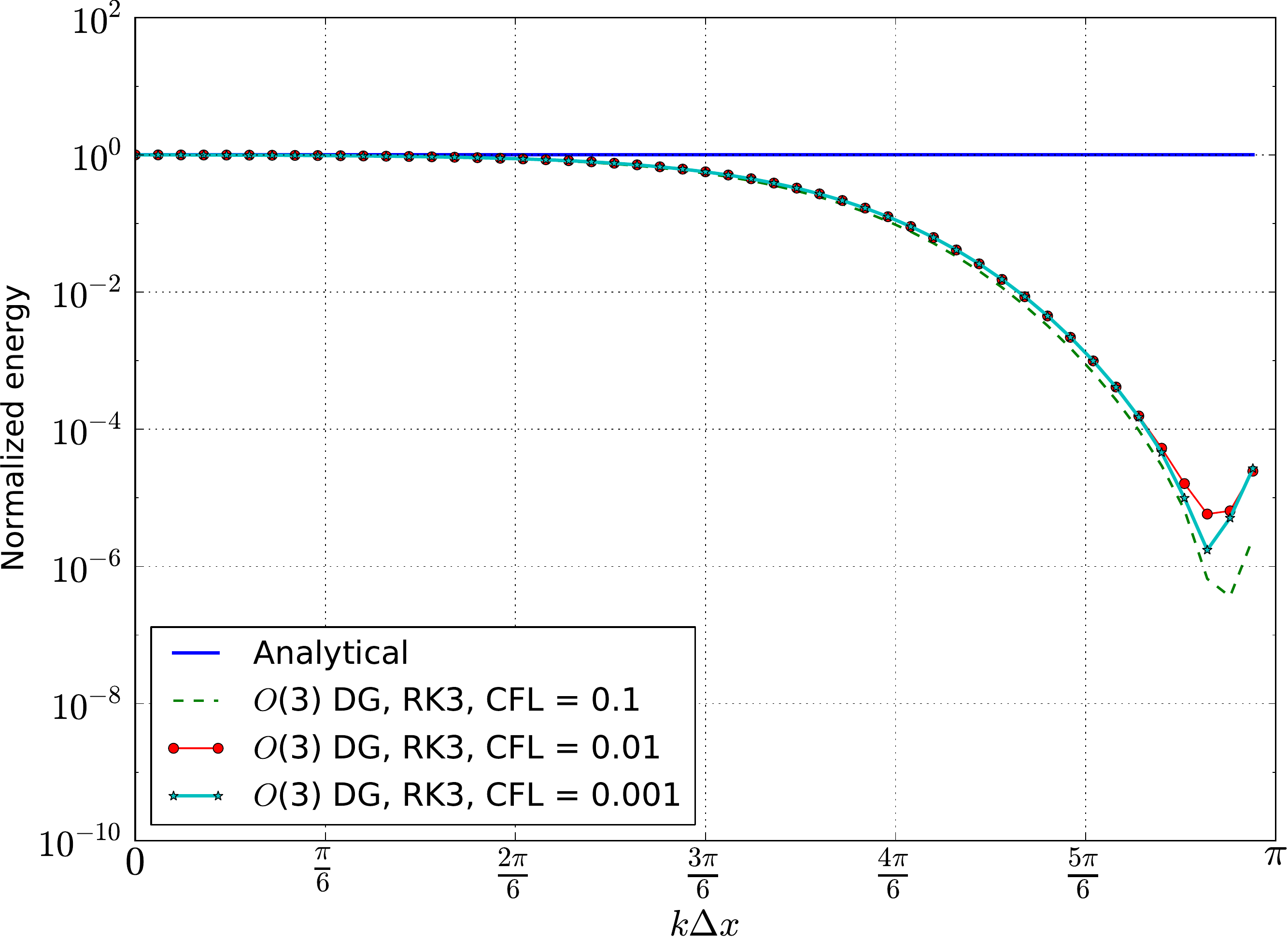}}
		\end{center}
		\caption{\label{dtdg3rk3}$O(3)DG, RK3$}
	\end{subfigure}%
	\newline
	\begin{subfigure}[t]{0.5 \textwidth}
		\begin{center}{\includegraphics[width=\textwidth]{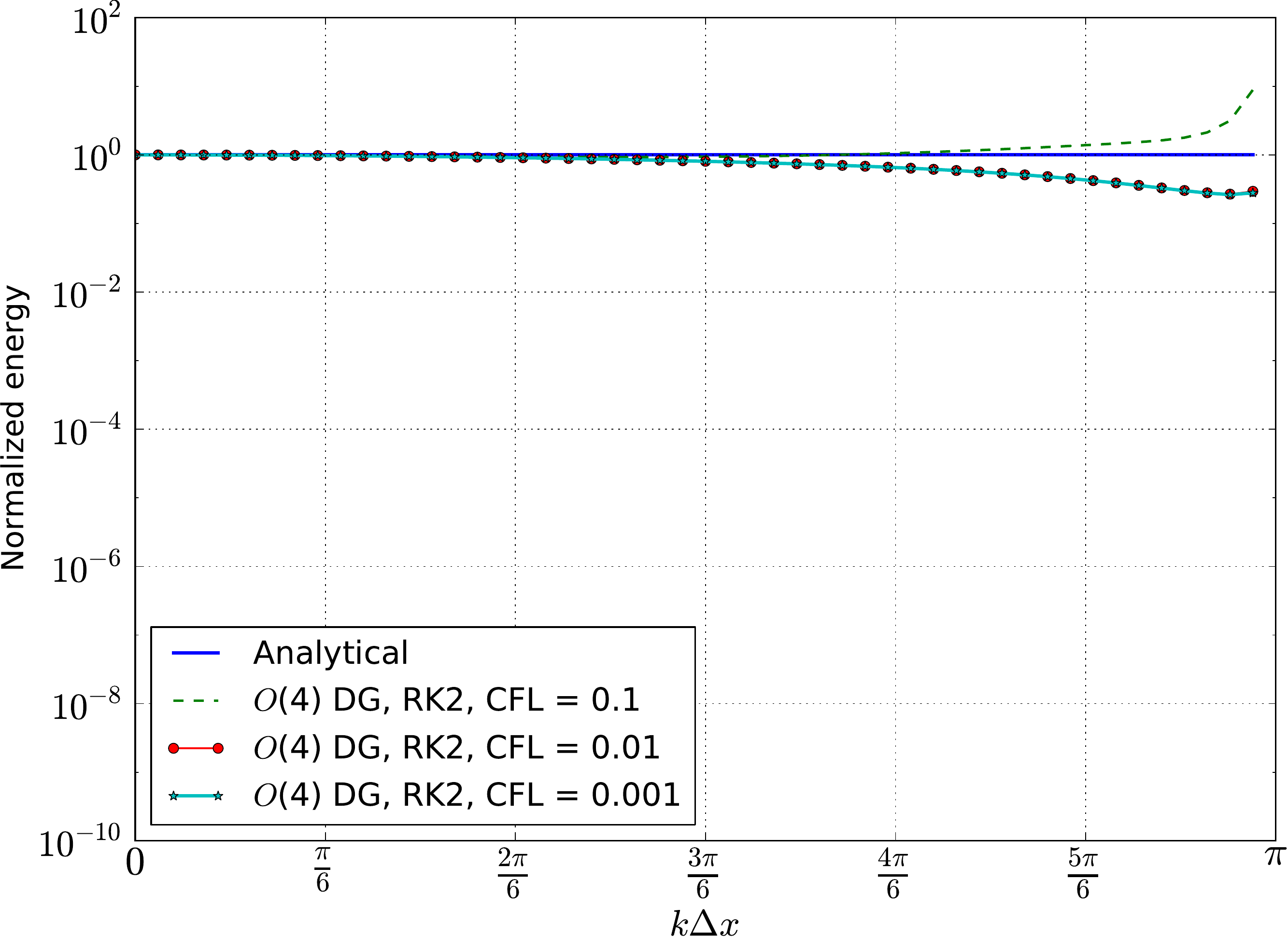}}
		\end{center}
		\caption{\label{dtdg4rk2}$O(4)DG, RK2$}
	\end{subfigure}%
	\begin{subfigure}[t]{0.5 \textwidth}
		\begin{center}{\includegraphics[width=\textwidth]{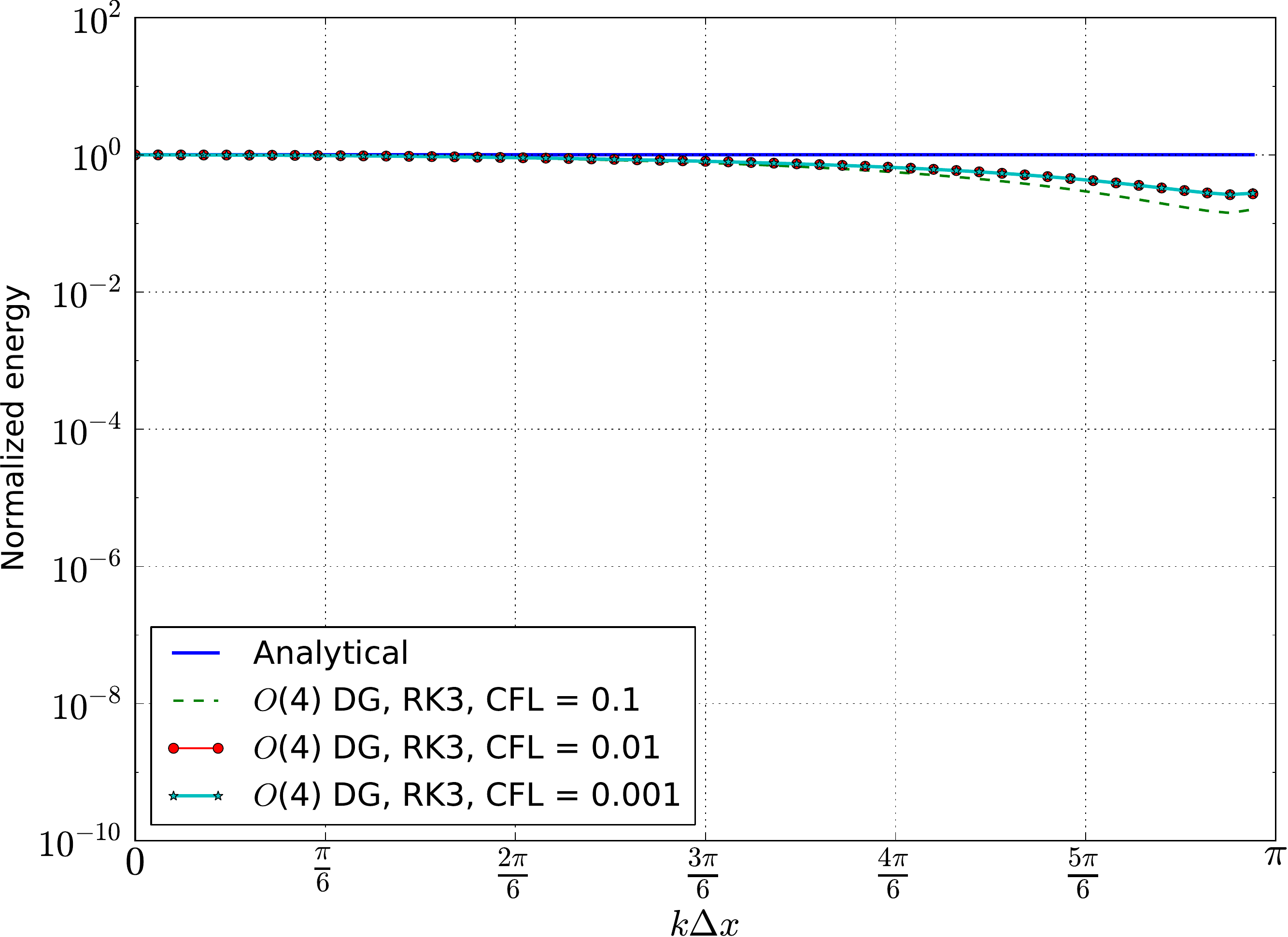}}
		\end{center}
		\caption{\label{dtdg4rk3}$O(4)DG, RK3$}
	\end{subfigure}%
	\caption{\label{dtdg} Effect of time-step size on discontinuous Galerkin schemes with 
	Runge Kutta time-stepping}
\end{figure}

To study the effect of time-step size on diffusion characteristics,
the scalar linear advection equation is solved using a semidiscrete DG scheme with Runge Kutta 
time-stepping.
Simulations are performed with Courant numbers (denoted as CFL)
equal to $0.1, 0.01$ and $0.001$.
For identical element sizes and a constant wavespeed, change in the Courant number  
proportionally affects the time-step size.
Results are shown in Fig.\ref{dtdg}.

	Results confirm that 
	if the spatial order of accuracy is equal to or lower than the order of the
		time integrator, the spatial error dominates. In this case, change in time-step
		has little or no effect on the power spectral plots. A small change in energy
		of higher Fourier modes is observed, however, this is several orders of magnitude
		smaller than the mean energy of the signal showing negligible effect on the overall 
		solution. For example,
		\begin{itemize}
			\item Second-order DG method with RK2 and RK3 time-stepping as shown in Figs.
				\ref{dtdg2rk2} and \ref{dtdg2rk3}.
			\item Third-order scheme with RK3 time-stepping as shown in Fig.\ref{dtdg3rk3}.
		\end{itemize}

	If the spatial order of accuracy is higher than the order of the time integrator,
		then the temporal error may be larger. In such a case, reducing the time-step size
		reduces temporal error. At small enough time-step size, the temporal errors become
		negligible. Further reduction in the time-step size
		does not have any effect on the PSD curve.
		This is also the reason for time-step 
		sizes to be kept very small when computing formal order of spatial 
		accuracy of a numerical scheme
		while using a lower-order time integration method \cite{Joshi2017}.
		Examples include:
		\begin{itemize}
			\item Third order DG method with RK2 time-stepping as shown in Fig.
				\ref{dtdg3rk2}. 
			\item Fourth order DG method with RK2 and RK3 time-stepping as shown in
				Figs.\ref{dtdg4rk2} and \ref{dtdg4rk3}.
		\end{itemize}

\section{Total energy of the signal}
For a diffusive numerical scheme, diffusion gets
added initially to higher Fourier modes and progressively shifts to lower modes.
For nonlinear numerical schemes such as ENO and WENO schemes, spurious modes can get created
due to nonlinearity and higher wavenumber modes show increased energy over time while 
energy in lower modes is seen to decrease. 
Thus, information from PSD curve are limited to dissipation added in individual Fourier modes.
Evolution of total energy of the signal becomes an important aspect 
for better understanding of dissipation characteristics of a numerical scheme.
In this section, we study effect of various simulation
parameters on total energy of the signal.

A regular Gaussian profile with $\alpha = \frac{1}{6}$ is taken as an initial condition.
Simulation is run till the Gaussian pulse travels once round the complete domain and re-centers
at the origin. 
Normalized energy of the signal (in \%) is plotted over simulation time.

\subsection{Effect of spatial order on total energy}

\begin{figure}[h!]
	\begin{subfigure}[t]{0.5 \textwidth}
		\begin{center}{\includegraphics[width=\textwidth]{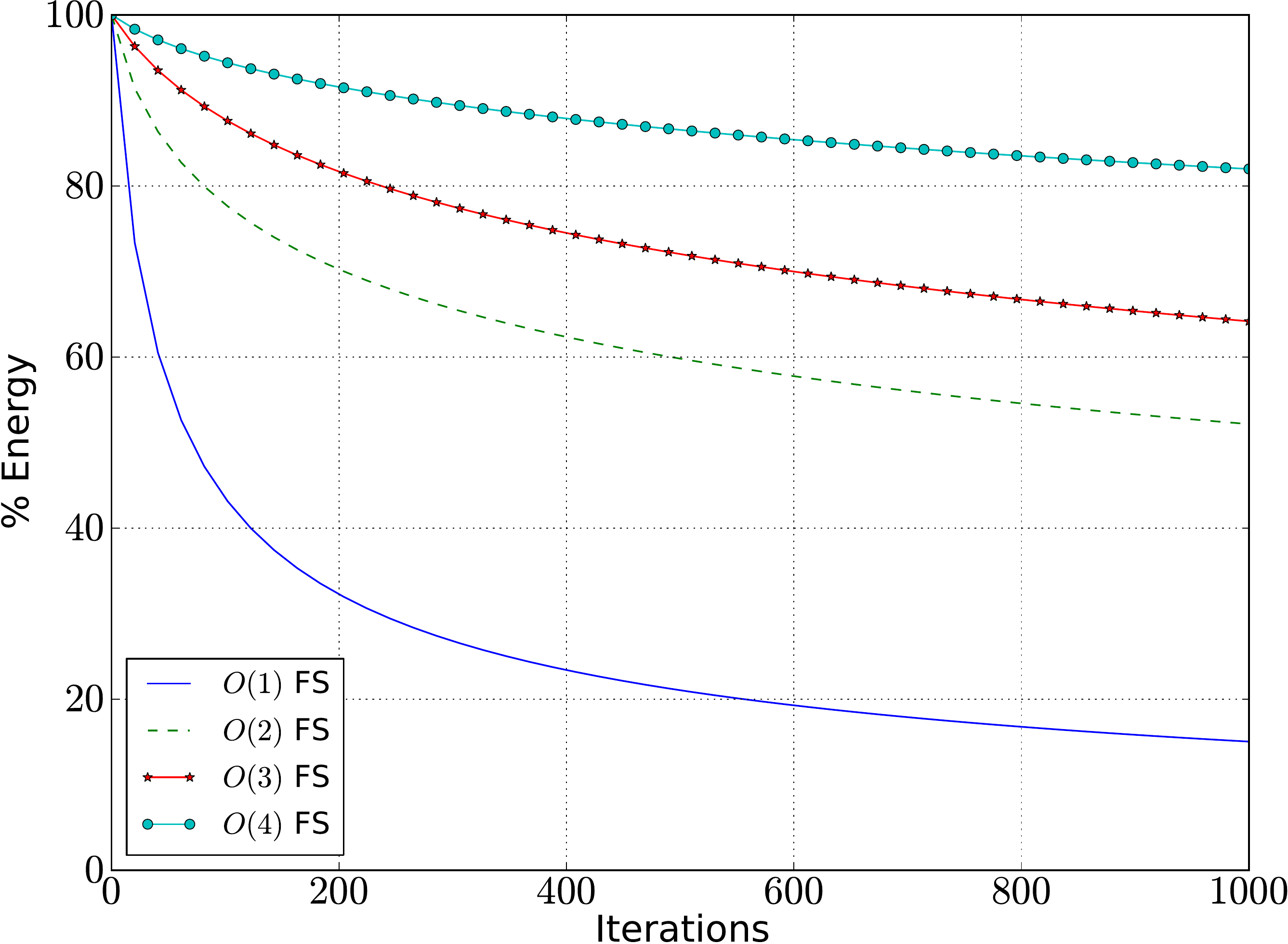}}
		\end{center}
		\caption{Fixed-stencil(FS) scheme}
	\end{subfigure}%
	\begin{subfigure}[t]{0.5 \textwidth}
		\begin{center}{\includegraphics[width=\textwidth]{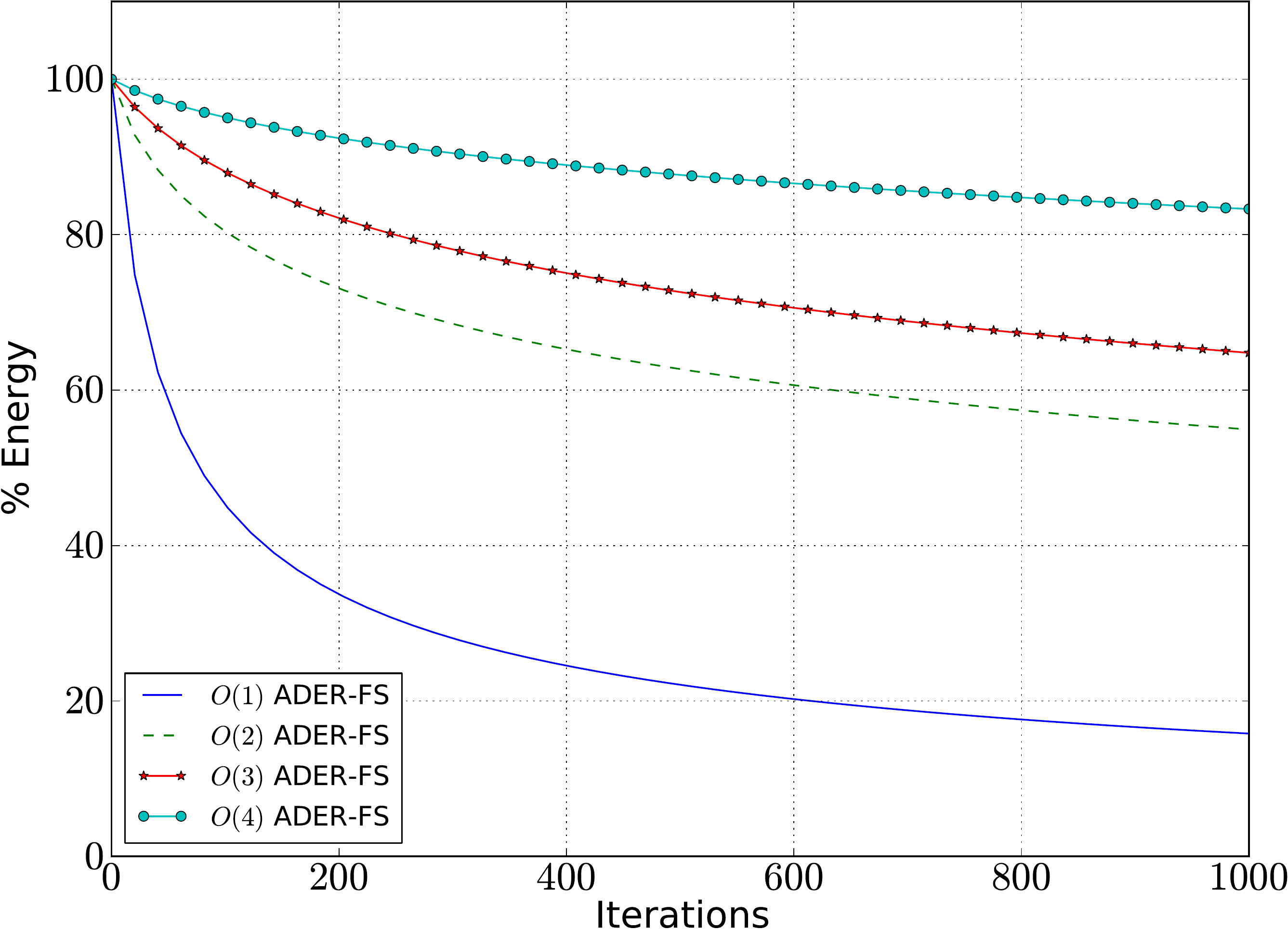}}
		\end{center}
		\caption{ADER-FS scheme}
	\end{subfigure}%
	\newline
	\begin{subfigure}[t]{0.5 \textwidth}
		\begin{center}{\includegraphics[width=\textwidth]{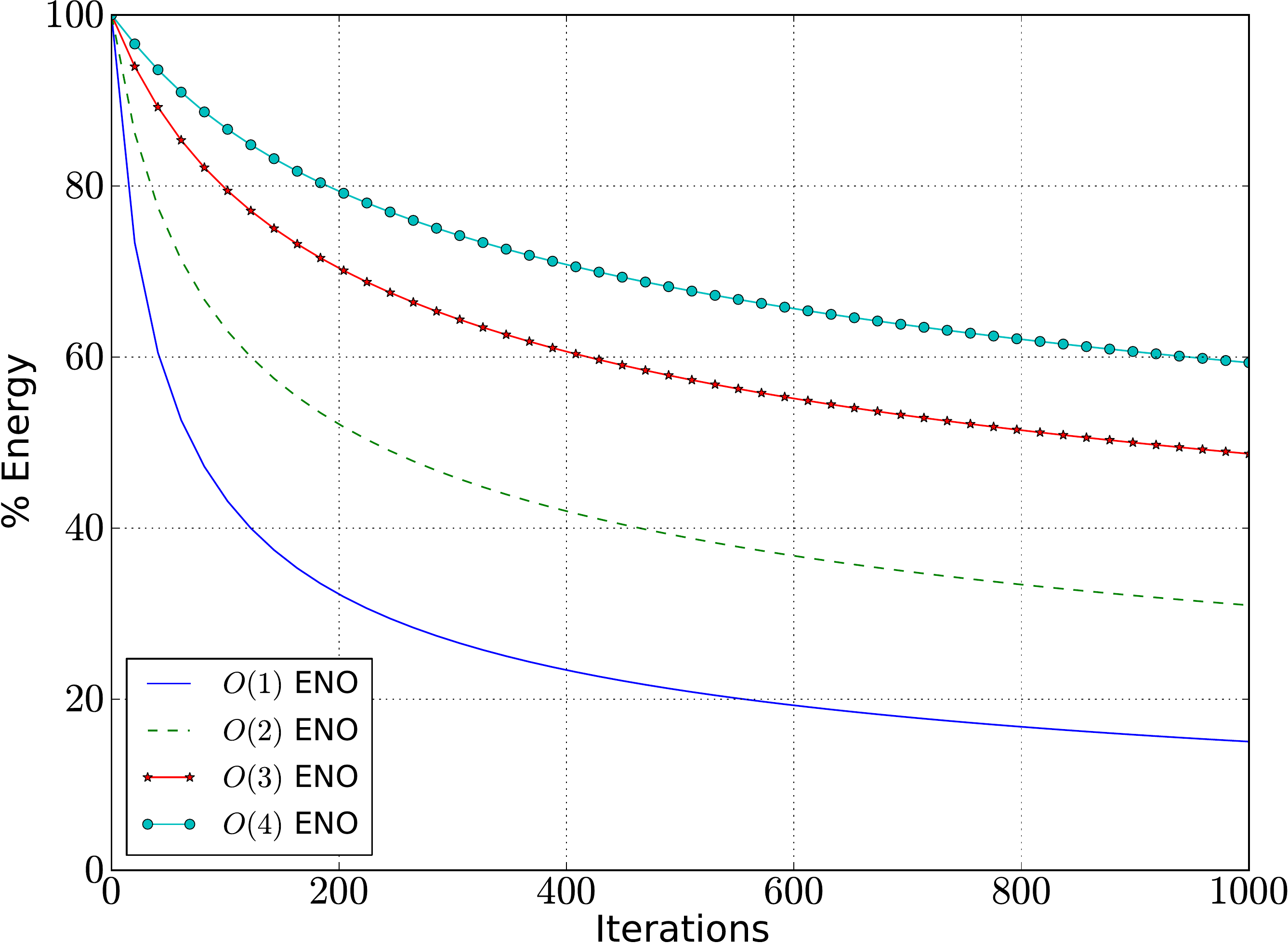}}
		\end{center}
		\caption{\label{teeno}ENO scheme}
	\end{subfigure}%
	\begin{subfigure}[t]{0.5 \textwidth}
		\begin{center}{\includegraphics[width=\textwidth]{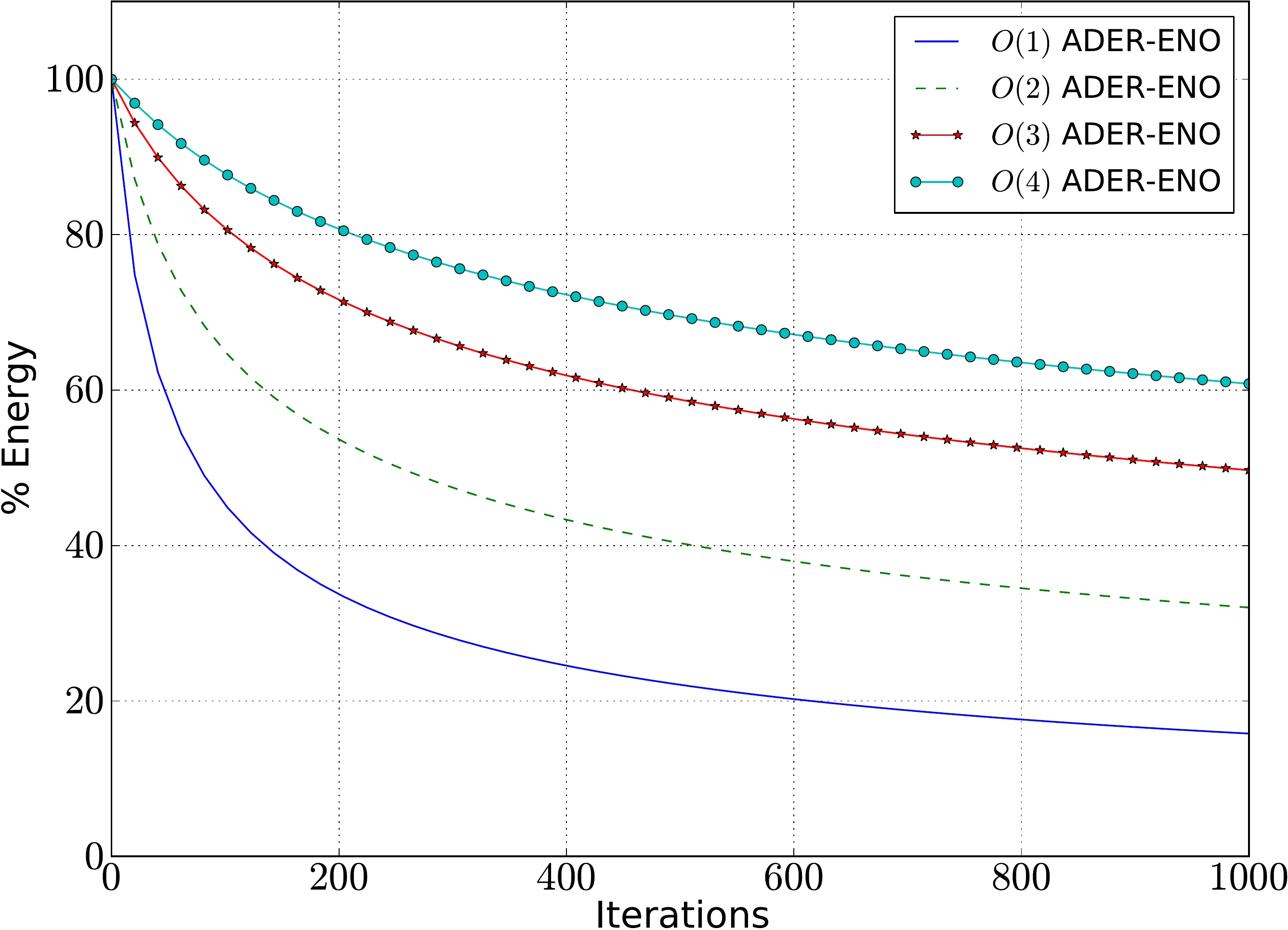}}
		\end{center}
		\caption{ADER-ENO scheme}
	\end{subfigure}%
	\newline
	\begin{subfigure}[t]{0.5 \textwidth}
		\begin{center}{\includegraphics[width=\textwidth]{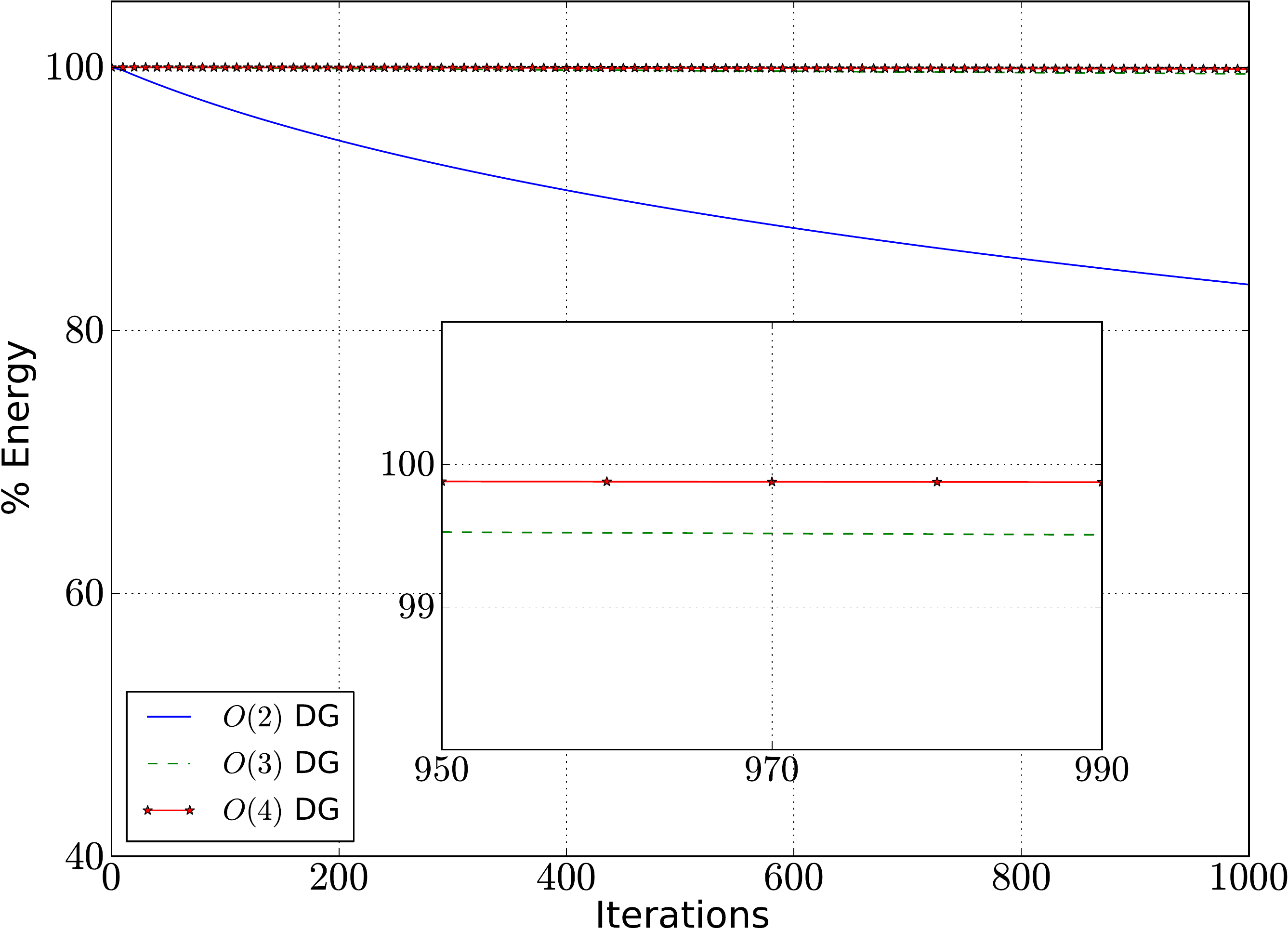}}
		\end{center}
		\caption{DG scheme}
	\end{subfigure}%
	\begin{subfigure}[t]{0.5 \textwidth}
		\begin{center}{\includegraphics[width=\textwidth]{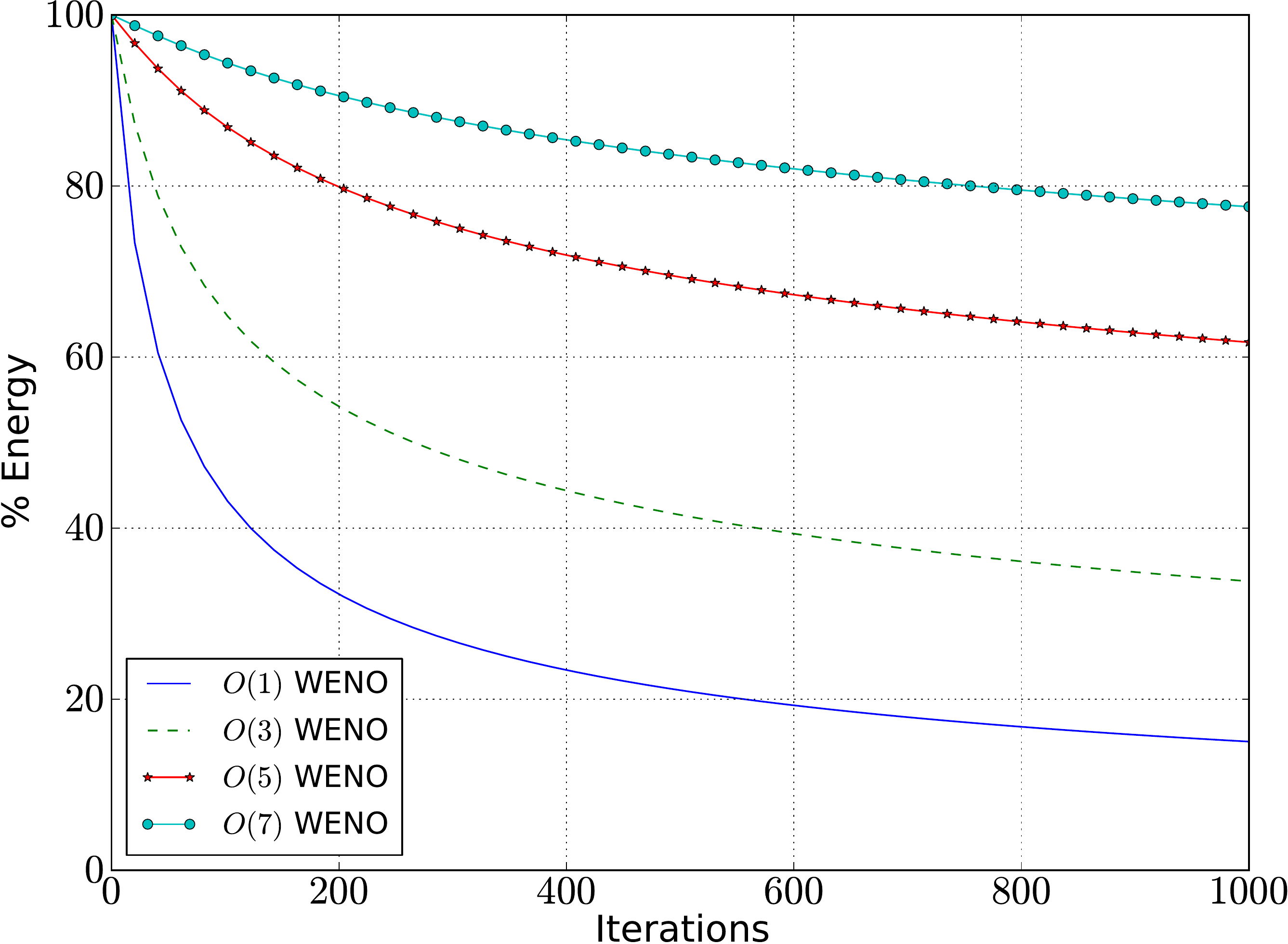}}
		\end{center}
		\caption{\label{teweno}WENO scheme}
	\end{subfigure}%
	\caption{\label{energy1} Energy of a signal over time }
\end{figure}

\begin{figure}[h!]
	\begin{center}
	\begin{subfigure}[t]{0.5\textwidth}
		\begin{center}{\includegraphics[width=\textwidth]{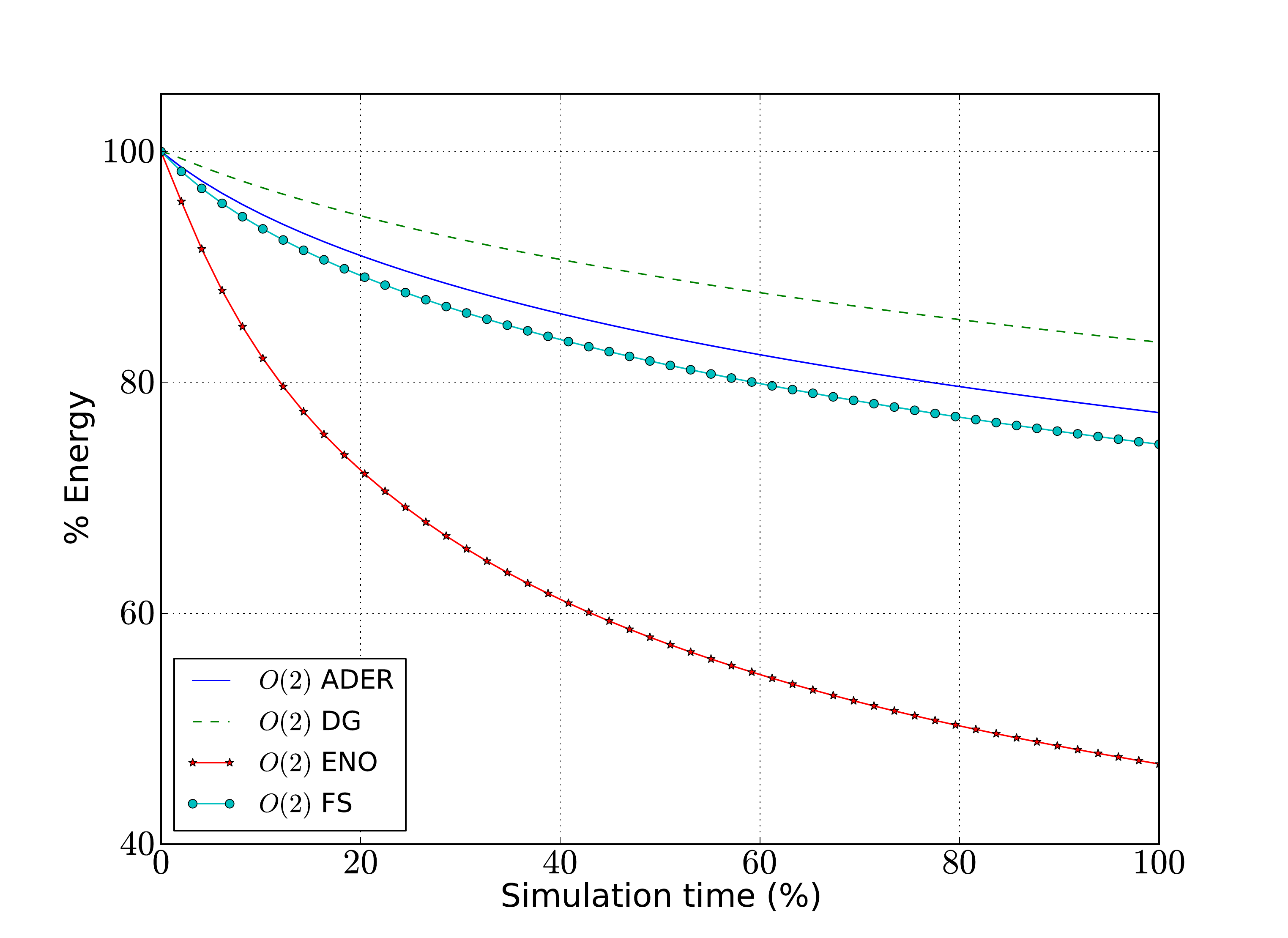}}
		\end{center}
		\caption{\label{energy2}Second-order}
	\end{subfigure}%
	\begin{subfigure}[t]{0.5\textwidth}
		\begin{center}{\includegraphics[width=\textwidth]{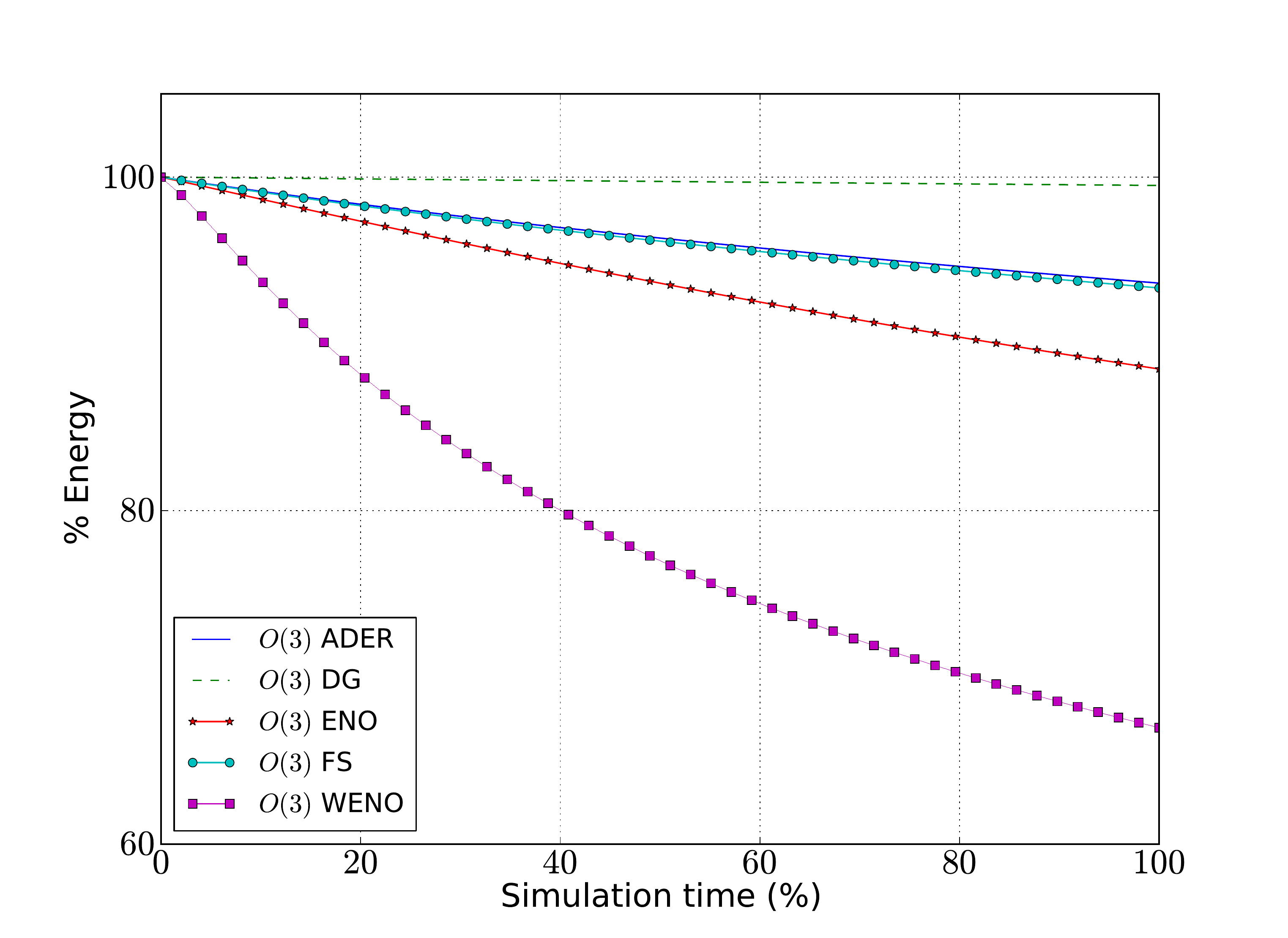}}
		\end{center}
		\caption{\label{energy3}Third-order}
	\end{subfigure}%
	\newline
	\begin{subfigure}[t]{0.5\textwidth}
		\begin{center}{\includegraphics[width=\textwidth]{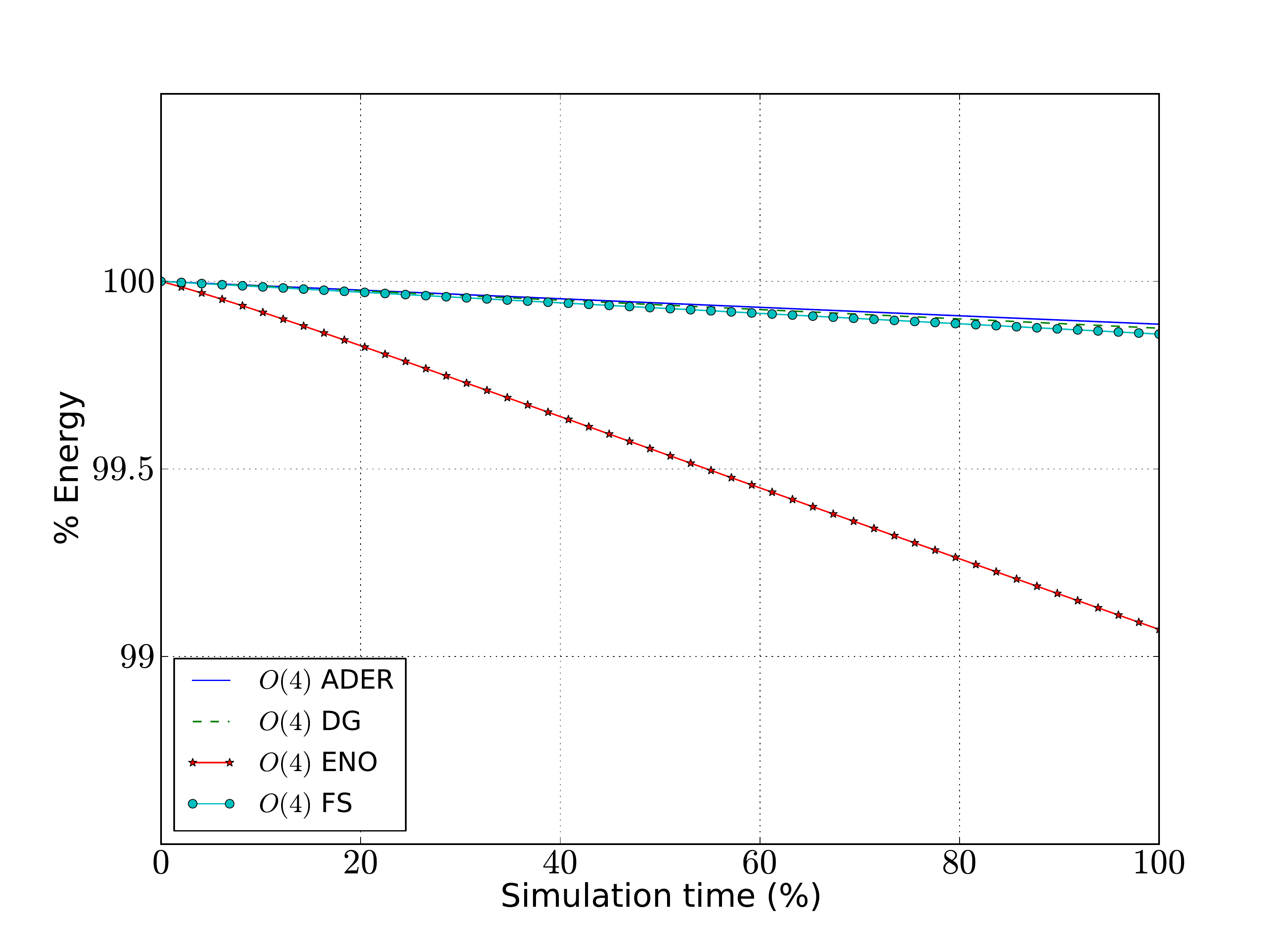}}
		\end{center}
		\caption{\label{energy4}Fourth-order}
	\end{subfigure}%
	\end{center}
	\caption{\label{energyall} Comparison of different numerical schemes with same DOFs}
\end{figure}

Figs.\ref{energy1}a-\ref{energy1}f show signal energy over time for various numerical schemes.
For computational problems involving propagating linear
waves, the ability of linear numerical schemes to preserve signal energy is important
similar to ability of numerical schemes to preserve TKE in turbulent flows \cite{Modesti2017,
Ladeinde2001}. 
Interestingly, numerical schemes with same formal spatial order of accuracy demonstrate 
differences in ability to preserve signal energy.

For a stable numerical scheme, 
	energy of the signal reduces over simulation time. 
	This is due to numerical diffusion added in the solution over time. 
From the results shown in Figs.\ref{energy1}a-\ref{energy1}f, 
higher-order schemes are found to be better at retaining total signal energy
	than the lower-order schemes.
Among all candidate schemes under consideration,
the DG scheme is seen to retain most signal energy over simulation time. 
	However, for the same number of finite elements, 
	higher-order DG schemes have more degrees of freedom (DOFs) than their
	finite volume counterparts. 

	Figs.\ref{energyall}a-\ref{energyall}c show results for various numerical schemes 
	with same number of overall DOFs. 
	It is observed that the DG scheme retains energy of the broadband signal 
	better than other schemes under consideration for the same number of  overall DOFs.
The ADER scheme using a Fixed Stencil (FS) reconstruction performs slightly better
	than a semidiscrete FS scheme of same order of spatial accuracy.
ENO and WENO schemes results in maximum loss of signal energy as seen from
	Figs.\ref{energy2},\ref{energy3} and \ref{energy4}.
	Thus, in spite of presence of spurious modes in the solution, the total energy
	reduces over computing time for ENO and WENO schemes.

\subsection{Effect of temporal order on total energy}
\begin{figure}[h!]
	\begin{subfigure}[t]{0.5 \textwidth}
		\begin{center}{\includegraphics[width=\textwidth]{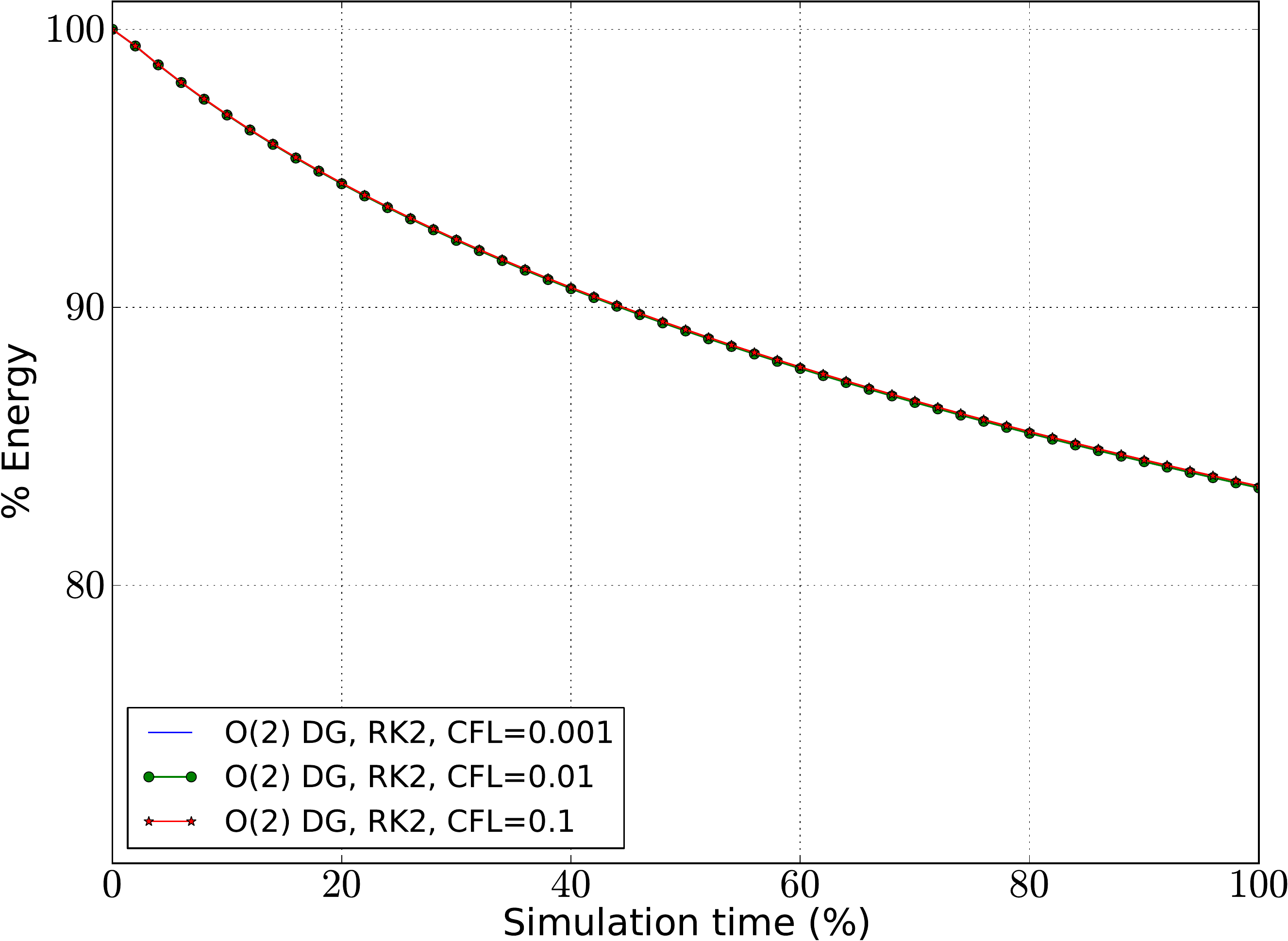}}
		\end{center}
		\caption{\label{energyo2rk2}$O(2)$ DG, RK2}
	\end{subfigure}%
	\begin{subfigure}[t]{0.5 \textwidth}
		\begin{center}{\includegraphics[width=\textwidth]{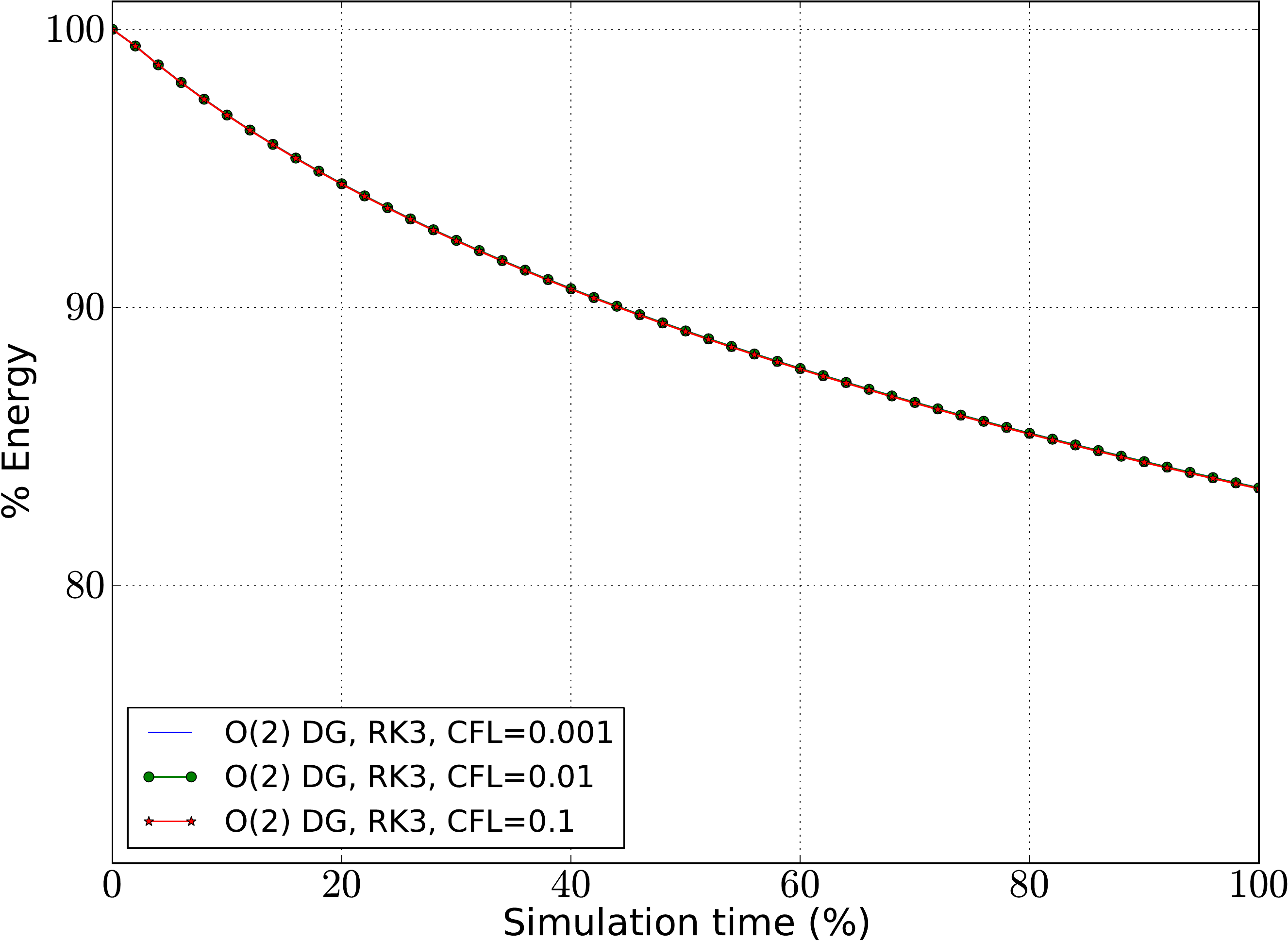}}
		\end{center}
		\caption{\label{energyo2rk3}$O(2)$ DG, RK3}
	\end{subfigure}%
	\newline
	\begin{subfigure}[t]{0.5 \textwidth}
		\begin{center}{\includegraphics[width=\textwidth]{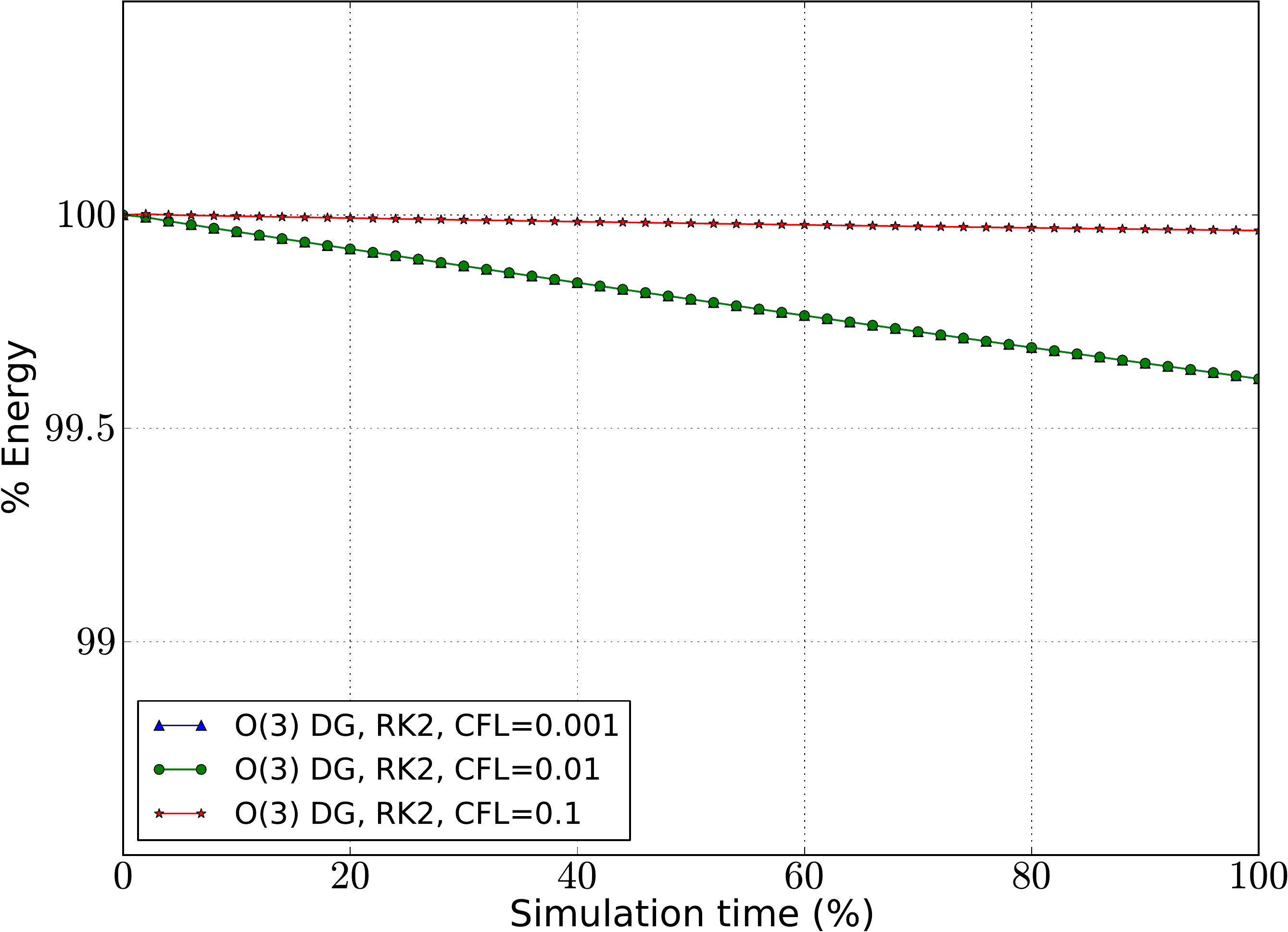}}
		\end{center}
		\caption{\label{energyo3rk2}$O(3)$ DG, RK2}
	\end{subfigure}%
	\begin{subfigure}[t]{0.5 \textwidth}
		\begin{center}{\includegraphics[width=\textwidth]{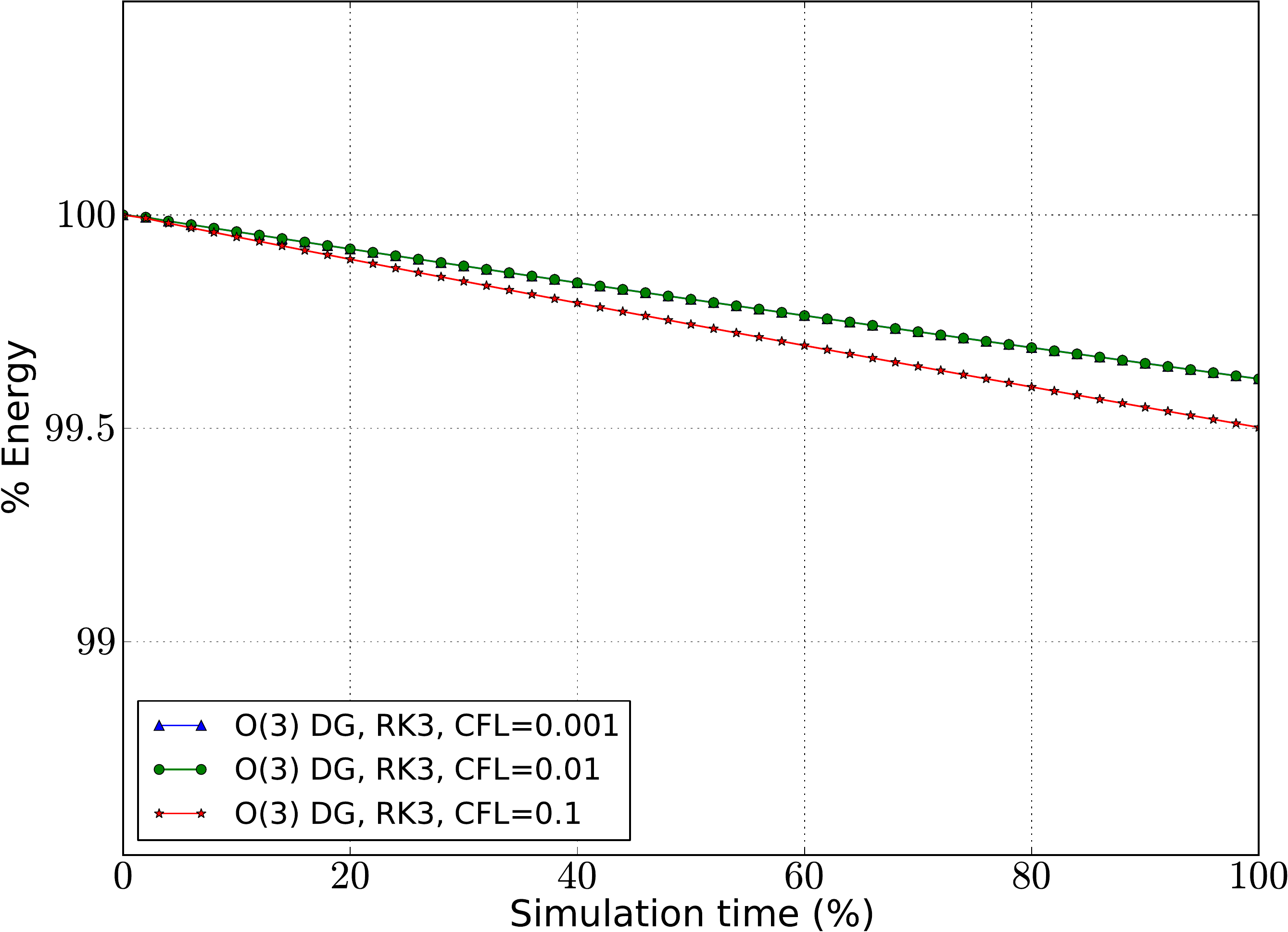}}
		\end{center}
		\caption{\label{energyo3rk3}$O(3)$ DG, RK3}
	\end{subfigure}%
	\newline
	\begin{subfigure}[t]{0.5 \textwidth}
		\begin{center}{\includegraphics[width=\textwidth]{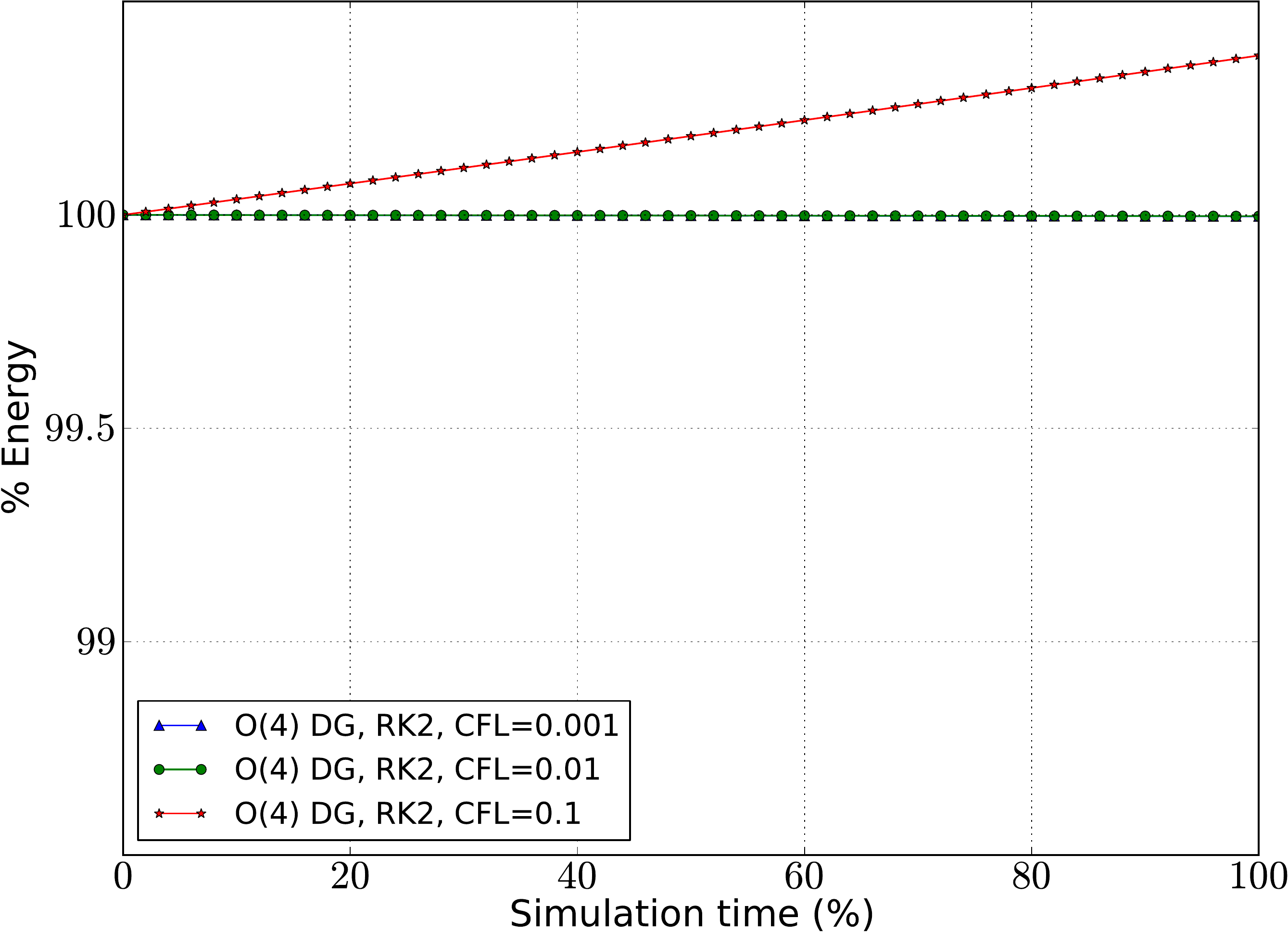}}
		\end{center}
		\caption{\label{energyo4rk2}$O(4)$ DG, RK2}
	\end{subfigure}%
	\begin{subfigure}[t]{0.5 \textwidth}
		\begin{center}{\includegraphics[width=\textwidth]{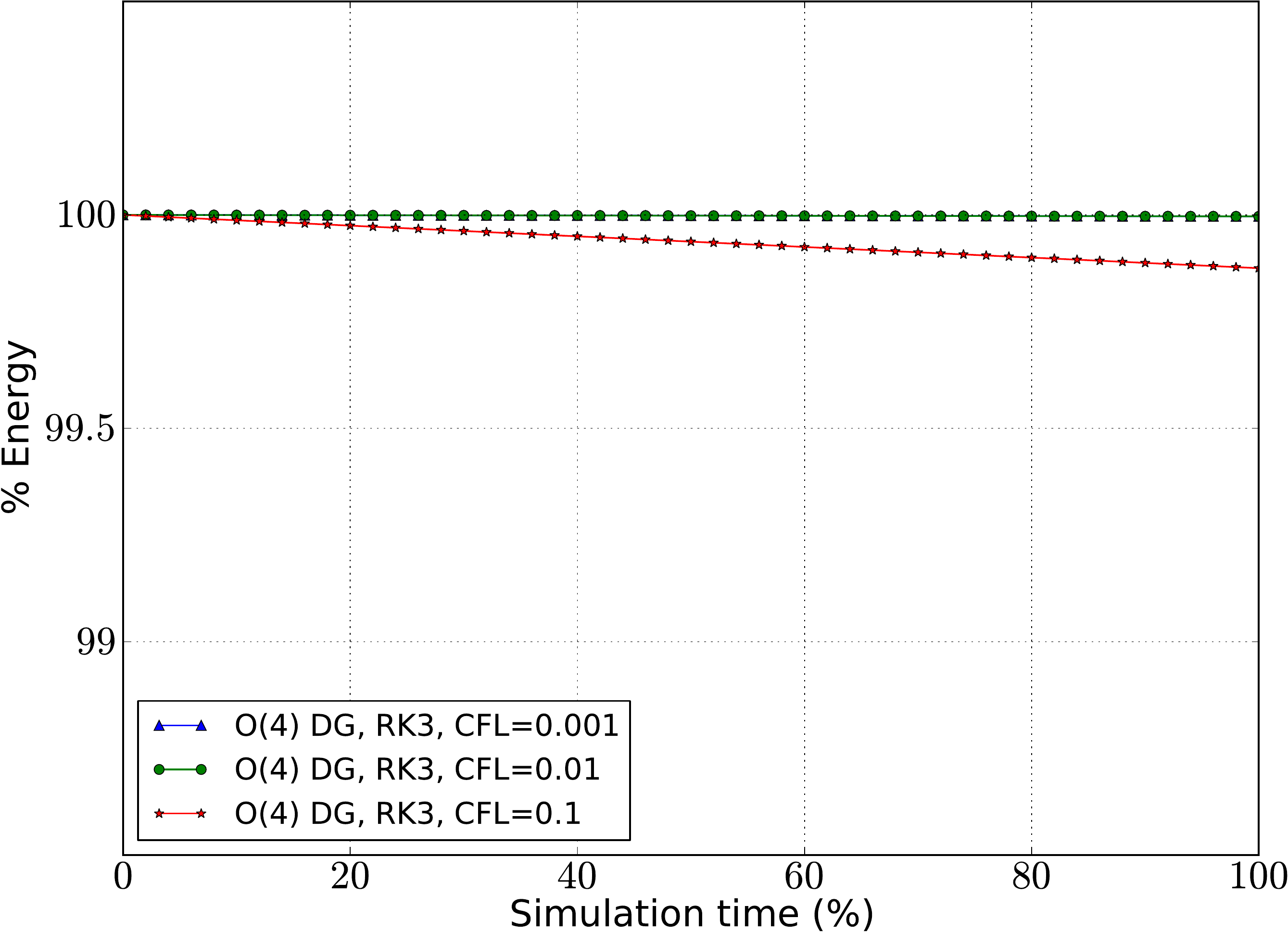}}
		\end{center}
		\caption{\label{energyo4rk3}$O(4)$ DG, RK3}
	\end{subfigure}%
	\caption{\label{energydtsize} Effect of time-step size on signal energy }
\end{figure}

In order to study effect of time-step size on energy of the signal, a semidiscrete discontinuous
Galerkin framework with Runge Kutta time-stepping is used. 
Three values of the Courant number ($0.1, 0.01$ and $0.001$) are used.
The simulation is run till the Gaussian pulse travels once round the complete domain and re-centers
at the origin.
Energy of the signal is computed at several intermediate time levels.
Plots of the total energy of the signal with simulation time are shown in Fig.\ref{energydtsize}.

If the spatial order of accuracy is lower than the temporal accuracy, then
  the overall error again mostly consists of spatial error. In such a case, reduction
  in time-step size has little or no effect on energy of the signal. 
  For example, the second-order DG scheme with RK3 time-stepping as shown in
  Fig.\ref{energyo2rk3}.
In case of both spatial and temporal orders of accuracy being similar,
  reduction in time-step size reduces temporal errors. After sufficient reduction in 
  the time-step size,
  spatial errors start dominating and further reduction in time-step size doesn't
  show changes in signal energy. Examples of this include a second-order DG scheme
  with RK2 time-stepping (Fig.\ref{energyo2rk2})
  and a third-order scheme with RK3 time-stepping (Fig.\ref{energyo3rk3}).
If the spatial order of accuracy is higher than the temporal order of accuracy,
then the overall error mostly consists of the temporal error. Reduction in 
time-step size in such a case results in reduction in the error initially. However,
after reaching a sufficiently small time-step, further reduction in dissipation error 
is not seen. This is observed in case of the third-order DG scheme with RK2
time integrator as shown in Fig.\ref{energyo3rk2} and a fourth order DG
scheme with RK2 and RK3 time integrators as seen in 
Figs.\ref{energyo4rk2} and \ref{energyo4rk3}.

It is interesting to see that the temporal error may add positive or negative 
  diffusion in the solution.
  For example, in case of a fourth-order DG scheme using a RK2 time-stepping 
  (Fig.\ref{energyo4rk2}),
  negative diffusion gets added to the solution indicating a growing instability.
For a higher-order scheme in space and time, such as the fourth-order DG 
scheme with RK3 time-stepping as shown in Fig.\ref{energyo4rk3}, 
change in the signal energy due to change in time-step size
  is negligible (less than $0.5\%$ over entire simulation length)
  due to smaller overall error as a result of higher-order accuracy.

\subsection{Onset of instability}
Instabilities can be detected very early through the application of the Fourier transform. 
For a conservative scheme, overall energy of the signal can increase 
only through formation and growth of spurious modes. 
To study this phenomenon, two numerical schemes are considered. 
\begin{enumerate}
	\item Second-order FV scheme with central flux differencing 
	\item First-order upwind FV scheme
\end{enumerate}
both with forward Euler time-stepping. The first example is that of an unconditionally unstable
numerical scheme \cite{leveque}. The second example consists of a conditionally stable numerical scheme.

\begin{figure}[H]
	\begin{subfigure}[t]{0.5\textwidth}
		\begin{center}{\includegraphics[scale=0.3]{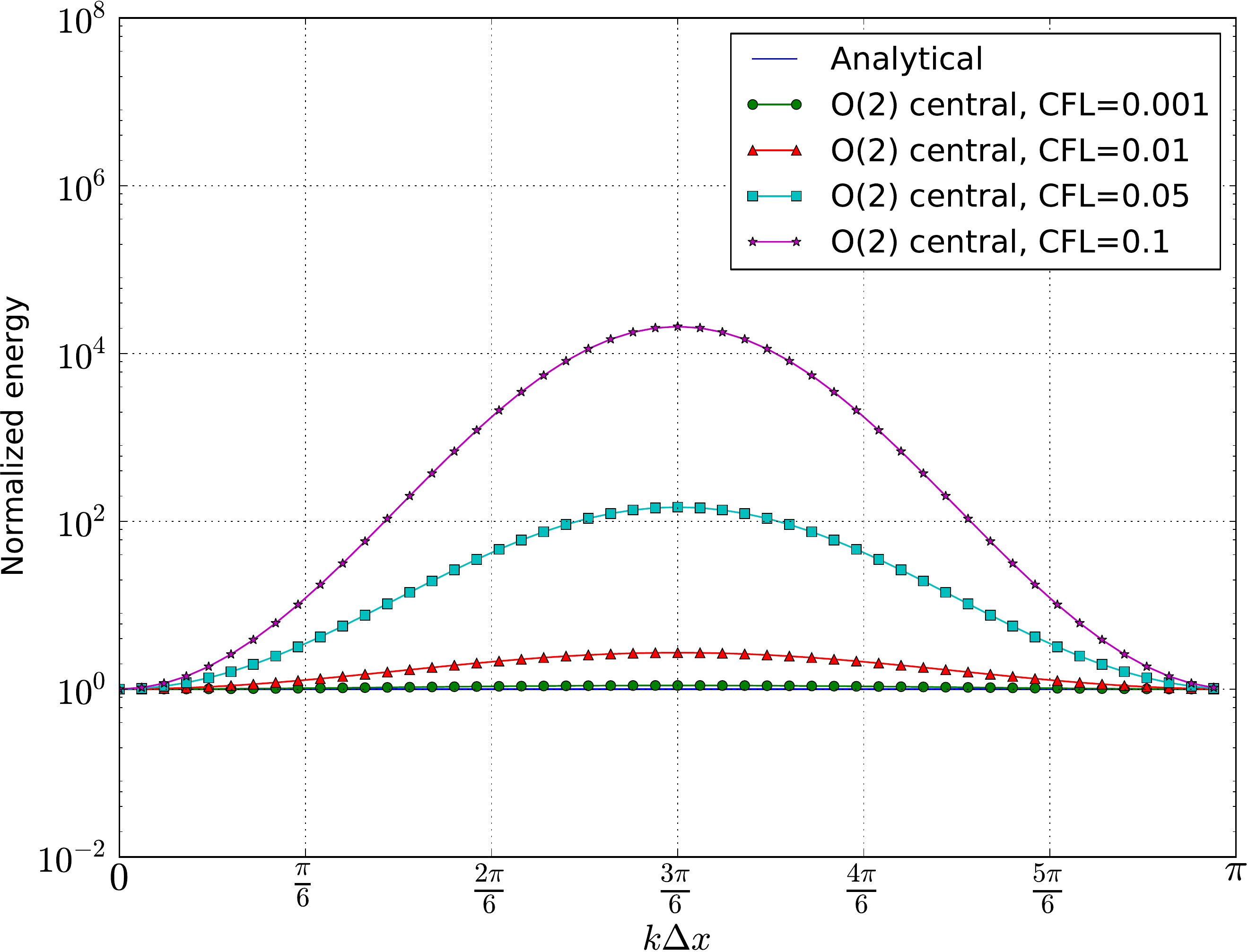}}
		\end{center}
	\caption{\label{inst1} Power spectral density}
	\end{subfigure}%
	\begin{subfigure}[t]{0.5\textwidth}
		\begin{center}{\includegraphics[scale=0.3]{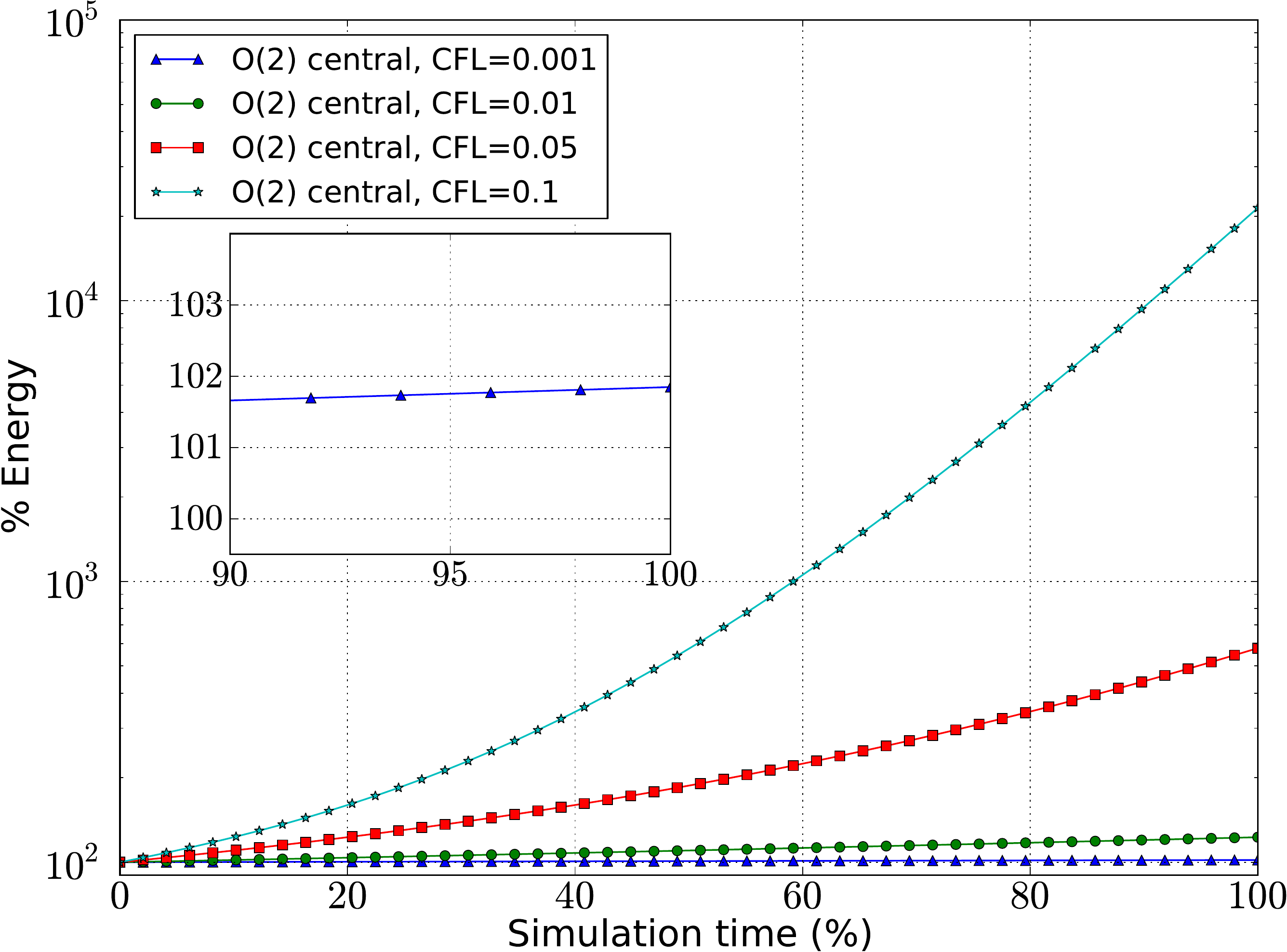}}
		\end{center}
	\caption{\label{inst2} Evolution of total energy}
	\end{subfigure}
	\caption{\label{insta} Second-order FV scheme with central flux differencing}
\end{figure}

		Fig.\ref{inst1} and \ref{inst2} show energy spectrum and evolution of total energy 
		for a second order 
		finite volume scheme using centered differencing. 
		Increase in energy of most of the modes are observed, as well as overall 
		energy of the signal is found to increase with time. 
		At very small time-steps, the energy added to the Fourier modes is also small 
		still indicating instability.

\begin{figure}[H]
	\begin{subfigure}[t]{0.5\textwidth}
		\begin{center}{\includegraphics[scale=0.3]{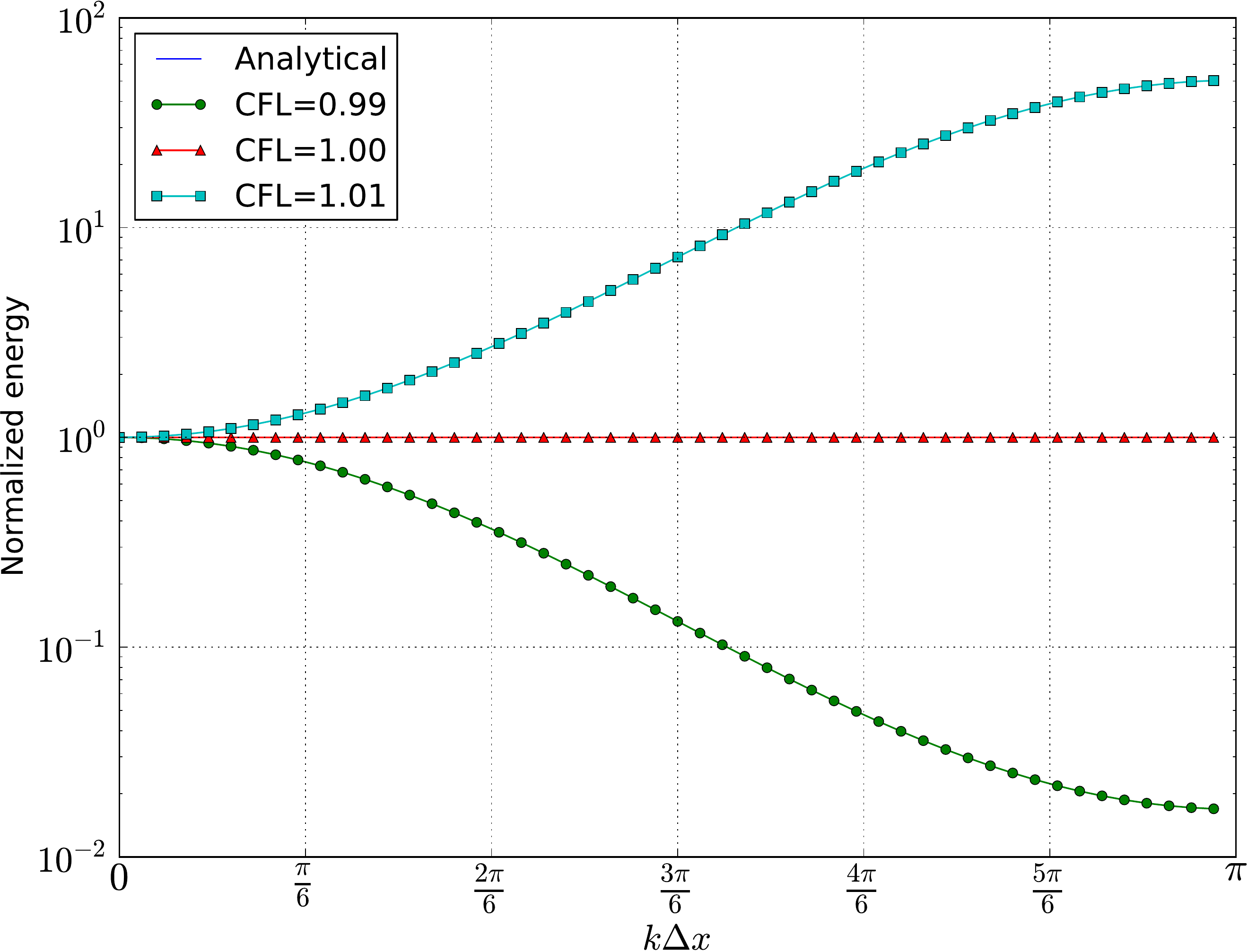}}
		\end{center}
	\caption{\label{inst3}  Power spectral density}
	\end{subfigure}
	\begin{subfigure}[t]{0.5\textwidth}
		\begin{center}{\includegraphics[scale=0.3]{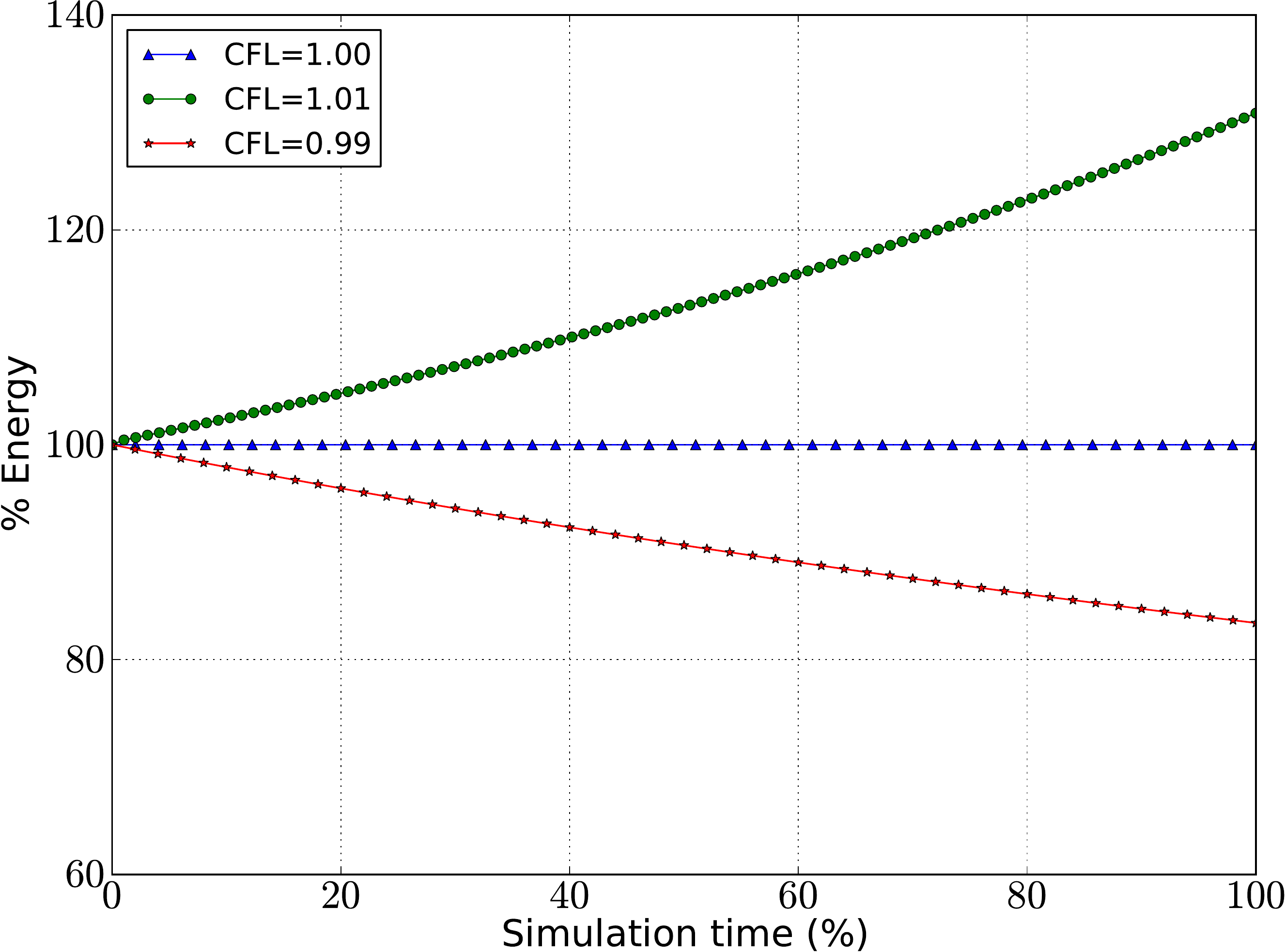}}
		\end{center}
	\caption{\label{inst4} Evolution of total energy}
	\end{subfigure}
	\caption{\label{instb} First-order upwind FV scheme}
\end{figure}

		Fig.\ref{inst3} and \ref{inst4} show energy spectrum and evolution of total signal
		energy for a first-order upwind 
		scheme with forward Euler time-stepping at different Courant numbers. 
		This scheme is unstable for Courant number greater than $1$. 
		Induction of anti-dissipation in some Fourier modes 
		as well as increase in overall energy of the signal over time 
		show a clear instability for such a condition (Courant number $= 1.01$). 
		For Courant number equal to $1$, 
		there is no dissipation or instability and for that less than $1$ ($0.99$ in this case)
		there is a positive dissipation added to the solution.

Increase in some of the Fourier modes does not always necessarily indicate instabilities. 
For example in nonlinear schemes such as the ENO and WENO schemes, the energy
corresponding to higher modes increases during the simulation implying addition of negative
dissipation in those modes. 
This can be seen in Fig. \ref{enowenotvsf}. 
However, overall signal energy still reduces with time as seen in Figs.\ref{teeno} and \ref{teweno},
which indicates that the cumulative dissipation added to the solution is positive.

\begin{figure}[H]
	\begin{center}
		\begin{center}{\includegraphics[scale=0.5]{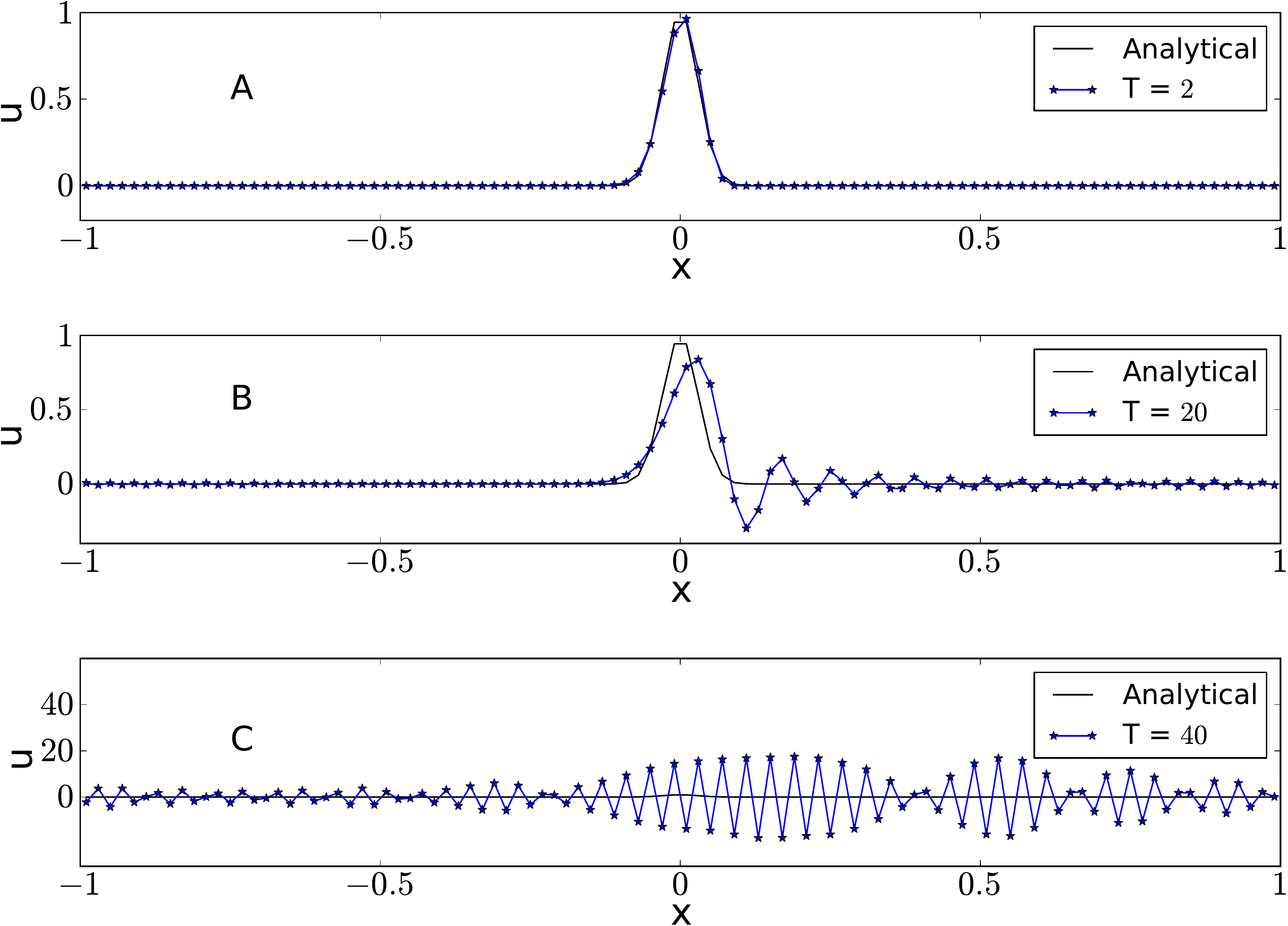}}
		\end{center}
	\caption{\label{inst5}  Time domain signal, fourth-order DG scheme}
	\end{center}
\end{figure}
\begin{figure}[H]
	\begin{center}
		\begin{center}{\includegraphics[scale=0.5]{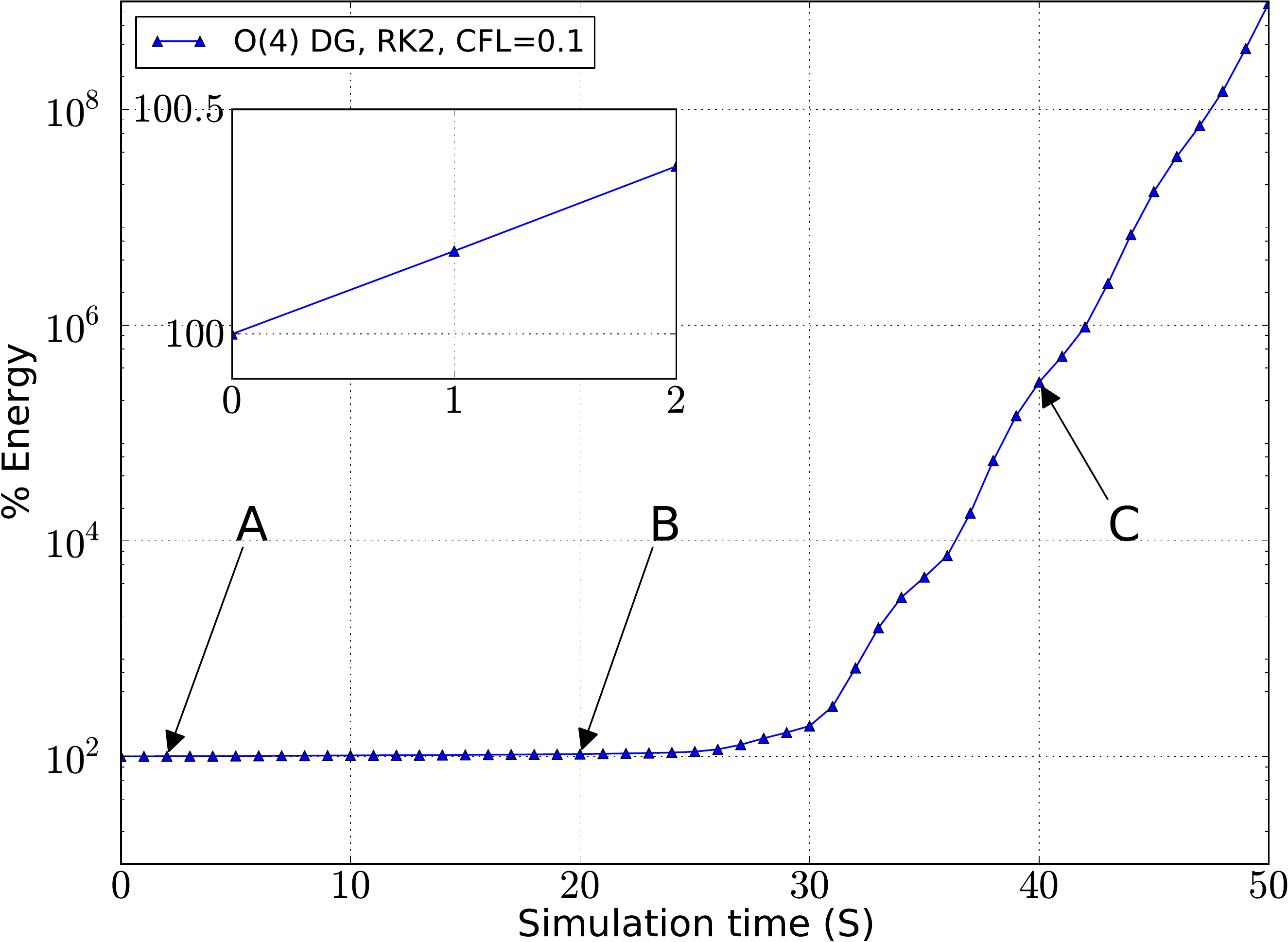}}
		\end{center}
	\caption{\label{inst6} Evolution of total energy, fourth-order DG scheme}
	\end{center}
\end{figure}

Consider the fourth-order DG scheme with RK2 timestepping.
The simulation is run with Courant number equal to $0.1$. 
The time evolution of the total energy of the signal for simulation-time $T = 2$ units
is shown in Fig.\ref{energyo4rk2}. 
Results imply addition of 
anti-dissipation in the solution, thus indicating onset of a numerical instability.
To verify this further, the simulation is run for a longer final-time $T$. 
The time domain results for $T=2$, $T=20$ and $T=40$ units are shown in 
Fig.\ref{inst5}.
The time-evolution of the total energy of the signal is shown in Fig.\ref{inst6}. 
Flags $A$, $B$ and $C$  in Fig.\ref{inst6} indicate the total energy corresponding to time $T=2, 
20$ and $40$ units respectively.
The evolution of the total energy for the first $2$ units of time
is shown in an inset in Fig.\ref{inst6}, which is also identical to Fig.\ref{energyo4rk2}.
It is noted from Fig.\ref{inst5} that, 
the time-domain signal does not show signs of an instability for time $T=2$ units.
The instability grows nonlinear over time and the noise becomes comparable to the signal
at around $T=20$ units. 
However, onset of an instability can be 
identified from the total-energy diagram (Fig.\ref{inst6}) within the first few iterations.

\section{Wave resolution}

\begin{figure}[h!]
	\begin{center}{\includegraphics[scale=0.45]{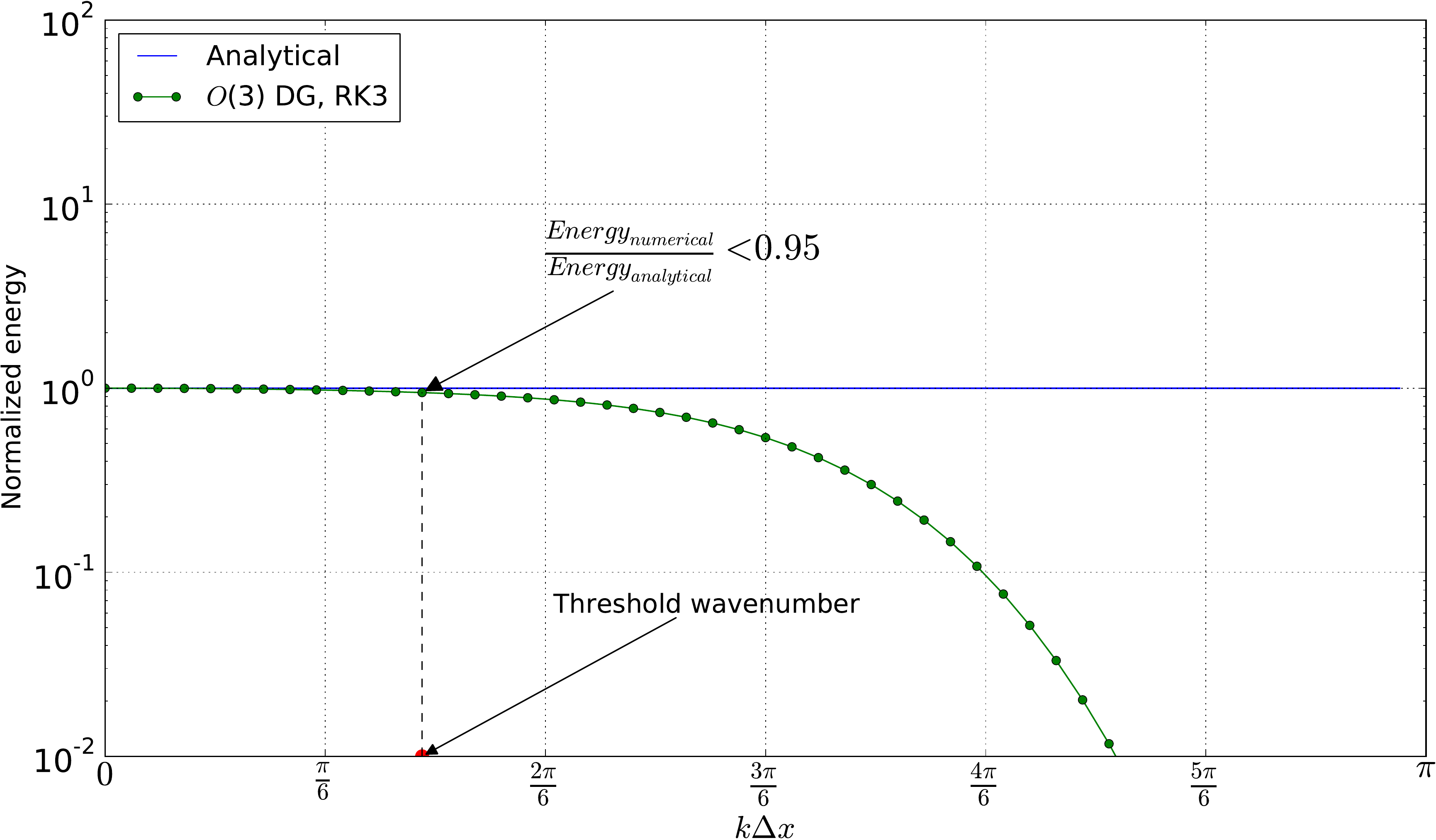}}
	\end{center}
	\caption{\label{twnbr} Maximum resolved wavenumber for a third-order DG scheme }
\end{figure}
\begin{figure}[h!]
	\begin{center}{\includegraphics[scale=0.45]{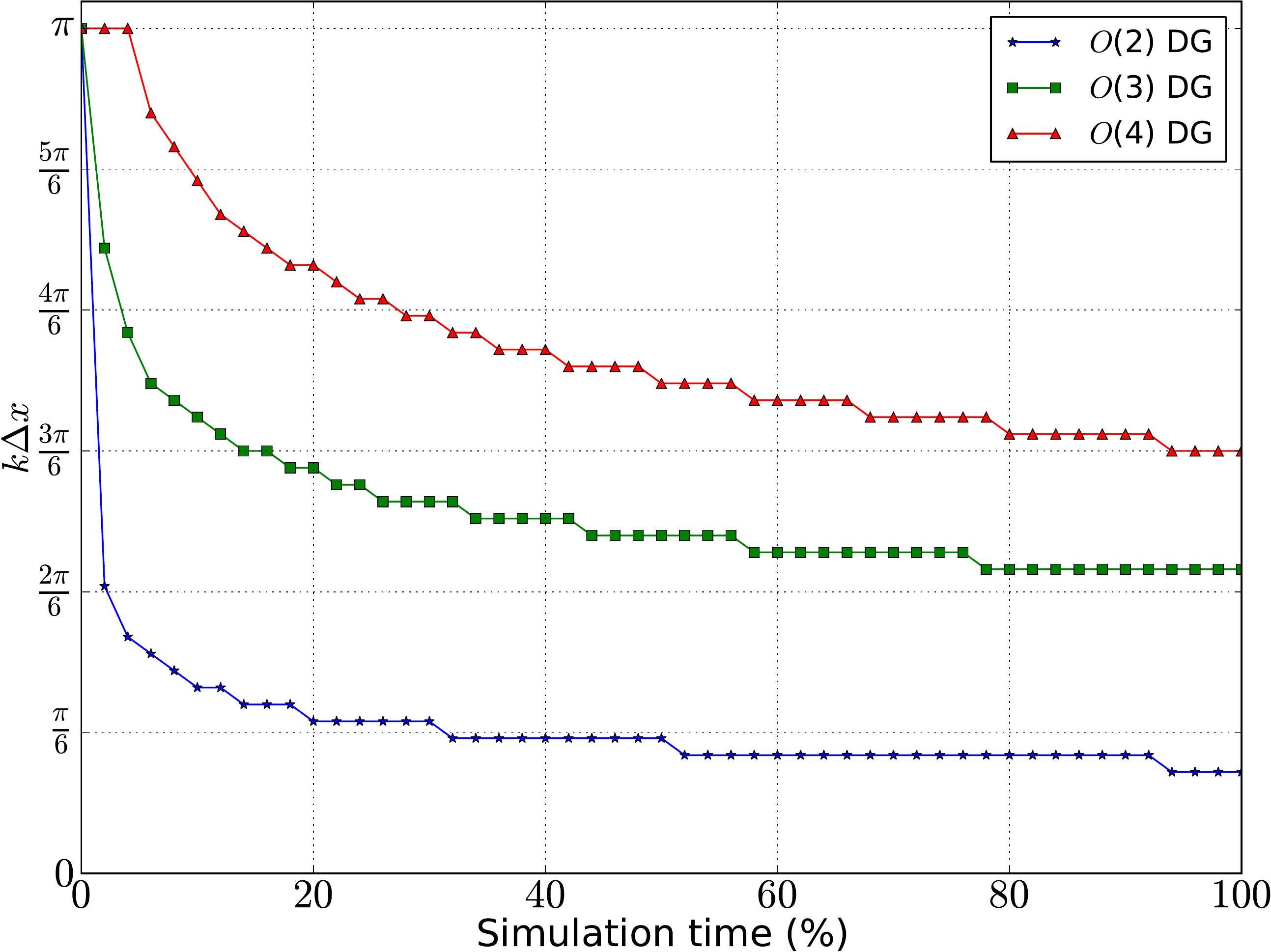}}
	\end{center}
	\caption{\label{th1} Maximum resolved wavenumber for DG schemes over simulation time}
\end{figure}

\begin{figure}[h!]
	\begin{center}{\includegraphics[scale=0.45]{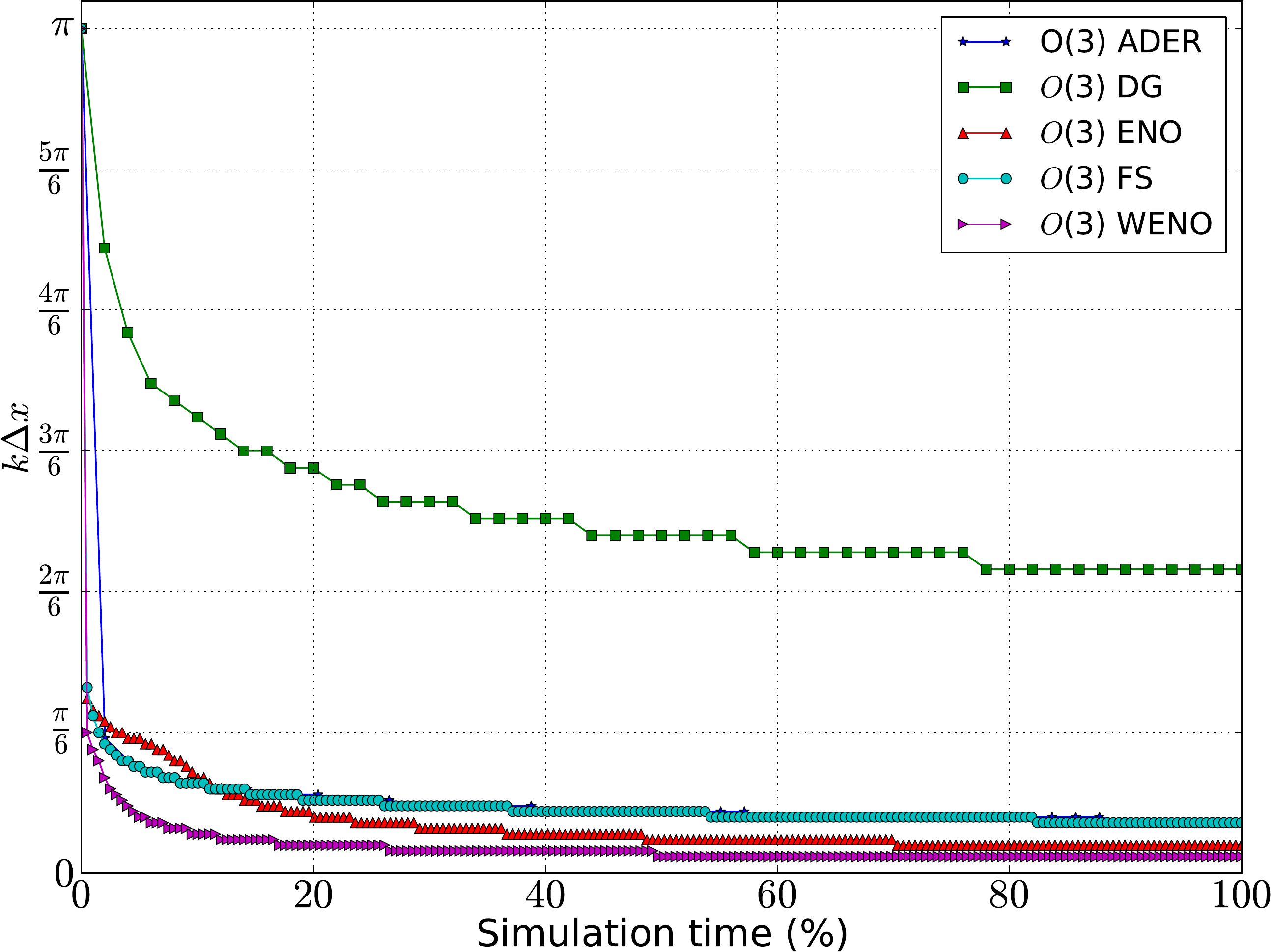}}
	\end{center}
	\caption{\label{th3}Maximum resolved wavenumber ($k\Delta x$) for third-order numerical schemes}
\end{figure}

\begin{figure}[h!]
	\begin{center}{\includegraphics[scale=0.45]{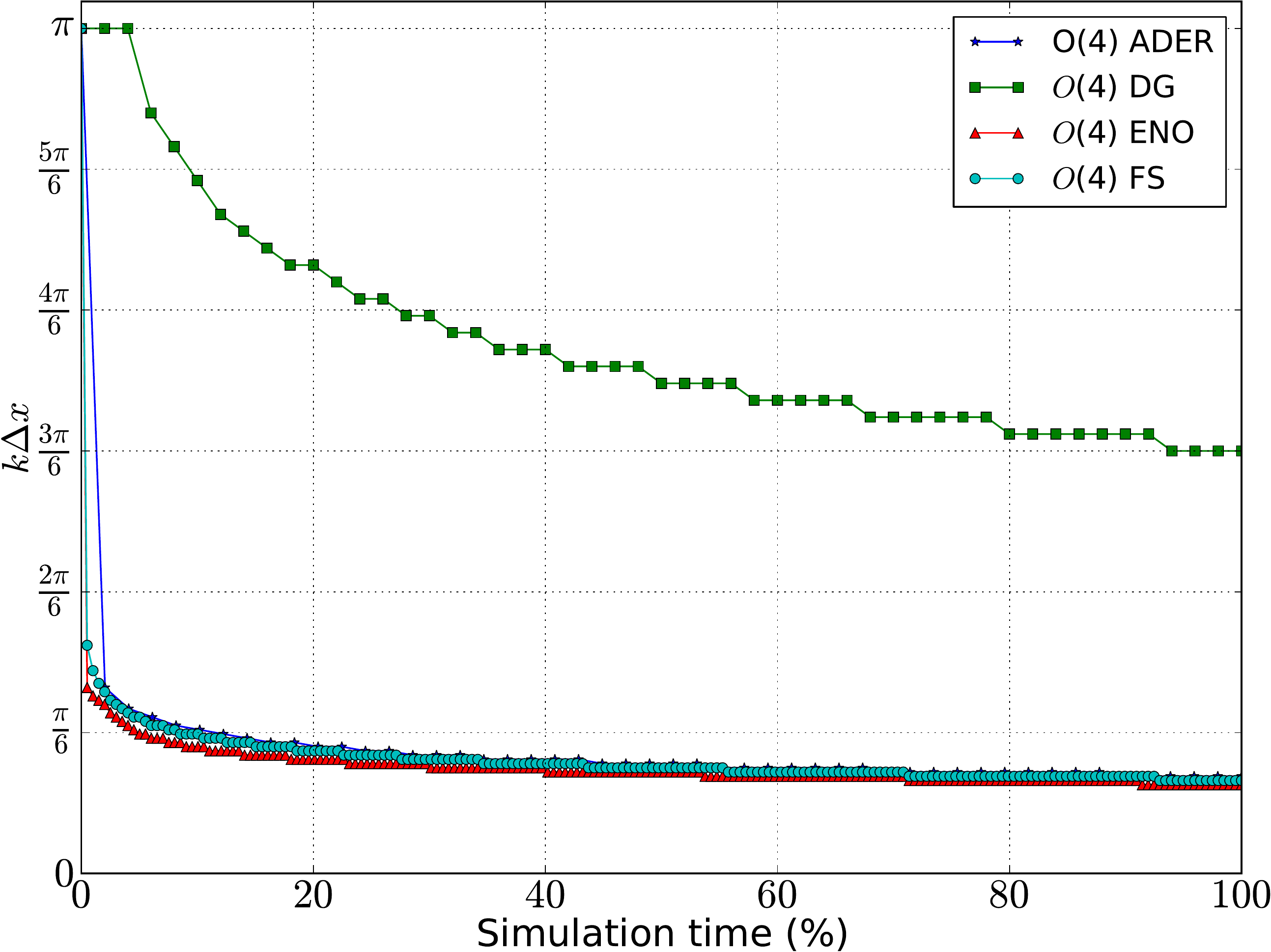}}
	\end{center}
	\caption{\label{th4}Maximum resolved wavenumber ($k\Delta x$) for fourth-order numerical schemes}
\end{figure}

Numerical experiments indicate that 
diffusion starts affecting higher wavenumbers first.
As the simulation progresses in time, diffusion progressively starts affecting lower wavenumbers.
For all numerical schemes, the numerical PSD curve deviates from the analytical curve at a 
threshold wavenumber. 
The threshold wavenumber can be arbitrarily defined. In this study, the threshold
wavenumber is defined as the wavenumber at which modal energy of the numerical solution 
is lower than $95\%$ of that of the analytical solution.
Fig.\ref{twnbr} shows the threshold wavenumber for a third-order DG scheme. 
The threshold wavenumber is found to vary with time.
For the given Gaussian signal, higher wavenumbers contain relatively less energy.
Thus, reduction in overall energy starts sooner if dissipation starts affecting
lower wavenumbers early.
A plot of threshold wavenumber versus simulation time directly correlates with
evolution of overall signal energy in time.
As expected this threshold wavenumber reduces with 
increase in simulation time. However, the point of deviation is different 
for different numerical schemes even if they have same formal order of accuracy.
 All wavenumbers lower than this threshold wavenumber are well resolved.

Fig.\ref{th1} shows the threshold wavenumber for DG schemes with third-order Runge Kutta
time-stepping. 
	 As expected, higher-order variants show higher threshold wavenumber.
Fig.\ref{th3} and Fig.\ref{th4} show threshold wavenumber as a function of time for
different numerical schemes.
Due to a limited number of DOFs, the nearest resolved wavenumber is selected in all simulations.
All schemes have identical simulation parameters including same number of overall DOFs.
	DG schemes resolve higher wavenumbers for same number of DOFs as compared to 
		other schemes under consideration. This property, along with ability to retain
		signal energy over long simulation time makes DG schemes an excellent 
		candidate for simulations involving traveling linear waves over large computational
		domain.
	ADER scheme with fixed stencil (FS) reconstruction show similar diffusion
		characteristics as a semidiscrete FV scheme with FS reconstruction.
	ENO and WENO schemes start showing diffusion addition in lower wavenumbers
		earlier than rest of the candidate schemes. Thus, 
		ENO and WENO schemes may not be a perfect choice for traveling linear waves 
		over a long computational domain.

\section{Conclusion}
In this paper, a new technique for qualitative analysis of diffusion characteristics of 
numerical schemes is presented. 
The frequency representation of the numerical solution is obtained using 
the DFT technique. Numerical schemes are compared based on modal energy content and 
time evolution of total energy of the broadband signal. 
The analysis is applicable to semi-discrete as well as space-time coupled linear and non linear 
schemes used for simulating propagating linear waves.
The analysis yields information on
properties of a numerical scheme like conservation, ability to preserve signal energy
and overall diffusion characteristics.
	
Numerical schemes with same formal spatial order of 
accuracy may have very different diffusion characteristics.
Diffusion is seen to get added initially to higher wavenumber modes and then
progressively shift to the lower end of the Fourier spectrum. 
Consequently, the threshold wavenumber at which the numerical PSD curve deviates
from the analytical curve shifts to lower end of the Fourier spectrum over time. 
Numerical instabilities can be identified at an early stage through this analysis. 
For a stable numerical scheme, time-step size is seen to have a 
small effect on amount of diffusion added
to the solution.

Some popular higher-order accurate numerical schemes were analyzed using this technique.
Discontinuous Galerkin (DG) schemes are found to best preserve maximum signal energy 
over time. On the other hand, 
nonlinear schemes such as ENO and WENO schemes show  
maximum loss of total energy of the signal.
Nonlinear schemes are also characterized by presence of spurious modes in the frequency domain.
The space-time coupled ADER methods are seen to perform slightly better than 
their semidiscrete counterparts using a similar reconstruction method.

\section*{Acknowledgement}
We would like to thank Dr. Praveen Chandrashekar (TIFR-CAM, Bangalore) 
for the discontinuous-Galerkin code.

\bibspacing=\dimen 100
\bibliographystyle{unsrt}
\bibliography{Reference-FFT_journal}

\end{document}